\documentclass[12pt]{article}
\usepackage{hyperref} 
\hypersetup{colorlinks=true, allcolors=black} 
\usepackage{amsfonts,amsmath,amsxtra,amsthm}
\usepackage{amssymb}
\usepackage{bm}
\usepackage[utf8]{inputenc}
\usepackage{lscape}
\usepackage[normalem]{ulem}
\usepackage[russian, english]{babel}
\usepackage{xcolor}
\usepackage{pst-circ}
\usepackage{tikz, subfig, pgfplots, ifthen}
\usetikzlibrary{decorations.markings}
\usetikzlibrary{arrows.meta}
\usepackage{tcolorbox}
\usepackage{mathrsfs}
\usepackage{cancel}

\def\a{\alpha}

\def\C{\mathbb{C}}
\def\b{\beta}

\def\beq{\begin{equation}}
\def\eeq{\end{equation}}
\def\beqq{\begin{equation*}}
\def\eeqq{\end{equation*}}

\def\bmm{\bm{m}}

\def\bs{\begin{split}}
	\def\es{\end{split}}

\def\bi{\bm{i}}
\def\bj{\bm{j}}
\def\bl{{\boldsymbol{\lambda}}}

\def\bll{\boldsymbol{l}}
\def\bmm{\bm{m}}
\def\bo{\boldsymbol{\omega}}
\def\bu{\boldsymbol{u}}
\def\bv{\boldsymbol{v}}
\def\bx{\boldsymbol{x}}
\def\by{\boldsymbol{y}}
\def\bz{\boldsymbol{z}}
\def\bk{\boldsymbol{k}}

\def\bbx{\underline{\boldsymbol{x}}}
\def\bby{\underline{\boldsymbol{y}}}
\def\bbz{\underline{\boldsymbol{z}}}

\def\bw{\boldsymbol{w}}

\def\const{2\pi\imath}
\def\d{\partial}
\def\e{\bar{e}}

\def\Im{\operatorname{Im}}
\def\g{\gamma}

\def\k{\bar{k}}
\def\K{{K}}

\def\l{\lambda}

\def\n{\mathfrak{n}}
\def\nn{[n]}

\def\o{\omega}
\def\p{\mathbf{p}}
\def\Ph{{\Phi}}
\newcommand{\Pochab}[3]{[#1|\bo]_{#2,#3}}
\newcommand{\Pocha}[2]{[#1|\o_1]_{#2}}
\newcommand{\Pochb}[2]{[#1|\o_2]_{#2}}
\def\R{\mathbb{R}}
\def\Re{\mathrm{Re}\,}
\def\S{S_2}
\def\Res{\operatorname{Res}}

\def\tq{\tilde{q}}

\def\tS{\tilde{S}}
\def\tx{\tilde{x}}
\def\ty{\tilde{y}}

\def\ve{\varepsilon}
\def\vf{\varphi}

\def\x{\bm{x}}
\def\X{{\cal U}}
\def\XL{\operatorname{U}}

\def\YR{\operatorname{V}}
\def\z{\bm{z}_n}
\def\Z{\mathbb{Z}}

\newtheorem{lemma}{Lemma}
\newtheorem{proposition}{Proposition}
\newtheorem{corollary}{Corollary}
\newtheorem{theorem}{Theorem}

\newcommand{\rf}[1]{(\ref{#1})}

\newcommand{\LL}[1]{\Lambda_{#1}}

\newcommand{\Q}[1]{Q_{#1}}

\begin{document}
\-\vspace{-2cm}

\begin{center}
{\bf \large   Baxter operators in  Ruijsenaars hyperbolic system  I.\\[4pt] Commutativity of $Q$-operators}
\bigskip

{\bf  N. Belousov$^{\dagger\times}$, S. Derkachov$^{\dagger\times}$, S. Kharchev$^{\bullet\ast}$, S. Khoroshkin$^{\circ\ast}$
}\medskip\\
$^\dagger${\it Steklov Mathematical Institute, Fontanka 27, St. Petersburg, 191023, Russia;}\smallskip\\
$^\times${\it National Research University Higher School of Economics, Soyuza Pechatnikov 16, \\St. Petersburg, 190121, Russia;}\smallskip\\
$^\bullet${\it National Research Center ``Kurchatov Institute'', 123182, Moscow, Russia;}\smallskip\\
$^\circ${\it National Research University Higher School of Economics, Myasnitskaya 20, \\Moscow, 101000, Russia;}\smallskip\\
$^\ast${\it Institute for Information Transmission Problems RAS (Kharkevich Institute), \\Bolshoy Karetny per. 19, Moscow, 127994, Russia}
\end{center}

\begin{abstract}
\noindent We introduce Baxter $Q$-operators for the quantum Ruijsenaars hyperbolic system. We prove that they represent a commuting family of integral operators and also commute with Macdonald difference operators, which are gauge equivalent to the Ruijsenaars Hamiltonians of the quantum system. The proof of commutativity of the Baxter operators uses a hypergeometric identity on rational functions that generalize Ruijsenaars kernel identities.
\end{abstract}

\tableofcontents

\section{Introduction}

\subsection{Ruijsenaars system} In this paper we develop the theory of Baxter operators for relativistic hyperbolic Ruijsenaars system~\cite{R1}. This model is parametrized by three positive constants
$\o_1, \o_2$ (``periods'') and $g$ (coupling constant), subject to the relation
\beq\label{I3a} 0<g<\o_1+\o_2.	\eeq
The dual coupling constant
\beq\label{I3b} g^\ast=\o_1+\o_2-g\eeq
is used as well everywhere.

The Ruijsenaars system is governed by commuting symmetric difference operators
\beq \label{I2} H_r(\bx_n,g|\bo)=\sum_{\substack{I\subset[n] \\ |I|=r}}
\prod_{\substack{i\in I \\ j\notin I}}
\frac{\sh^{\frac{1}{2}}\frac{\pi}{\o_2}\left(x_i-x_j-\imath g\right)}
{\sh^{\frac{1}{2}}\frac{\pi}{\o_2}\left(x_i-x_j\right)}
\cdot T^{-\imath\o_1}_{I,x}\cdot
\prod_{\substack{i\in I \\ j\notin I}} \frac{\sh^{\frac{1}{2}}\frac{\pi}{\o_2}\left(x_i-x_j+\imath g\right)}
{\sh^{\frac{1}{2}}\frac{\pi}{\o_2}\left(x_i-x_j\right)}
\eeq
acting on meromorphic functions of $n$ complex variables analytic in the strip
\beqq |\Im x_i|<\o_1+\ve, \qquad \ve>0. \eeqq
Here and in what follows we denote tuples of $n$ variables as
\beq \bx_n=(x_1,\ldots,x_n). \eeq
The sum in \eqref{I2} is taken over all subsets
\beq\label{I0} I \subset [n] = \{1, \dots, n\}\eeq
of cardinality $r$. By $T^{a}_{x_i}$
we denote shift operators
\beq \label {I3}
T^{a}_{x_i}:=e^{a\d_{x_i}}, \qquad (T^{a}_{x_i} f)(x_1,\ldots,x_i,\ldots,x_n)=
f(x_1,\ldots,x_i+a,\ldots,x_n),
\eeq
with $T^a_{I,x}$ being their product
\begin{align}\label{Ta}
T^{a}_{I,x}=\prod_{i\in I} e^{a\d_{x_{i}}}
\end{align}
for any subset $I\subset[n]$.

The Ruijsenaars  operators \rf{I2} are closely related to Macdonald operators
\beq \label{I2a}
M_r(\bx_n;g|\bo)=\sum_{\substack{I\subset[n] \\ |I|=r}}
\prod_{\substack{i\in I \\ j\notin I}}
\frac{\sh\frac{\pi}{\o_2}\left(x_i-x_j- \imath g\right)}
{\sh\frac{\pi}{\o_2}\left(x_i-x_j\right)}
\cdot T^{-\imath\o_1}_{I,x} .
\eeq
Namely, denote by $\mu(z|\bo)$ the function
\beq\label{I4a}\mu(z|\bo)=S_2(\imath z|\bo)
S_2(-\imath z+g^\ast|\bo)\eeq
and by $\mu(\bx_n|\bo)$
the product
\beq\label{I5}
\mu(\bx_n|\bo)=\prod_{\substack{i,j=1 \\ i\neq j}}^n\mu(x_i-x_j).\eeq
Here $S_2(z|\bo)$ is the double sine function, its definition and key properties are given in Appendix \ref{AppendixA}. Then
\beq\label{I4}
\sqrt{\mu(\bx_n|\bo)} \,
M_r(\bx_n;g|\bo)\, \frac{1}{\sqrt{\mu(\bx_n|\bo)}}=
H_r(\bx_n,g|\bo).
\eeq 
Note that the function $\mu(\bx_n|\bo)$ is non-negative (assuming real constants $g, \o_1, \o_2$), since 
\begin{equation}
\begin{aligned}
\mu(x|\bo) \mu(-x|\bo) &= S_2(\imath x|\bo) S_2(-\imath x |\bo) S_2(\imath x+g^* |\bo) S_2(-\imath x + g^* |\bo) \\[6pt]
&= \bigl| S_2(\imath x|\bo) S_2(\imath x+g^* |\bo) \bigr|^2.
\end{aligned}
\end{equation} 
Ruijsenaars operators are symmetric with respect to the pairing
  \beqq (\vf,\psi)=\int_{\R^n}\vf(\bx_n)\bar{\psi}(\bx_n)d\bx_n, \eeqq
  	 while the
Macdonald operators are symmetric with respect to the pairing
\beq\label{I5a} (\vf,\psi)=\int_{\R^n}\vf(\bx_n)\bar{\psi}(\bx_n)\mu(\bx_n)d\bx_n .\eeq
\subsection{Kernel function and kernel identities}
Unlike the original Ruijseenars' setting, we do not suppose that the periods $\o_1, \o_2$ and the coupling constant $g$ are real positive. Instead, we assume everywhere that all of them are complex numbers with positive real parts
\beq\label{I0a} \Re \o_1>0, \qquad \Re \o_2>0,\qquad  \Re g>0. \eeq
We also require the condition
\beq\label{I0b} 0<\Re g<\Re\o_1+\Re\o_2.\eeq
Further we usually fix the periods $\bo=(\o_1,\o_2)$ and for brevity skip them in the notation, for example we use the symbol $S_2(z)$ for the double sine function
$$S_2(z):=S_2(z|\bo).$$
Denote by $\K(z)$ the following function of a complex variable
\beq\label{I6} \K(z)=\S^{-1}\left(\imath z+\frac{g^\ast}{2}\right)\S^{-1}\left(-\imath z+\frac{g^\ast}{2}\right).\eeq
In terms of the Ruijsenaars hyperbolic Gamma function
\beq\label{I6a} G(z)=S_2\left(\imath z+\frac{\o_1+\o_2}{2}\right)\eeq
using reflection formula for the double sine function \eqref{trig4b} it can be written as
\beq\label{I6b}
\K(z)=G\left(z-\frac{\imath g}{2}\right)G^{-1}\left(z+\frac{\imath g}{2}\right).
\eeq
Also let
\beqq\bz_n=(z_1,\ldots,z_n),\qquad \by_n=(y_1,\ldots, y_n), \qquad z_i,y_i\in\C\eeqq
be two tuples of $n$ complex variables. The Ruijsenaars {\bf kernel function} $\K(\bz_n,\by_n)$  is
defined as a product
\beq\label{I7} \K(\bz_n,\by_n)=\prod_{i,j=1}^n \K(z_i-y_j).
\eeq
The kernel function $K(\bz_n,\by_n)$ satisfies the relations
\beq \label{I8} \left(M_r(\bz_n;g)- M_r(-\by_n;g)\right)K(\bz_n,\by_n)=0, \qquad r=1, \dots, n,\eeq
that is $K(\bz_n,\by)$ is a zero value eigenfunction for commuting difference operators
\beqq M_r(\bz_n;g)- M_r(-\by_n;g),
\eeqq
see \cite{R0}. The relations \rf{I8} are the corollary of the trigonometric version of {\bf kernel function identity} \cite{R0}, valid for any tuples $\bz_n$ and $\by_n$ of $n$ complex variables and arbitrary parameter~$\a$:
\beq\begin{split}\label{I9}
\sum_{\substack{I\subset[n] \\ |I|=r}}\prod_{i\in I}\left(\prod_{j\in [n]\setminus I}\frac{\sin(z_i-z_j-\a)}{\sin(z_i-z_j)}\prod_{a=1}^{n}\frac{\sin(z_i-y_a+\a)}{\sin(z_i-y_a)}\right)
=\\ \sum_{\substack{A\subset[n] \\ |A|=r}}\prod_{a\in A}\left(\prod_{b\in [n]\setminus A}
\frac{\sin(y_a-y_b+\a)}{\sin(y_a-y_b)}\prod_{i=1}^{n}\frac{\sin(z_i-y_a+\a)}{\sin(z_i-y_a)}\right).
\end{split}\eeq
In the following we also use the kernel function with the second argument being a tuple of $n-1$ complex variables
\beq \label{I9a}\K(\bz_n,\by_{n-1})=\prod_{i=1}^n \prod_{j=1}^{n-1}\K(z_i-y_j).
\eeq

\subsection{Baxter $Q$-operators} Let us introduce the family of Baxter $Q$-operators $\Q{n}(\l)$ parameterized by $\lambda \in \mathbb{C}$ as the integral operators
\beq\label{I14}\begin{split} &\left(\Q{n}(\l)f\right)(\bz_n)= \int_{ \R^{n}}Q(\bz_n,\by_{n};\l) f(\by_{n}) d\by_{n}
\end{split}\eeq
with the kernel
\beq\label{I15} Q(\bz_n,\by_{n};\l)= e^{\const \l(\bbz_n-\bby_n)}
\K(\bz_n,\by_{n})\mu(\by_{n})\,, \qquad z_j, y_j \in \R.
\eeq
Here and in what follows we denote the sum of tuple components as
$$\bbz_{n}=z_1+\ldots +z_{n}.$$
The $Q$-operator maps functions of  $n$ real variables to functions of $n$ real variables.

The following theorem is a simple consequence of kernel function identities \rf{I9}, its proof is given in Section \ref{BM}.
 \begin{theorem}\label{theoremBM}  Under  the condition
 	\beq\label{I15a} 0<\Re g<\Re \o_2\eeq
 	the operators $\Q{n}(\l)$ commute with Macdonald operators $M_r(\bz_n;g|\bo)$
 \beq\label{MQ}	M_r(\bz_n;g) \, \Q{n}(\l)= \Q{n}(\l) \,	M_r(\bz_n;g),\qquad r=1,\ldots, n.\eeq
 	\end{theorem}

Assume in addition to \rf{I0a}, \rf{I0b} that
\beq\label{I17a} \nu_g = \Re\frac{g}{\o_1\o_2}>0.\eeq
Due to the bounds for the functions $K(y)$ and $\mu(y)$ \eqref{B3} proven in Appendix \ref{AppendixB}, the product of two $Q$-operators
\beqq \Q{n}(\l) \, \Q{n}(\rho) \eeqq
is a well defined integral operator with the kernel $\Q{n}(\bz_n,\bw_n;\l,\rho)$ given by absolutely convergent integral
\beq \label{I17}
\Q{n}(\bz_n,\bw_n;\l,\rho)=\int_{\R^n}Q(\bz_n,\by_n;\l)Q(\by_n,\bw_n;\rho)d\by_n
\eeq
and the domain that consists of fast decreasing functions $f(\bw_n)$, see Proposition \ref{propB1} and remark after it in Appendix \ref{AppendixB}. The main result of this paper is the commutativity of Baxter $Q$-operators, its proof is given in Sections \ref{sec:Qcomm} and \ref{AppX}.

\begin{theorem}\label{theorem1}
Under the conditions \rf{I0a}, \rf{I0b}, \rf{I17a} Baxter operators commute
\beq\label{I18} \Q{n}(\l) \, \Q{n}(\rho)=\Q{n}(\rho) \, \Q{n}(\l).\eeq
 The kernels of the operators in both sides of \rf{I18} are analytic functions of $\l,\rho$ in the strip
\beq |\Im (\l-\rho) | < \Re \frac{g}{\o_1 \o_2}. \eeq
\end{theorem}

{\bf Remark}. Both sides of the relation \rf{I18} depend analytically on all the parameters $\l, \rho, g, \bo$ in the region of absolute convergence of the integrals.

As it was observed for other integrable systems (see lectures \cite{S} for review), the classical analog of the Baxter $Q$-operator is a special canonical transformation called Backlund transformation. In the paper \cite{KS} V. Kuznetsov and E. Sklyanin proposed a general scheme that relates the kernel of a $Q$-operator and generating function of the corresponding Backlund transformation. In the work \cite{HR1} M. Hallnäs and S. Ruijsenaars showed that in the certain classical limit the kernel and measure functions $K(\bz_n, \by_n), \mu(\by_n)$ contained in the $Q$-operator kernel \eqref{I15} give rise to the Backlund transformation for the classical Ruijsenaars system. In this context the $Q$-operator we defined is a quantum counterpart of this transformation.

We also note that for the case of two particles $n = 2$ (and real constants $\omega_1, \omega_2,g$) such an operator first implicitly appeared in the work \cite{HR5}. Moreover, the commutativity of $Q$-operators in this particular case also follows from the results of \cite{HR5}.

\subsection{Hypergeometric identities} The proof of Theorem \ref{theorem1} consists of residue calculation of the integrals  \rf{I17}. This includes the proof of cancellation of higher order poles and the equality of sums of ordinary poles. The latter is equivalent to certain identity for basic hypergeometric series, which resembles duality transformation theorem for multiple hypergeometric series by Y. Kajihara and M.~Noumi, see \cite{KN}.

Let $q$ and $t$ be formal variables. Denote by $(z;q)_k$ the $q$-analog of the Pochhammer symbol,
\beq\label{I19}(z;q)_k=(1-z)(1-qz)\cdots(1-q^{k-1}z).\eeq
Let $\bu=(u_1,\ldots,u_n)$ and $\bv=(v_1,\ldots, v_n)$ be two tuples of $n$ variables.
\begin{theorem}\label{theorem2} For any integer $K$ we have the following equality of rational functions
\beq\label{I20}\begin{split}  &\sum_{|\bk| =K}\prod_{i=1}^n\frac{(qt;q)_{k_i}}{(q;q)_{k_i}}	
	\times \prod_{\substack{i,j=1 \\ i\not=j}}^n
	\frac{(t^{-1}q^{-k_j}u_i/u_j;q)_{k_i}}{(q^{-k_j}u_i/u_j;q)_{k_i}}
	\times
	\prod_{a,j=1}^n\frac{(tu_j/v_a;q)_{k_j}}{(u_j/v_a;q)_{k_j}}
	=\\
	&\sum_{|\bk|=K}\prod_{a=1}^n\frac{(qt;q)_{k_a}}{(q;q)_{k_a}}	
	\times \prod_{\substack{a,b=1 \\ a\not=b}}^n
	\frac{(t^{-1}q^{-k_a}v_a/v_b;q)_{k_b}}{(q^{-k_a}v_a/v_b;q)_{k_b}}
	\times
	\prod_{a,j=1}^n\frac{(tu_j/v_a;q)_{k_a}}{(u_j/v_a;q)_{k_a}}.
\end{split}\eeq
\end{theorem}
 Here the sum on both sides of the equality is taken over $n$ tuples of non-negative integers with total sum equal to $K$
\beq \bk=(k_1,\ldots, k_n), \qquad  k_i\geq0,\qquad k_1+\ldots+k_n=K.\eeq
Note that the kernel function identity \rf{I9} is a particular limit of the hypergeometric identity \rf{I20}, see \cite{BDKK2} for details.

After our work was completed, O. Warnaar and H. Rosengren communicated to us that an elliptic analog of this identity was proven by different methods in the papers \cite[Corollary 4.3]{LSW}, \cite[eq. (6.7)]{HLNR}. 

\subsection{Further results}
Denote by $\LL{n}(\l)$ the integral operator
\begin{align}\label{I11} \left(\LL{n}(\l)f\right)(\bx_n)&= d_{n - 1}(g) \int_{\R^{n-1}}\Lambda(\bx_n,\by_{n-1};\l) f(\by_{n-1}) d\by_{n-1}
\end{align}
with the kernel
\beq\label{I10} \Lambda(\bx_n,\by_{n-1};\l)= e^{\const \l(\bbx_n-\bby_{n-1})}
\K(\bx_n,\by_{n-1})\mu(\by_{n-1})
\eeq
and the constant
\begin{equation}
d_{n - 1}(g) = \frac{1}{(n - 1)!} \left[ \sqrt{\omega_1 \omega_2} S_2(g) \right]^{-n + 1}.
\end{equation}
The operator $\LL{n}(\l)$ maps functions of  $n-1$ real variables to functions of $n$ real variables. M. Hallnäs and S. Ruijsenaars \cite{HR2} proved that for real periods $\bo$ under the condition
\beq\label{I13a}
0<\Re g<\o_2
\eeq
the function
\beq \label{I12}\Psi_{\bl_n}(\bx_n; g| \bm{\omega})=\LL{n}(\l_n) \, \LL{n-1}(\l_{n-1}) \, \cdots \, \LL{2}(\l_2) \,e^{\const \l_1x_1}
\eeq
is given by absolutely convergent integral and represents the joint eigenfunction of Macdonald operators
\beq \label{I13} M_r(\bx_n;g)\Psi_{\bl_n}(\bx_n; g)=e_r \bigl(e^{{2\pi \lambda_1\o_1}}, \dots, e^{{2\pi \lambda_n\o_1}} \bigr)\Psi_{\bl_n}(\bx_n; g), \qquad r = 1, \dots, n.
\eeq
Here $e_r(z_1,\ldots,z_n)$ is $r$-th elementary symmetric function,
\beqq e_r(z_1,\ldots,z_n)=\sum_{1\leq i_1<i_2<\ldots< i_r\leq n}z_{i_1}\cdots z_{i_r}. \eeqq

In the next  paper \cite{BDKK} we show that the operators \rf{I10} can be obtained in the certain limit from Baxter $Q$-operators, so that the commutativity of $Q$-operators imply commutation relations between $\Lambda$-operators and $Q$-operators, and between $\Lambda$-operators themselves. These relations allow to derive important properties of the eigenfunction. In particular, we show that the eigenfunction \rf{I12}
 \begin{enumerate}
 	\item  enjoys duality property
 	\begin{equation}
 	\Psi_{\bl_n}(\bx_n; g|\bo) = \Psi_{\bx_n}(\bl_n, \hat{g}^*|\hat{\bo}),
 	\end{equation}
 	and consequently admits another iterative integral representation given by Mellin-Barnes type of integrals over spectral parameters $\lambda_j$. Here we denoted
 \beq\label{I27a} \hat{a}=\frac{a}{\o_1\o_2}\eeq
 for any $a\in\C$, so that
 \beq
	\hat{\bo}=\left(\frac{1}{\o_2},\frac{1}{\o_1}\right) , \qquad \hat{g} = \frac{g}{\o_1 \o_2},\qquad \hat{g}^*=\hat{\o}_1 + \hat{\o}_2-\hat{g}=\frac{g^*}{\o_1\o_2};
 \eeq
 	\item   is symmetric function of the coordinates $x_j$, as well as of the spectral variables $\l_j$;
 	\item  is an eigenfunction of the Baxter $Q$-operator with the eigenvalue
 	\beq
 		\prod_{j = 1}^n \hat{K}(\lambda-\lambda_j)= \prod_{j = 1}^n\S^{-1}\Bigl(\imath(\lambda-\lambda_j) +\frac{\hat{g}}{2}\Big|\hat{\bo}\Bigr)\,\S^{-1}\Bigl(-\imath (\lambda-\lambda_j)+\frac{\hat{g}}{2}\Big|\hat{\bo}\Bigr);
 	\eeq
 	
 		\item  is a solution of bispectral problem
 	 for Macdonald operators
 	$M_r(\bx_n;g|\bo)$ and \break $M_s(\bl_n;\hat{g}^*|\hat{\bo})$:
 			\beq \label{I28}
 			\begin{split} M_r(\bx_n;g|\bo)\Psi_{\bl_n}(\bx_n; g)&=e_r \bigl(e^{{2\pi \lambda_1\o_1}}, \dots, e^{{2\pi \lambda_n\o_1}} \bigr)\Psi_{\bl_n}(\bx_n; g),\\[4pt]	
 				M_s(\bl_n;\hat{g}^*|\hat{\bo})\Psi_{\bl_n}(\bx_n; g)&=e_s \bigl(e^{\frac{2\pi x_1}{\o_2}}, \dots, e^{\frac{2\pi x_n}{\o_2}} \bigr)\Psi_{\bl_n}(\bx_n; g),
 			\end{split}
 			\eeq
 			if  $ \Re g < \Re\omega_2$ and $ \Re \hat{g}^*<\Re \hat{\o}_2$.
 			
 	\end{enumerate}
For the hyperbolic Calogero-Sutherland model, which represents a non-relativistic limit of the Ruijsenaars system, the latter result was established in \cite{KK1} and \cite{KK2}.

\setcounter{equation}{0}
\section{Baxter and Macdonald operators commute}\label{BM}
	Theorem \ref{theoremBM} follows from  the kernel function identity and from the invariance of the measure  $\mu(\bx_n)d\bx_n$ with respect to the Macdonald operators.
	
	We present here two proofs of this theorem. Both work for complex periods $\o_1, \o_2$ and coupling constant $g$. The first prove is direct analytical. The second one is its short algebraic reformulation. It exploits symmetry properties of the Macdonald operators with respect to symmetric bilinear pairing \rf{BM3a} in a way analogous to
	\cite[Chapter VI, §9]{M}.  We describe here both proofs since the first one allows to visualize the appearing restriction on the coupling constant $g$, while the second outlines the responsible algebraic properties.
	\medskip
	
	{\bf I. Direct proof}.
	Let $\Phi(\bz_n)$ be a function of $n$ complex variables $\bz_n=(z_1,\ldots,z_n)$ analytic in a strip
	\beq\label{BM0}\Pi_\ve \colon \quad -\ve-\Re \o_1<\Im z_i<\Re \o_1+\ve.\eeq
	We are going to prove the equality
	\beq \label{BM1}  	M_r(\bz_n;g) \, \Q{n}(\l) \, \Phi(\bz_n)= \Q{n}(\l) \, 	M_r(\bz_n;g) \, \Phi(\bz_n),\qquad r=1,\ldots, n.\eeq
	Explicitly it looks as
	\beq\label{BM2}\begin{split}
	M_r(\bz_n;g)\int\limits_{\mathbb{R}^n} \mu(\by_{n})\,K(\bz_n,\by_{n})\,
	e^{\const\l(\bbz_n-\bby_{n})}\,\Phi(\by_n)\,d\by_{n}=\\
	\int\limits_{\mathbb{R}^n} \mu(\by_{n})\,K(\bz_n,\by_{n})\,e^{\const\l(\bbz_n-\bby_{n})}\,
M_r(\by_n;g)\,\Phi(\by_n)\,d\by_{n}.\end{split}\eeq
	Here we assume convergence of the corresponding integrals and $\bz_n \in \mathbb{R}^n$.
	Note the important property of the integration contour $\R^n$ in the integral \rf{BM2}:
	it separates two series of poles of the kernel function
	\beq\label{t04}
	\imath y_i=\imath z_j+\frac{g^\ast}{2} +m\o_1+k\o_2 \qquad \text{and}\qquad
	\imath y_i=\imath z_k-\frac{g^\ast}{2} -m\o_1-k\o_2, \qquad \ m,k\geq 0,
	\eeq
	and two series of poles of the measure function
	\beq \label{t04a} \imath y_i=\imath y_j+g +m\o_1+k\o_2 \qquad\text{and}\qquad
	\imath y_i=\imath y_k-g -m\o_1-k\o_2,\qquad \ m,k\geq 0,
	\eeq
	see \eqref{A1a}, \eqref{A1b} for the poles and zeros of the double sine function.
	
	The left hand side of \rf{BM1} looks as
	\beq \label{BM3} M_r(\bz_n;g) \, \Q{n}(\l) \, \Phi(\bz_n)=
		\sum_{\substack{I \subset \nn \\ |I|=r}}\;\prod_{\substack{i\in I \\[2pt] j\in\nn\setminus I}}
		\frac{\sh\frac{\pi}{\o_2}\left(z_i-z_j-\imath g\right)}
		{\sh\frac{\pi}{\o_2}\left(z_i-z_j\right)}
		\cdot T^{-\imath\o_1}_{I,z} \Q{n}(\l)\Phi(\bz_n),\eeq
where the shift operator $T^{-\imath\o_1}_{I,z}$ is defined in \eqref{Ta}.
	Consider the summand corresponding to subset
	$I=\{i_1,i_2,\ldots, i_r\}$. Denote this summand by $J_{I}$:
	\beq\label{t05}
		J_{I}=	\frac{\sh\frac{\pi}{\o_2}\left(z_i-z_j- \imath g\right)}
		{\sh\frac{\pi}{\o_2}\left(z_i-z_j\right)}
		\cdot T^{-\imath\o_1}_{I,z} \int\limits_{\R^n} \mu(\by_{n}) \K(\bz_n,\by_{n}) e^{\const \l(\bbz_n-\bby_n)}
		\, \Phi(\by_{n}) \, d\by_{n}.
	\eeq
	Shifts  act nontrivially on the kernel $K(\bz_n,\by_{n})$ and
	exponent $ e^{\const \l(\bbz_n-\bby_{n})}$. By \rf{I6},
	\rf{I7} we have
	\beq\label{p8}\begin{split}
		&\prod_{i\in I}
		T_{z_{i}}^{-\imath\o_1}K(\bz_n,\by_{n})=\prod_{i\in I}\prod_{a=1}^{n}
		\frac{\sh\frac{\pi}{\o_2}(z_i-y_a-\imath\frac{g^\ast}{2})}
		{\sh\frac{\pi}{\o_2}(z_i-y_a-\imath\frac{g^\ast}{2}-\imath g)}K(\bz_n,\by_{n}),\\[10pt]
		&\prod_{i\in I}T_{z_{i}}^{-\imath\o_1}e^{\const\l(\bbz_n-\bby_{n})}=
		\,e^{{2\pi r\l\o_1}}\cdot
		e^{\const\l(\bbz_n-\bby_{n})},
	\end{split} \eeq
where in the first formula we have transformed the right hand side
to the form similar to~\eqref{I9}.

The operator $T_{z_{i}}^{-\imath\o_1}$ shifts $z_i$ and we have to shift the integration contour in \rf{t05}, so that the conditions \rf{t04} are satisfied with the replacement of $\imath z_i$ by $\imath z_i+\o_1$. That is the shifted contour should separate set of poles
	\beq \label{p9a}\begin{aligned} \imath y_{a}&=\imath z_{j}+\frac{g^\ast}{2}+m\o_1+k\o_2, &\quad m,k\geq 0, &\qquad j\not\in I,\\[3pt]
		\imath y_{a}&=\imath z_{j}+\frac{g^\ast}{2}+(m+1)\o_1+k\o_2,&\quad m,k\geq 0,&\qquad j\in I,\\	
	\end{aligned} \eeq
	from
	\beq \label{p9}\begin{aligned} \imath y_{a}&=\imath z_{j}-\frac{g^\ast}{2}-m\o_1-k\o_2, &\quad m,k\geq 0, &\qquad j\not\in I,
		\\[3pt] \imath y_{a}&=\imath z_{j}-\frac{g^\ast}{2}-(m-1)\o_1-k\o_2, &\quad m,k\geq 0, &\qquad j\in I,\\	
	\end{aligned} \eeq
	and also separate two series of poles \rf{t04a} of the measure functions.
	For  this we can use the contour
	\beq \label{p11}
	C \colon \ \Im y_a=-c, \qquad -\Re \frac{g^\ast}{2}+\Re \o_1<c<\Re \frac{g^\ast}{2},\qquad a=1,\ldots, n
	\eeq
	which exists provided
	\beq\label{p11a} \Re g^\ast>\Re \o_1,\qquad\text{or equivalently}\qquad \Re g<\Re \o_2.\eeq
	Since the contour $C$ does not depend on a set $I$, we can permute integration and summation procedures, so that
	\beq \label{p12}\begin{split}&	M_r(\bz_n;g) \, \Q{n}(\l) \, \Phi(\bz_n)=
		\sum_{\substack{I\subset\nn\\|I| = r}}J_{I}=e^{2\pi r\l \o_1}\times\\&\int\limits_{C} S_r(\bz_n,\by_{n})
		\mu(\by_{n}) K(\bz_n,\by_{n})
		e^{\const\l(\bbz_n-\bby_{n})}
		\,\Ph(\by_{n})\,d\by_{n}, 	
	\end{split}
	\eeq
	where
	\beqq
	S_r(\bz_n,\by_{n})=\sum_{\substack{I\subset\nn \\ |I| = r}}\prod_{i\in I}\left(\prod_{j\in \nn\setminus I}
	\frac{\sh\frac{\pi}{\o_2}\left(z_i-z_j-\imath  g\right)}
	{\sh\frac{\pi}{\o_2}\left(z_i-z_j\right)}
	\prod_{a=1}^{n}
	\frac{\sh\frac{\pi}{\o_2}(z_i-y_a-\imath\frac{g^\ast}{2})}
	{\sh\frac{\pi}{\o_2}(z_i-y_a-\imath\frac{g^\ast}{2}-\imath g)}\right)
	\eeqq
Define  similar sum
	\beqq
	\tS_r(\by_{n},\bz_n)=\sum_{\substack{A\subset\nn \\ |A|=r}}\prod_{a\in A}\left(\prod_{b\in \nn\setminus A}
	\frac{\sh\frac{\pi}{\o_2}\left(y_a-y_b+ \imath g\right)}
	{\sh\frac{\pi}{\o_2}\left(y_a-y_b\right)}
	\prod_{i=1}^{n}
	\frac{\sh\frac{\pi}{\o_2}(z_i-y_a-\imath\frac{g^*}{2})}
	{\sh\frac{\pi}{\o_2}(z_i-y_a-\imath\frac{g^\ast}{2}-\imath g)}\right)
	\eeqq
One can  see that the sum $S_r(\bz_n,\by_{n})$ is obtained from the
left hand side of the kernel function identity \rf{I9} by the  change of variables
\begin{align}
z_k \to \frac{\imath \pi}{\o_2} z_k, \qquad y_a \to \frac{\imath \pi}{\o_2}
\Bigl(y_a+\imath\frac{g^\ast}{2}+\imath g\Bigr), \qquad \alpha \to \frac{\imath \pi}{\o_2}\imath g
\end{align}
and there is the same correspondence between $\tS_r(\by_{n},\bz_n)$ and the
right hand side of \rf{I9}. It implies the equality
	\beq\label{p15c}
	S_r(\bz_n,\by_{n})=\tS_{r}(\by_{n},\bz_n).
	\eeq
	Thus  we rewrite \rf{p12} as
	\beq \label{p16}\begin{split}
		&M_r(\bz_n;g) \, \Q{n}(\l) \, \Phi(\bz_n)=\\[4pt]&e^{{2\pi r\l\o_1}}\int\limits_{C}\tS_{r}(\by_{n},\bz_n,)
		\mu(\by_{n}) K(\bz_n,\by_{n}) e^{\const \l(\bbz_n-\bby_{n})}	
\,\Ph(\by_{n})\, d\by_{n}	
	\end{split}
	\eeq
	and apply to each occurring summand the same procedure in opposite direction.
	Namely, for any subset $A \subset \nn$ of cardinality $r$ in the integral
	\beq\label{p16a}\begin{split}
		J'_{A}=\int\limits_{C}&\prod_{a\in A}\left(\prod_{b\in \nn\setminus A}
		\frac{\sh\frac{\pi}{\o_2}\left(y_a-y_b+\imath  g\right)}
		{\sh\frac{\pi}{\o_2}\left(y_a-y_b\right)}
		\prod_{i=1}^{n}
		\frac{\sh\frac{\pi}{\o_2}(z_i-y_a-\imath\frac{g^\ast}{2})}
		{\sh\frac{\pi}{\o_2}(z_i-y_a-\imath\frac{g^\ast}{2}-\imath g)}\right) \times \\[6pt]
		&e^{\const \l(\bbz_n-\bby_{n})}\mu (\by_{n}) K(\bz_n,\by_{n}) \,	\Ph(\by_{n}) \,d\by_{n}\end{split}\eeq
	we perform the change of integration variables
	\beq\label{p16b}
	y_a\to y_a-\imath\o_1,\qquad a\in A.
	\eeq
We have
	\begin{align}
	 &	\prod_{a\in A}
		T_{y_{a}}^{-\imath\o_1}K(\bz_n,\by_{n})=\prod_{a\in A}\prod_{i=1}^{n}
		\frac{\sh\frac{\pi}{\o_2}(y_a-z_i-\imath\frac{g^\ast}{2})}
		{\sh\frac{\pi}{\o_2}(y_a-z_i-\imath\frac{g^\ast}{2}-\imath g)}K(\bz_n,\by_{n}),\notag
		\\[8pt]
	&	\prod_{a\in A}T_{y_{a}}^{-\imath\o_1}
\mu(\by_{n})= \prod_{\substack{a\in A \\[2pt] b\in [n] \setminus A}}
		\frac{\sh\frac{\pi}{\o_2}(y_a-y_b-\imath\o_1)}{\sh\frac{\pi}{\o_2}(y_a-y_b)}\cdot
		\frac{(-1)\sh\frac{\pi}{\o_2}(y_a-y_b-\imath g)}{\sh\frac{\pi}{\o_2}(y_a-y_b-\imath g^\ast)}\,
		\mu(\by_{n}),  \label{y}
		\\[6pt]
	&	\prod_{a\in A}
		T_{y_{a}}^{-\imath\o_1}e^{\const\l(\bbz_n-\bby_{n})}=
		\,e^{-{2\pi r \l\o_1}}\cdot
		e^{\const\l(\bbz_n-\bby_{n})}. \notag
	\end{align}
	Using \rf{y} and the relations
	\beq \label{p18}\begin{split}
		\prod_{a\in A}
		T_{y_{a}}^{-\imath\o_1}\prod_{a\in A}\left(\prod_{b\in \nn\setminus A}
		\frac{\sh\frac{\pi}{\o_2}\left(y_a-y_b+\imath  g\right)}
		{\sh\frac{\pi}{\o_2}\left(y_a-y_b\right)}
		\prod_{i=1}^{n}
		\frac{\sh\frac{\pi}{\o_2}(z_i-y_a-\imath\frac{g^\ast}{2})}
		{\sh\frac{\pi}{\o_2}(z_i-y_a-\imath\frac{g^\ast}{2}-\imath g)}\right)=\\
		\prod_{a\in A}\left(\prod_{b\in \nn\setminus A}
		\frac{(-1)\sh\frac{\pi}{\o_2}\left(y_a-y_b-\imath  g^\ast\right)}
		{\sh\frac{\pi}{\o_2}\left(y_a-y_b-\imath\o_1\right)}
		\prod_{i=1}^{n}
		\frac{\sh\frac{\pi}{\o_2}(z_i-y_a+\imath\frac{g^\ast}{2}+\imath g)}
		{\sh\frac{\pi}{\o_2}(z_i-y_a+\imath\frac{g^\ast}{2})}\right)
	\end{split}\eeq
	we see that
	\beq \label{p19}\begin{split}
		J'_{A}=\int\limits_{\tilde{C}} & d\by_{n} \,
		\mu(\by_{n}) K(\bz_n,\by_{n}) \, e^{\const \l(\bbz_n-\bby_{n})} \times\\
		&\prod_{\substack{a\in A \\[2pt] b \in \nn\setminus A}}
		\frac{\sh\frac{\pi}{\o_2}\left(y_a-y_b-\imath  g\right)}
		{\sh\frac{\pi}{\o_2}\left(y_a-y_b\right)}
		\prod_{a\in A}
		T_{y_{a}}^{-\imath\o_1}
		\Ph(\by_{n}),
	\end{split}\eeq
	where the contour $\tilde{C}$ is the deformation of the contour $C$ according to the change of variables \rf{p16b}.
	In the assumption $\Re g^\ast>\Re \o_1$ we may choose again $\tilde{C}=\R^{n}$ provided the conditions \rf{t04a} on separation of the poles of the measure are not spoiled during the move of the contour. Note that zeros of the measure $\mu(\by_{n})$ cancel the poles of the hyperbolic sine functions in the last line of \rf{p19}.
	On the other hand, zeros of the sine functions
	$$\sh\frac{\pi}{\o_2}\left(y_a-y_b- \imath g\right)$$
	cancel poles
	$$y_b=y_a-\imath g -\imath p \o_2,\qquad p\geq 0$$
	of the measure function, so that the first pole which we can meet during the move of the contour is
	$$y_b=y_a-\imath g -\imath \o_1$$
	and its shift does not touch the real plane. Then we can deform the contour $\tilde{C}$ to its original position $\mathbb{R}^n$. Summing up \rf{p19} with the integration contour replaced by $\mathbb{R}^n$ we arrive at the statement of  Theorem \ref{theoremBM}. \hfill{$\Box$}
	\medskip
	
	{\bf II. Algebraic version.} In the space of functions $\vf(\bz_n)$ analytical in the strip $\Pi_\ve$ \rf{BM0} and satisfying the bound
	\beqq \vf(\bz_n)=O(|\bz_n|^{-1}),\qquad  \Re z\to \infty,\qquad z\in \Pi_\ve\eeqq
	introduce the symmetric bilinear pairing
	\beq\label{BM3a} (\vf,\psi)= \int_{\R^n}\vf(\by_n)\psi(-\by_n)\mu(\by_n)d\by_n.\eeq
	Denote by $\tau_y$ the operator that changes the sign of argument in a function
	\beqq  \tau_y \vf(\by_n)=\vf(-\by_n). \eeqq
	Then we can rewrite this pairing as
	\beq\label{BM4} (\vf,\psi)= \int_{\R^n}\vf(\by_n)\tau_y [\psi(\by_n)]\mu(\by_n)d\by_n.\eeq
	The eigenvalue property \rf{I8} of the Ruijsenaars kernel function $K(\bz_n,\by_n)$
	can be written~as
	\beq\label{BM5}  M_r(\bz_n;g)K(\bz_n,\by_n)=\tau_y M_r(\by_n;g)\tau_yK(\bz_n,\by_n).\eeq
	However, if we want to use the relation \rf{BM5} for the operator with a kernel containing
	$K(\bz_n,\by_n)$, we should impose the condition \rf{p11a} in order to have correctly defined shift operators.
	Macdonald operators are symmetric with respect to the pairing  \rf{BM3} (compare with \cite[Chapter VI, §9, eq.(9.4)]{M})
	\beq\label{BM6} \left(M_r(\by_n;g) \, \vf(\by_n),\psi(\by_n)\right)=
	\left(M_r(\by_n;g) \, \psi(\by_n),\vf(\by_n)\right).
	\eeq
	Then the left hand side of the relation \rf{BM1} can be written as
	\beq\label{BM7}\begin{split} & M_r(\bz_n;g)\big(K(\bz_n,\by_n) \, e^{\const\l\bbz_n},\ e^{\const\l\bby_n} \, \tau_y \,\Phi(\by_n)  \big)=\\[8pt]
		&e^{{2\pi r\l\o_1}} \, e^{{2\pi \imath \l}\bbz_n} \big(M_r(\bz_n;g) K(\bz_n,\by_n),\ e^{{2\pi \imath \l}\bby_n} \, \tau_y\,\Phi(\by_n) \big).\end{split}	\eeq
	Using \rf {BM5}  we rewrite \rf{BM7} as
	\beq\label{BM9}\begin{split}
		&e^{{2\pi r\l\o_1}}e^{{2\pi \imath \l}\bbz_n} \big(\tau_yM_r(\by_n;g)\tau_yK(\bz_n,\by_n),\ e^{{2\pi \imath \l}\bby_n} \, \tau_y\,\Phi(\by_n)  \big)=\\[8pt]
		&e^{{2\pi r\l\o_1}}e^{{2\pi \imath \l}\bbz_n} \big(M_r(\by_n;g)\tau_yK(\bz_n,\by_n),\ e^{-{2\pi \imath \l}\bby_n} \, \Phi(\by_n)   \big).
	\end{split}	\eeq
	Next applying \rf{BM6} we have
	\beq\label{BM10}\begin{split}
		&e^{{2\pi r\l\o_1}}e^{{2\pi \imath \l}\bbz_n} \big(\tau_yK(\bz_n,\by_n),\ M_r(\by_n;g)e^{-{2\pi \imath \l}\bby_n} \, \Phi(\by_n)   \big)=\\[8pt]
		&e^{{2\pi \imath \l}{}\bbz_n} \big(\tau_yK(\bz_n,\by_n),\ e^{-{2\pi \imath \l\bby_n}{}}M_r(\by_n;g) \Phi(\by_n)   \big)=\\[8pt]
		&e^{{2\pi \imath \l}{}\bbz_n} \big(K(\bz_n,\by_n),\ e^{{2\pi \imath \l\bby_n}{}}\tau_y \,M_r(\by_n;g) \Phi(\by_n)   \big).
	\end{split}	\eeq
	The last line of \rf{BM10} coincides with the right hand side of \rf{BM1}. \hfill{$\Box$}

\setcounter{equation}{0}
\section{Commutativity of Baxter operators}\label{sec:Qcomm}
The commutativity of $Q$-operators
\begin{equation}\label{J1}
Q_{n}(\rho) \, Q_{n}(\rho') = Q_{n}(\rho') \, Q_{n}(\rho)
\end{equation}
follows from the equality of the kernels
\beq \label{J3}
Q_{n}(\bm{z}_n, \bm{x}_n ; \rho, \rho' )= Q_{n}( \bm{z}_n, \bm{x}_n ; \rho', \rho )
\eeq
of their products
\begin{equation} \label{J2}
Q_{n}(\bm{z}_n, \bm{x}_n ; \rho, \rho' ) = \int_{\mathbb{R}^n} d\bm{y}_n \, Q( \bm{z}_n, \bm{y}_n;\rho ) Q(\bm{y}_n, \bm{x}_n; \rho' ), \qquad z_j, x_j \in \mathbb{R}.
\end{equation}
One can further note that  the variables $\bz_n$ and $\bx_n$ enter the equality \rf{J3} in a similar way. Thus we combine them into a common $2n$ array
$$\bz_{2n}= \{z_1,\ldots, z_n,x_1,\ldots, x_n\}$$
and set
\beq\label{J4} Q_{n}( \bz_{2n}; \l) = \int_{ \R^n}e^{\const\l\bby_n}\prod_{a=1}^{2n}\prod_{i=1}^n K(z_a-y_i) \mu(\by_n)d\by_n .\eeq
Then the equality  \rf{J3} takes the form of the following integral identity
\beq\label{J5}
Q_{n}(\bz_{2n}; \l)=e^{\const\l\bbz_{2n}}Q_{n}(\bz_{2n}; -\l ),
\eeq
where we put
$$\l=\rho'-\rho.$$
 Under the condition \rf{I17a}
both integrals in \rf{J5} absolutely converge uniformely on compact subsets of the parameters, see Proposition \ref{propB1} in Appendix \ref{AppendixB}. Thus, both sides of the
equality \rf{J5} are analytic functions of all the parameters therein.  Having in mind these analyticity properties we prove the equality \rf{J5} step by step following the plan below.
\begin{enumerate}
	\item We prove that for complex periods $\o_1,\o_2$ with $\Re \o_i > 0$ and $\o_1 / \o_2 \not\in \mathbb{R}$ both integrals may be calculated by residues technique for big enough negative values of $\Re \lambda$.
	In other words, one can find a sequence of contours that in the limit encircle all poles in the corresponding half plane (for each integration variable) and such that the integrals over encircling contours tend to zero.
	
	\item Next we assume that the real parameters $z_i$ are generic. Under this assumption we prove that the sum of residues over higher order poles vanishes. For this we accumulate the vanishing properties of the integration measure $\mu(\by_n)$ into Lemma~\ref{lemmaD2}, which says that the sum of $4^r$ integrand values over the points with interchanged coefficients at the periods gives zero of order $2r$. This lemma is used to describe the result of $k$ successive integrations computed by residues and to show that after each integration higher order poles vanish.
	
	\item At this stage we are left with the sum of simple poles on both sides; each of them is a product of one-dimensional residues integrals over shifted parameters $z_i$. Their sums decompose into the sums of $\binom{2n}{n}$ series, depending of which parameters $z_i$ enter the residue calculations. The equalities of corresponding series reduce to certain identities on rational functions which we prove separately. The identities generalize Ruijsenaars' kernel function identity and could be treated as certain duality transformations for multivalued basic hypergeometric series \cite{KN}.
	
	\item Finally, due to analiticity of the statement \rf{J5} we conclude that it is valid first for all values of the parameters $z_i$, $\Re\l$ and as well for $\o_1/\o_2 \in \R$ (in particular, for real values of the periods $\o_1, \o_2$).
\end{enumerate}

\subsection{Estimates of integrals over encircling contours} \label{sectioneEstimates}
In the end of this subsection we prove that the integrals in $Q$-commutativity identity \eqref{J5} can be calculated by residues in the case $\o_1 / \o_2 \not\in \R$.

Denote by $\sigma_i$ the arguments of the periods~$\o_i$, $|\sigma_i|<\pi/2$. Since the double sine function is invariant under permutation of $\o_1$, $\o_2$, suppose for definiteness that $\sigma_1 \geq \sigma_2$. Let $D_+$ and $D_-$ be the cones of poles \eqref{A1a} and zeros \eqref{A1b} of the double sine function $S_2(z|\bo)$:
\beqq D_+=\{ z\colon \sigma_2< \arg z<\sigma_1\},\qquad D_-=\{ z\colon \pi+\sigma_2< \arg z<\pi+\sigma_1\},\qquad D=D_+\cup D_-.\eeqq

In the first step of our plan we consider contour in the limit encircling big interval in the real line to a big semicircle in the corresponding half plane. This contour contains three different parts and inside each part the needed bounds are obtained in different ways. In the part of the contour close to the real plane we apply the bound given in Proposition~\ref{propB2} in Appendix \ref{AppendixB}. In the next part of the contour which lies in the regular region $\C\setminus D$ we use the general statements about at most exponential growth of the functions $K$ and $\mu$, see \rf{A12}, \rf{A16}.

A subtle point is the estimate of the integrand along the part of the contour lying in ``forbidden'' areas $D_+\cup D_-$ and passing between poles of the double sine function. This estimate is not possible for purely real periods or for periods whose ratio is real.
Therefore, for the estimates in the area $D_+ \cup D_-$ we assume
\begin{equation}
\Im \frac{\o_1}{\o_2} > 0.
\end{equation}
Then we use an infinite product representation of the double sine function,
\beq \label{J10}S_2(z|\bo)=e^{\frac{\pi\imath}{2}B_{2,2}(z|\bo)}\vf(z|\bo)=e^{-\frac{\pi\imath} {2}B_{2,2}(z|\bo)}\vf'(z|\bo)\eeq
where
\beq\label{J12}\vf(z|\bo)=\frac{r(z|\bo)}{s(z|\bo)}=\frac{\prod\limits_{m=0}^\infty\left(1-q^{2m}
	e^{\frac{ 2\pi iz}{\omega_2}}\right)}
{\prod\limits_{m=1}^\infty\left(1- \tq^{2m} 	e^{\frac{ 2\pi iz}{\omega_1}}\right)},
\eeq
\beq\label{J12a}
\vf'(z|\bo)=\frac{r'(z|\bo)}{s'(z|\bo)}=
\frac{\prod\limits_{m=0}^\infty\left(1-\tq^{2m}
	e^{-\frac{ 2\pi iz}{\omega_1}}\right)}
{\prod\limits_{m=1}^\infty\left(1-q^{2m}
	e^{-\frac{ 2\pi iz}{\omega_2}}\right)}
\eeq
with
\beq\label{J13}
q=e^{\pi\imath\frac{\o_1}{\o_2}},\qquad \tq=e^{-\pi\imath\frac{\o_2}{\o_1}}
\eeq
and $B_{2,2}(z|\bo)$ is a particular multiple Bernoulli polynomial \eqref{A2}. For any real $t_0$ and $\ve$, $0<\ve<1$ denote by $\Pi_{1,+}(t_0,\ve)$ the strip in the complex plane of the variable $z$, bounded from one side
\beq \label{J14} \Pi_{1,+}(t_0,\ve)=\{ z=t\o_1+\theta \o_2\ |\  t>t_0,\ \ve<\theta<1-\ve\}.\eeq
\begin{lemma}\label{lemmas1} The function $\vf(z|\bo)$ is restricted and bounded from zero in any strip  $\Pi_{1,+}(t_0,\ve)$,
	\beq 0<C_1< |\vf(z|\bo)|<C_2 \qquad\text{for}\qquad z\in \Pi_{1,+}(t_0,\ve).\eeq
\end{lemma}
{\bf Proof}. Let $\o_1/\o_2=\a+i\b,\ \b>0$ and assume first that $t_0>0$. Consider  the nominator of $\vf(z|\bo)$.
It could be written as
\beq\begin{split}\label{J15}
	r(z|\bo)=&	 \prod\limits_{m\geq 0}\left(1-q^{2m}e^{2\pi\imath\frac{t\o_1+\theta\o_2}{\o_2}}\right)=\\
	\prod\limits_{m\geq 0}	\left(1-e^{2\pi\imath\big((m+t)\frac{\o_1}{\o_2}+\theta\big)}\right)=&
	\prod\limits_{m\geq 0}	\left(1-e^{-2\pi\b(m+t)}e^{2\pi\imath(\a(m+t)+\theta)}\right).
\end{split}\eeq
We have for $t>0$
\beq\label{J15b} |e^{-2\pi\b(m+t)}e^{2\pi\imath(\a(m+t)+\theta)}|=e^{-2\pi\b(m+t)}<1.\eeq
Due to inequality
\beq\label{J15a} 1-|a|<|1-a|<1+|a| \eeq
we get the following bound for the nominator $r(z|\bo)$ of $\vf(z|\bo)$
\beqq \xi_\beta(t_0)<|r(z|\bo)|<\eta_\b(t_0)\eeqq
where
\beqq \xi_\b(t_0)=\prod\limits_{m\geq 0}\left(1-e^{-2\pi\beta(m+t_0)}\right),\qquad
\eta_\b(t_0)=\prod\limits_{m\geq 0}\left(1+e^{-2\pi\beta(m+t_0)}\right).\eeqq
Both these infinite products are converging products not equal to zero. In order to extend the desired bound for
negative values  of $t_0$ we note that this extension adds finite product of factors, each of them is bounded from zero due to the restriction on $\theta$. Let
$$\frac{\o_2}{\o_1}=c-\imath d,\qquad d>0.$$
Consider  the denominator $s(z|\bo)$ of $\vf(z|\bo)$. It looks as
\beq\label{J16}s(z|\bo)=\prod\limits_{m\geq 1}\left(1-\tq^{2m}e^{2\pi\imath\frac{z}{\o_1}}\right)=
	\prod\limits_{m\geq 1}\left(1-e^{-2\pi d(m-\theta)}e^{2\pi\imath(-c(m-\theta)+t)}\right).
\eeq
For $\theta<1$ we have
\beqq |e^{-2\pi d(m-\theta)}e^{2\pi\imath(-c(m+\theta)-t)}|= e^{-2\pi d(m-\theta)}<1.\eeqq
Again, using inequality \rf{J15a} we get the bound
\beqq \xi'_d(\ve)<|s(z|\bo)|<\eta'_d(\ve)\eeqq
where
\beq\label{J16a} \xi'_d(\ve)=\prod\limits_{m\geq 0}\left(1-e^{-2\pi d(m+\ve)}\right),\qquad
\eta'_d(\ve)=\prod\limits_{m\geq 0}\left(1+e^{-2\pi d(m+\ve)}\right)\eeq	
are convergent infinite products.	\hfill{$\Box$}

{\bf Remarks.} Similar arguments give the two-sided bound for the values of the functions $\vf(z|\bo)$ and $\vf'(z|\bo)$ in generic strips.

\begin{itemize}
	\item[\bf 1.] The function $\vf(z|\bo)$ and its inverse are restricted in strips
	\begin{align}
	&\Pi_{1,+}(t_0,m,\ve)=\{ z=t\o_1+\theta \o_2\ |\  t>t_0,\ m+\ve<\theta<m+1-\ve\}\,,\\
	&\Pi_{2,-}(t_0,m,\ve)=\{ z=-t\o_2-\theta \o_1\ |\  t>t_0,\ m+\ve<\theta<m+1-\ve\}
	\end{align}
	where $m\in\Z,\ 0<\ve<1$.
	
	\item[\bf 2.] The function $\vf'(z|\bo)$ and its inverse are restricted in any strip
	\begin{align}
	&\Pi_{2,+}(t_0,m,\ve)=\{ z=t\o_2+\theta \o_1\ |\  t>t_0,\ m+\ve<\theta<m+1-\ve\}\,,\\
	&\Pi_{1,-}(t_0,m,\ve)= \{ z=-t\o_1-\theta \o_2\ |\  t>t_0,\ m+\ve<\theta<m+1-\ve\}
	\end{align}
	where $m\in\Z,\ 0<\ve<1$.
\end{itemize}

\begin{figure}[h]
	\centering
	\begin{tikzpicture}[scale = 0.8, line width=0.4mm, line cap=round]
	\def\a{0.6};
	\def\b{1.1};
	\def\d{0.9};
	\def\e{0.2};
	\def\t{1.2};
	\def\m{0};
	\def\n{2};
	\def\l{2};
	\def\p{3};
	
	\draw[gray!30!white, ->] (-6.5,0) -- (6.5, 0) node[xshift = 0.4cm, yshift = -0.3cm] {\footnotesize \textcolor{black}{$\Re z$}};	
	\draw[gray!30!white, ->] (0, -5.5) -- (0, 5.5) node[xshift = 0.4cm, yshift = 0.3cm] {\footnotesize \textcolor{black}{$\Im z$}};
	
	\draw (0,0) node[xshift = -0.2cm, yshift = -0.25cm] {\footnotesize $0$};
	\draw (\a,\d) node[xshift = -0.1cm, yshift = -0.25cm] {\footnotesize $\omega_1$};
	\draw (\b,0) node[yshift = -0.3cm] {\footnotesize $\omega_2$};
	
	\begin{scope}
	\clip (-6.55,-5) rectangle (6.55,5);
	\foreach \k in {0,...,6}{
		\foreach \m in {0,...,6}{
			\draw[gray!50!black, fill=gray!50!black] (\k*\a + \m*\b, \k*\d) circle (0.6pt);
			\draw[gray!50!black, fill=gray!50!black] (-\k*\a - \m*\b, -\k*\d) circle (0.6pt);
		}
	}
	\end{scope}
	
	\draw[draw opacity = 0, fill = gray!20!white] (5*\a + \m*\b + \e*\b, 5*\d) -- (\t*\a + \m*\b + \e*\b, \t*\d) -- (\t*\a + \m*\b + \b - \e*\b, \t*\d) -- (5*\a + \m*\b +\b - \e*\b, 5*\d);
	
	\draw[draw opacity = 0, fill = gray!20!white] (4.5*\b + \n*\a + \e*\a, \n*\d + \e*\d) -- (\t*\b + \n*\a + \e*\a, \n*\d + \e*\d) -- (\t*\b + \n*\a + \a - \e*\a, \n*\d + \d - \e*\d) -- (4.5*\b + \n*\a +\a - \e*\a, \n*\d + \d - \e*\d);

	\draw[gray, dashed, line width = 0.2mm] (5*\a + \m*\b + \e*\b, 5*\d) -- (\t*\a + \m*\b + \e*\b, \t*\d) -- (\t*\a + \m*\b + \b - \e*\b, \t*\d) -- (5*\a + \m*\b +\b - \e*\b, 5*\d) node[xshift = -0.2cm, yshift = 0.35cm] {\footnotesize \color{black} $\Pi_{1,+}(t_0, \m, \ve)$};
	
	\draw[gray, dashed, line width = 0.2mm] (4.5*\b + \n*\a + \e*\a, \n*\d + \e*\d) -- (\t*\b + \n*\a + \e*\a, \n*\d + \e*\d) -- (\t*\b + \n*\a + \a - \e*\a, \n*\d + \d - \e*\d) -- (4.5*\b + \n*\a +\a - \e*\a, \n*\d + \d - \e*\d) node[xshift = 1.cm, yshift = -0.35cm] {\footnotesize \color{black} $\Pi_{2,+}(t_0, \n, \ve)$};

	\draw[draw opacity = 0, fill = gray!20!white] (-5*\a - \l*\b - \e*\b, -5*\d) -- (-\t*\a - \l*\b - \e*\b, -\t*\d) -- (-\t*\a - \l*\b - \b + \e*\b, -\t*\d) -- (-5*\a - \l*\b - \b + \e*\b, -5*\d);
	
	\draw[draw opacity = 0, fill = gray!20!white] (- 4*\b - \p*\a - \e*\a, - \p*\d - \e*\d) -- (- \t*\b - \p*\a - \e*\a, - \p*\d - \e*\d) -- (- \t*\b - \p*\a - \a + \e*\a, - \p*\d - \d + \e*\d) -- ( - 4*\b - \p*\a - \a + \e*\a, - \p*\d - \d + \e*\d);
	
	\draw[gray, dashed, line width = 0.2mm] (-5*\a - \l*\b - \e*\b, -5*\d) -- (-\t*\a - \l*\b - \e*\b, -\t*\d) -- (-\t*\a - \l*\b - \b + \e*\b, -\t*\d) -- (-5*\a - \l*\b - \b + \e*\b, -5*\d) node[xshift = 0.2cm, yshift = -0.4cm] {\footnotesize \color{black} $\Pi_{1,-}(t_0, \l, \ve)$};
	
	\draw[gray, dashed, line width = 0.2mm] (- 4*\b - \p*\a - \e*\a, - \p*\d - \e*\d) -- (- \t*\b - \p*\a - \e*\a, - \p*\d - \e*\d) -- (- \t*\b - \p*\a - \a + \e*\a, - \p*\d - \d + \e*\d) -- ( - 4*\b - \p*\a - \a + \e*\a, - \p*\d - \d + \e*\d) node[xshift = -1.cm, yshift = 0.3cm] {\footnotesize \color{black} $\Pi_{2,-}(t_0, \p, \ve)$};
	
	\end{tikzpicture}
	\caption{The parameter $m \in \mathbb{Z}$ controls a place of the strip $\Pi_{j, \pm}$ in~the~$(\mathbb{Z}\omega_1, \mathbb{Z}\omega_2)$-lattice. The value of $t_0$ determines where strip starts, and the parameter $\ve$ defines its thickness}
\end{figure}
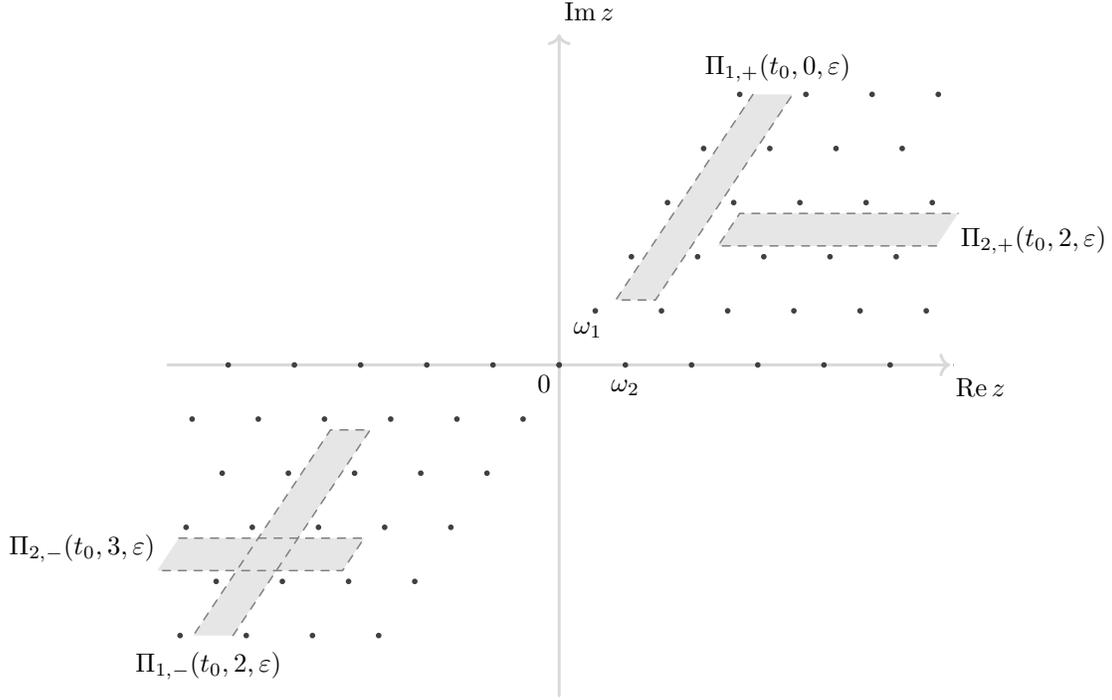

Next we consider the function $\vf(z|\bo)\vf^{-1}(z+g|\bo)$.
Assume that
\beq\label{J17} g\not\in \Z\o_1+\Z\o_2.\eeq
First of all note that for any $t_0\in \R$ there exist $0<\ve'<\ve$, $t'_0\in\R$ and $N_0\in \Z$ such that the strip
$ \{ z=-g+t\o_1+\theta \o_2\ |\  t>t_0,\ \ve'<\theta<1-\ve'\}$ is inside the strip $\Pi_{1,+}(t'_0,N_0,\ve)$.
 Then for any $N\in\Z_+$ the ratio $\vf(z|\bo)\vf^{-1}(z+g|\bo)$ has no poles and no zeros in the strip $\Pi_{1,+}(t_0,N,\ve)$ and is restricted in this strip. We now show that its bound is not more than exponential on $N$.
\begin{lemma}\label{lemmas2} There exist real $a,b$ and $C_1, C_2>0$, such that
	\beqq\label{J18} C_1e^{aN}<\left|\vf(z|\bo)\vf^{-1}(z+g|\bo)\right|<C_2e^{bN}\eeqq
	in the strip $\Pi_{1,+}(t_0,N,\ve)$.
\end{lemma}
{\bf Proof}. First of all note that the bounds for the product $r(z|\bo)$ do not depend on $N$ for any strip $\Pi_{1,+}(t_0,N,\ve)$ {(the bound \eqref{J15b} doesn't depend on the range of $\theta$)}. We just have to estimate the ratio $s(z|\bo)s^{-1}(z+g|\bo)$ in the strip $\Pi_{1,+}(t_0,N,\ve)$. This product can be divided into two parts: one is an infinite convergent product
\beq\label{J19}s_\infty(z|\bo)=\frac{\prod\limits_{m\geq N}\left(1-\tq^{2m}e^{2\pi\imath\frac{z}{\o_1}}\right)}{\prod\limits_{m\geq N+N_0}\left(1-\tq^{2m}e^{2\pi\imath\frac{z+g}{\o_1}}\right)},\eeq
which after the change of the product indices can be evaluated in the strip $\Pi_{1,+}(t_0,N,\ve)$ independently of $N$
\beq \frac{\xi'_d(\ve)}{\eta'_d(\ve)}<|s_\infty(z|\bo)|< \frac{\eta'_d(\ve)}{\xi'_d(\ve)},\eeq
where $\xi'_d(\ve)$ and $\eta'_d(\ve)$ are given in \rf{J16a}; and another is a finite product
\beq\label{J20}\begin{split}
	s_0(z|\bo)=&\frac{\prod\limits_{1\leq m < N}\left(1-\tq^{2m}e^{2\pi\imath\frac{z}{\o_1}}\right)}{\prod\limits_{1\leq m < N+N_0}\left(1-\tq^{2m}e^{2\pi\imath\frac{z+g}{\o_1}}\right)}=\\
	& {\frac{\tq^{-N(N-1)}e^{2\pi\imath (N - 1)\frac{z'}{\o_1}}}{\tq^{-(N+N_0)(N+N_0-1)}e^{2\pi\imath (N+N_0-1)\frac{z''}{\o_1}}} }
	\cdot
	\frac{\prod\limits_{1\leq m < N}\left(\tq^{2m}e^{-2\pi\imath\frac{z'}{\o_1}}-1\right)}{\prod\limits_{1\leq m < N+N_0}\left(\tq^{2m}e^{-2\pi\imath\frac{z''}{\o_1}}-1\right)},
\end{split}		\eeq
where both $z' = z - N\o_2$ and $z'' = z+g - (N + N_0)\o_2$ are now in the strip $\Pi_{1,+}(t_0,0,\ve)= \Pi_{1,+}(t_0,\ve)$. The second fraction can be bounded from both sides independently of $N$ with a help of analogous infinite product, that is
\beq \label{J21} \frac{\xi'_d(\ve)}{\eta'_d(\ve)}<
\left|\frac{\prod\limits_{1\leq m < N}\left(\tq^{2m}e^{-2\pi\imath\frac{z'}{\o_1}}-1\right)}{\prod\limits_{1\leq m < N+N_0}\left(\tq^{2m}e^{-2\pi\imath\frac{z''}{\o_1}}-1\right)}\right|< \frac{\eta'_d(\ve)}{\xi'_d(\ve)}\eeq
The estimate of the first one is also pure exponential
\beq\label{J22}C_1e^{2\pi d N(2\ve-1 - N_0)}< |\tq|^{2N_0N+N_0(N_0-1)}e^{2\pi d(N - 1)(\theta'-\theta'')}e^{-2\pi d N_0}<
C_2e^{2\pi d N(1-2\ve-N_0)}\eeq
This ends the proof of Lemma \ref{lemmas2}. \hfill{$\Box$}

Analogous statement holds for the function $\vf'(z|\bo)$ in corresponding strips.
\medskip

For each $0<\ve<1$ and positive integers $M,N$ denote by  $\Pi_{1,+}(t_0,N,M,\ve)$ the bounded open region
\beq\label{J23} \Pi_{1,+}(t_0,M,N,\ve)=\{ z=t\o_1+\theta \o_2\ |\  t_0<t<N,\ M+\ve<\theta<M+1-\ve\}.
\eeq
Lemma \ref{lemmas2}, its analog for the function  $\vf'(z|\bo)$ and the relations \rf{J10}  immediately imply the following corollary.
\begin{corollary}\label{corJ1} The ratio $S_2(z|\bo)S_2^{-1}(z+g|\bo)$ admits a two sided exponential bound in the region $\Pi_{1,+}(t_0,M,N,\ve)$
	\beq\label{J24} C_1e^{a(N+M)}<|S_2(z|\bo)S_2^{-1}(z+g|\bo)|< C_2e^{b(N+M)}
	\eeq
	for some real $a,b$ and $C_1,C_2>0$.	
\end{corollary}	
{\bf Remark}. Analogous exponential bounds hold for the regions
\beq\label{J25}\begin{split}
	\Pi_{2,+}(t_0,M,N,\ve)=&\{ z=t\o_2+\theta \o_1\ |\  t_0<t<N,\ M+\ve<\theta<M+1-\ve\} \\	
	\Pi_{1,-}(t_0,M,N,\ve)=&\{ z=-t\o_1-\theta \o_2\ |\  t_0<t<N,\ M+\ve<\theta<M+1-\ve\}\\
	\Pi_{2,-}(t_0,M,N,\ve)=&\{ z=-t\o_2+\theta \o_1\ |\  t_0<t<N,\ M+\ve<\theta<M+1-\ve\}	
\end{split}
\eeq
The proofs are similar.
\begin{corollary} \label{corollary5}  For big negative values of the real part of the parameter $\lambda$ the integral in the left hand side of \rf{J5} can be computed by residues calculation, moving the integration contours to the {lower} half plane; and  the integral in the right hand side of \rf{J5} can be also computed by residues calculation, moving the integration contours to the {upper} half plane.
\end{corollary}
{\bf Proof}. The residue calculation means that the initial straight contour in the integral in the left hand side of \rf{J5} is enclosed by a contour where for instance
\beqq |y_1|\gg|y_2|\gg\ldots \gg|y_n|\gg 1, \qquad \Im y_i<0 \eeqq
and we argue that the integral over this enclosing contour tends to zero when the contour grows.
For each variable $y_i$ its integration contour is either lies in the regular region for all occuring function
$S(\imath y_i-a_j)S^{-1}(\imath y_i+g-a_j)$ or in the cones of singularities of these functions.  In the part of the contour close to the real line we apply the bound given in Proposition \ref{propB2} in Appendix \ref{AppendixB}. In the next part of the contour which lies in the regular region $\C\setminus D$, using  the general statements about at most exponential growth, see \rf{A12}, \rf{A16}, we suppress the integrand by fast decreasing  exponent $e^{{2\pi \imath \lambda}\by_n}$ with sufficiently big negative values of $\Re\lambda$. Inside the irregular cone $D$ we put the contour into proper regions $\Pi_{k,\pm}(t_i,M_i,N_i,\ve_i)$, $k=1,2$ for sufficiently large $M_i$ and $N_i$. In the proper regions all functions grow at most exponentially, see Corollary \ref{corJ1} and remark after it, therefore we suppress them by fast decreasing exponent $e^{{2\pi \imath \lambda}\by_n}$ as well. The same procedure for the integral in the right hand side of \rf{J5}.

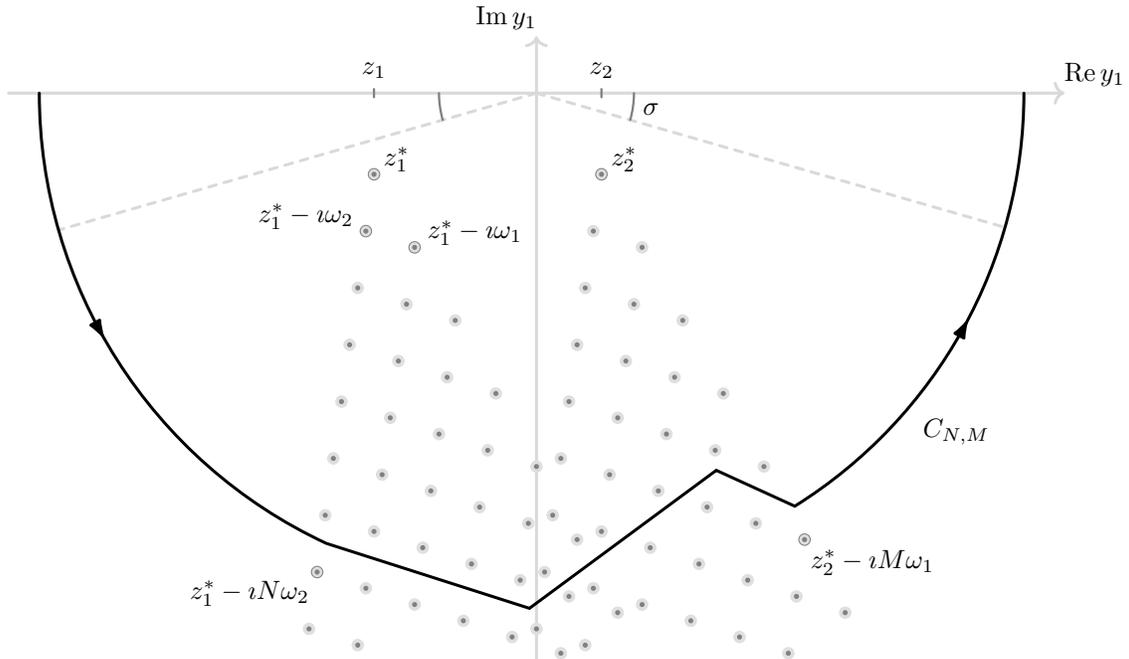
\begin{figure}[h]
	\centering
	\begin{tikzpicture}[scale = 1.08, line width=0.4mm, line cap=round]
	\tikzset{->-/.style={decoration={
				markings,
				mark=between positions 0.16 and 0.9 step 0.69 with {\arrow{Latex[round,length=3mm,width=2mm]}}},postaction={decorate}
	}}
	
	\draw[gray!30!white, ->] (-6.5,0) -- (6.5, 0) node[xshift = 0.4cm, yshift = 0.25cm] {\footnotesize \textcolor{black}{$\Re y_1$}};	
	\draw[gray!30!white, ->] (0, -7) -- (0, 0.7) node[xshift = -0.4cm, yshift = 0.25cm] {\footnotesize \textcolor{black}{$\Im y_1$}};
	
	\def\a{0.9 };
	\def\b{0.5};
	\def\c{0.7};
	\def\d{-0.1};
	\def\g{1};
	\def\x{-2};
	\def\z{0.8};
	\def\r{6};
	\def\m{1.02};
	\def\s{16};
	
	\draw[gray, line width=0.3mm] (\x,-0.05) -- (\x,0.05) node[yshift = 0.25cm] {\footnotesize \color{black} $z_1$};
	\draw[gray, line width=0.3mm] (\z,-0.05) -- (\z,0.05) node[yshift = 0.25cm] {\footnotesize \color{black} $z_2$};
	
	\begin{scope}
	\clip (-6.55,-7) rectangle (6.55,0);
	\foreach \k in {0,...,8}{
		\foreach \m in {0,...,7}{
			\draw[gray!25!white,fill=gray!25!white, line width=0.1mm] (\x + \m*\b + \k*\d, -\g - \m*\a - \k*\c) circle (2pt);
			\draw[gray, fill=gray] (\x + \m*\b + \k*\d, -\g - \m*\a - \k*\c) circle (0.35pt);
			
			\draw[gray!25!white,fill=gray!25!white, line width=0.1mm] (\z + \m*\b + \k*\d, -\g - \m*\a - \k*\c) circle (2pt);
			\draw[gray, fill=gray] (\z + \m*\b + \k*\d, -\g - \m*\a - \k*\c) circle (0.35pt);
		}
	}
	\end{scope}
	
	\draw[gray, line width=0.1mm] (\x, -\g) circle (2pt) node[xshift = 0.3cm, yshift = 0.2cm] {\footnotesize \color{black} $z_1^*$};
	\draw[gray, line width=0.1mm] (\z, -\g) circle (2pt) node[xshift = 0.3cm, yshift = 0.2cm] {\footnotesize \color{black} $z_2^*$};
	\draw[gray, line width=0.1mm] (\x + \b, -\g - \a) circle (2pt) node[xshift = 0.8cm, yshift = 0.2cm] {\footnotesize \color{black} $z_1^* - \imath \o_1$};
	\draw[gray, line width=0.1mm] (\x + \d, -\g - \c) circle (2pt) node[xshift = -0.8cm, yshift = 0.2cm] {\footnotesize \color{black} $z_1^* - \imath \o_2$};
	
	\draw[gray, line width=0.1mm] (\z + 5*\b, -\g - 5*\a) circle (2pt) node[xshift = 0.9cm, yshift = -0.3cm] {\footnotesize \color{black} $z_2^* - \imath M \o_1$};
	\draw[gray, line width=0.1mm] (\x + 7*\d, -\g - 7*\c) circle (2pt) node[xshift = -0.9cm, yshift = -0.28cm] {\footnotesize \color{black} $z_1^* - \imath N \o_2$};

	\draw[white, fill=white] (\x + 2*\b + 6*\d, -\g - 2*\a - 6*\c) circle (2.3pt);
	\draw[white, fill=white] (\z + 2*\b + 6*\d, -\g - 2*\a - 6*\c) circle (2.3pt);
	
	\draw[gray!30!white, dashed] (0, 0) -- ({\r*cos(\s)},{-\r*sin(\s)});
	\draw[gray!30!white, dashed] (0, 0) -- ({-\m*\r*cos(\s)},{-\m*\r*sin(\s)});
	\draw[gray, line width = 0.3mm] (0.2*\r,0) arc (0:-\s:0.2*\r) node[xshift = 0.28cm, yshift = 0.15cm] {\footnotesize \color{black} $\sigma$};
	\draw[gray, line width = 0.3mm] (-0.2*\r,0) arc (180:180+\s:0.2*\r);
	
	\draw[->-] ({-\m*\r}, 0) arc (180:180+65:\m*\r) -- +(2.5, -0.8) -- +(4.8, 0.9) -- ({\r*cos(-58)},-{\r*sin(58)}) arc (360-58: 360: \r) node[xshift = -0.9cm, yshift = -4.5cm] {\footnotesize $C_{N,M}$};
	
	\end{tikzpicture}
	\caption{The contour in the case $n = 1$. We denoted $z_j^* = z_j - \imath g^*/2$ and for clarity circled all labeled poles}
	\label{fig:contour}
\end{figure}

In the case $n = 1$ the $Q$-commutativity integral \eqref{J4} is one-dimensional
\begin{equation}
Q_1(z_1, z_2; \l) = \int_\mathbb{R} dy_1 \, K(z_1 - y_1)K(z_2 - y_1) \, e^{2\pi \imath \l y_1},
\end{equation} 
and the corresponding enclosing contour $C_{N,M}$ that appears when we calculate it by residues
\begin{equation}
\int_\mathbb{R} = \int_{C_{N,M}} + \; \sum \Res
\end{equation}
is shown in Figure \ref{fig:contour}. There are two sequences of poles in the lower half-plane
\begin{equation}
y_1 = z_j - \frac{\imath g^*}{2} - \imath m^1 \o_1 - \imath m^2 \o_2, \qquad j = 1,2, \quad  m^i \geq 0.
\end{equation}
Small circles around poles in Figure \ref{fig:contour} are restricted regions: the broken line stays away from the poles at the distance more than their radii (consequently, we have fixed exponent parameters $a,b$ from Corollary \ref{corJ1}). The angle $\sigma$ is determined from the condition \eqref{B11}, so that we have exponentially decreasing bound near the real line given in Proposition~\ref{propB2}. The integers $N, M$ are chosen such that the contour $C_{N,M}$ passes right above the pole $z_1-\imath g^*/2 - \imath N \o_2$ from the left and the pole $z_2 - \imath g^*/2 - \imath M \o_1$ from the right.
\hfill{$\Box$}

\subsection{Reduction to simple poles}
\subsubsection{Chains of integrals and double zeros Lemma}
Denote by $F$ the integrand of the left hand side of $Q$-commutativity relation \rf{J4}
\beq \label{D1}F(\by_{n},\bz_{2n})= e^{\const\l\bby_n}\prod_{a=1}^{2n}\prod_{i=1}^n K(y_i-z_a) \prod_{\substack{i, j = 1 \\ i \not= j}}^n\mu(y_i - y_j).\eeq
We integrate this function over $y_j$ by residues in the order of increasing indices. Let $G_m$ be the result of $m$ successive integrations
\beq\label{D4}  G_m(y_{m+1},\ldots,y_n,\bz_{2n})=\int_\R dy_m \cdots \int_\R dy_1 \, F(\by_{n},\bz_{2n}).\eeq
Moving contours to the lower half-plane we meet poles of the function $K(y_i - z_a)$
\beq\label{D4a}
\imath y_i=\imath z_a+\frac{g^\ast}{2} +m^1\o_1+m^2\o_2, \qquad
\imath y_i=\imath z_a-\frac{g^\ast}{2} -m^1\o_1-m^2\o_2
\eeq
and of the function $\mu(y_i - y_j)$
\beq \label{D4b}
\imath y_i=\imath y_j+g +m^1\o_1+m^2\o_2, \qquad
\imath y_i=\imath y_j-g -m^1\o_1-m^2\o_2
\eeq
where $m^1, m^2 \geq 0$. Below we prove (see Proposition \ref{propD1}) that the resulting function $G_m$ can be written solely in terms of two typical residue integrals with simple poles.

\begin{figure}[t]
	\centering
	\parbox{\textwidth}{
		\parbox{.5\textwidth}{
			\hspace{0.5cm}
			\vspace{0.5cm}
			\subfloat[$J_1$-integral]{
				\begin{tikzpicture}
				\def\l{1.2};
				\def\r{2};
				\draw (0, -0.25*\l) node[xshift = -3.7 cm] {$J_1(z_a | b,m^1,m^2)  = $};
				\draw[thick, fill] (0,0)  circle (\r pt) node[xshift = 0.5cm] {$y_{b}$} -- node[xshift = -1cm] {$m^1, m^2$} (0, -0.5*\l);
				\draw[thick, fill = white] (-\r pt, -0.5*\l cm - \r pt ) rectangle (\r pt, -0.5*\l cm + \r pt)  node[xshift = 0.45cm, yshift = - 2*\r pt]{$z_{a}$};
				\end{tikzpicture}}
			\vspace{0.5cm}
			
			\hspace{0.5cm}			
			\subfloat[{\centering Example: $i_0 = 4, \; \bi_3 = (2, 1, 3)$, $\bmm_3 = (1, 3, 2; 4, 0, 1)$}]{
				\begin{tikzpicture}
				\def\l{1.2};
				\def\r{2};
				\draw (0, -1.5*\l) node[xshift = -3 cm] {$I_3 \left(y_{4} | \bi_3, \bmm_3 \right)  = $};
				\draw[thick, fill] (0,0) -- node[xshift = -0.7cm] {$1, 4$} (0,-\l) circle (\r pt) node[xshift = 0.5cm] {$y_{2}$} -- node[xshift = -0.7cm] {$3,0$} (0,-2*\l) circle (\r pt) node[xshift = 0.5cm] {$y_{1}$} -- node[xshift = -0.7cm] {$2,1$} (0, -3*\l) circle (\r pt) node[xshift = 0.5cm] {$y_{3}$};
				\draw[thick, fill = white] (0,0) circle (\r pt) node[xshift = 0.5cm] {$y_{4}$};
				\draw (0,-3.2*\l) node[below,opacity=0] {};
				\end{tikzpicture}
			}
			
		}
		\parbox{.5\textwidth}{
			\subfloat[$I_k$-integral]{
				\begin{tikzpicture}
				\def\l{1.2};
				\def\r{2};
				\draw (0, -2.5*\l) node[xshift = -3.5 cm] {$I_k \left(y_{i_0} | \bi_k, \bmm_k \right)  = $};
				\draw[thick, fill] (0,0) -- node[xshift = -1cm] {$m^1_1, m^2_1$} (0,-\l) circle (\r pt) node[xshift = 0.5cm] {$y_{i_1}$} -- node[xshift = -1cm] {$m^1_2, m^2_2$} (0,-2*\l) circle (\r pt) node[xshift = 0.5cm] {$y_{i_2}$} -- (0, -2.5*\l) node[yshift = -0.4cm] {$\vdots$};
				\draw[thick, fill] (0, -3.3*\l) -- (0, -3.8*\l) circle (\r pt) node[xshift = 0.65cm] {$y_{i_{k - 1}}$} -- node[xshift = -1cm] {$m^1_k, m^2_k$} (0, -4.8*\l) circle (\r pt) node[xshift = 0.5cm] {$y_{i_k}$};
				\draw[thick, fill = white] (0,0) circle (\r pt) node[xshift = 0.5cm] {$y_{i_0}$};
				\end{tikzpicture}
			}
		}
	}
	\vspace{0.5cm}
	\caption{Chain integrals}
	\label{fig:chains}
\end{figure}
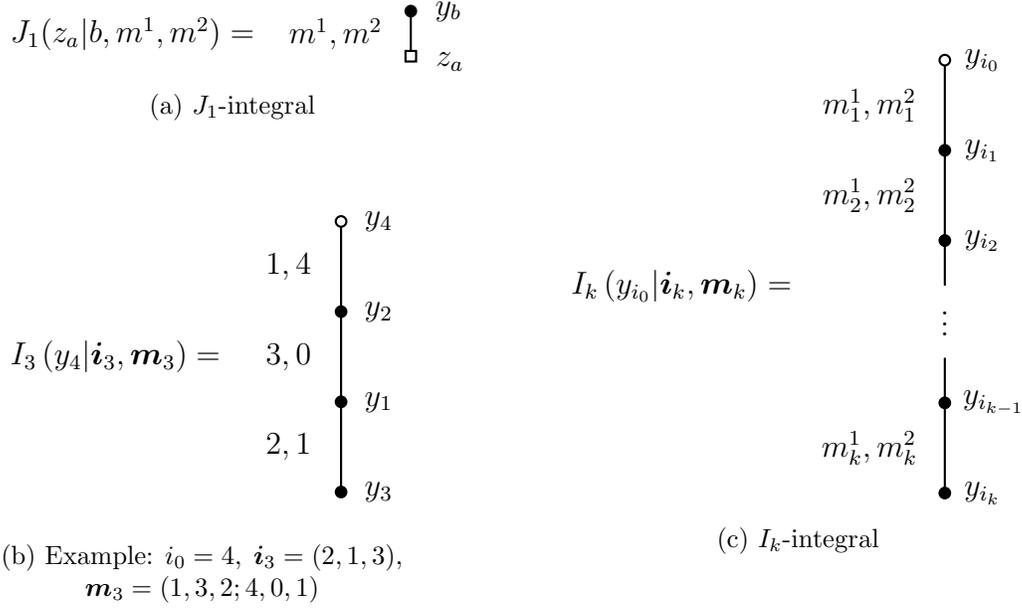

The first typical integral is $J_1(z_a| b,m^1,m^2)$. It depends on a complex parameter $z_a$, on the index $b$ of the variable $y_b$ and on the pair $(m^1,m^2)$ of non-negative integers. It is given by one-dimensional residue
\beq\label{D3} J_1(z_a| b,m^1,m^2)=-2\pi \imath\Res_{\imath y_b=\imath z_a+\frac{g^\ast}{2} +m^1\o_1+m^2\o_2}F(\by_{n},\bz_{2n}).
\eeq
Additionally, it is a function of all other parameters $z_c$ and variables $y_j$ different from $y_b$. This residue is nonzero due to the poles of functions $\K$ \eqref{D4a}.

The second typical integral is $k$-fold residue integral
$ I_k\left({y}_{i_0} | \bi_k,\bmm_k\right)$.
It depends on a complex valued variable ${y}_{i_0}$ with $i_0 \in [n] = \{1, \ldots, n\}$, on  a sequence $\bi_k$ of $k$ distinct indices
\beqq
\bi_k = (i_1,\ldots, i_k), \qquad i_a\in [n] \setminus \{i_0\}
\eeqq
corresponding to variables $y_{i_a}$ in \eqref{D1}, and on two non-negative sequences of $k$ integers
\beqq \bmm_k = (m^1_1,\ldots, m^1_k; m^2_1,\ldots, m^2_k), \qquad m_j^i \geq 0.\eeqq
We define $I_k \left({y}_{i_0}| \bi_k,\bmm_k\right)$ as the following $k$-fold residue integral
\beq \label{D2}\begin{split} I_k \left({y}_{i_0}| \bi_k,\bmm_k\right)= (-2\pi\imath)^k & \Res_{\imath y_{i_1}=\imath y_{i_0}+g+m^1_1\o_1+m^2_1\o_2} \;\cdots\\[6pt]
	\cdots \;&\Res_{\imath y_{i_k}=\imath y_{i_{k-1}}+g+m^1_{k} \o_1+m^2_{k}\o_2} \, F(\by_{n},\bz_{2n}).
\end{split}	\eeq
Additionally, it is a function of parameters $z_a$ and all variables $y_j$ which are not engaged in the integration procedure. It naturally refers to the point
\beq \label{D2a}
\begin{aligned}
	\imath y_{i_1} &= \imath y_{i_0}+g+m^1_1\o_1+m^2_1\o_2,\\[4pt]
	\imath y_{i_2} &= \imath y_{i_0}+2g+\bigl( m^1_1 + m^1_2 \bigr)\o_1 + \bigl( m^2_1 + m^2_2 \bigr) \o_2, \\[4pt]
	&\,\,\, \vdots \\[4pt]
	\imath y_{i_k} &= \imath y_{i_0}+kg+\bigl( m^1_1 + \ldots + m^1_k \bigr)\o_1+\bigl( m^2_1 + \ldots + m^2_k \bigr)\o_2.
\end{aligned}
\eeq
This residue is nonzero due to the poles of functions $\mu$ \eqref{D4b}. Note that, although in $G_m$ \eqref{D4} we integrate over $y_{i_a}$ in the order of increasing indices $i_a$, the multidimensional pole \eqref{D2a} is simple, so the definition \eqref{D2} doesn't depend on the order of residues.

We name the integrals $J_1(z_a| b,m^1,m^2)$ and  $ I_k\left({y}_{i_0} | \bi_k,\bmm_k\right)$ as {\bf chain integrals} and picture them as chains of vertices with the corresponding labels, see Figure \ref{fig:chains}. Note that length of the line in $J_1$-chain is twice smaller than in $I_k$-chains. This rule reflects difference between constants $g^*/2$ and $g$ in residue points of chain integrals.

The chain integrals are parametrized by the cycles in the space of corresponding integration variables.
It is natural to define the {\bf direct product} of chain integrals as the integrals over direct products of corresponding cycles. More precisely, assume that all the variables $y_{i_a}$ of the first chain integral including integration variables  and the free variable are different from those of the second chain integral. Then the direct product of these two chain integrals is defined as the integral over corresponding product of the contours. For instance, the direct product
{$$I_k \left({y}_{i_0}| \bi_k,\bmm_k\right)\times I_l \left({y}_{i'_0}| \bi'_l,\bmm'_l\right)$$ of two chain integrals is defined for disjoint sets $\bi_k$ and $\bi'_l$ and generic parameters ${y}_{i_0}$ and ${y}_{i'_0}$ as $k+l$ fold residue integral
\begin{align*} (-2\pi\imath)^{k+l}
&\Res_{\imath y_{i_{1}}=\imath y_{i_{0}}+g+m^1_{1}\o_1+m^2_{1}\o_2} \;\cdots \; \Res_{\imath y_{i_{k}}=\imath y_{i_{k-1}}+g+m^1_{k}\o_1+m^2_{k}\o_2} \\[4pt]
&\Res_{\imath y_{i'_{1}}=\imath y_{i'_{0}}+g+{m'}^1_{1}\o_1+{m'}^2_{1}\o_2} \; \cdots \; \Res_{\imath y_{i'_{l}}=\imath y_{i'_{l-1}}+g+{m'}^1_{l}\o_1+{m'}^2_{l}\o_2}
F(\by_{n},\bz_{2n})
\end{align*}
\color{black}
A direct product of integrals will be pictured simply by placing the corresponding chains next to each other, see Figure \ref{fig:chains-product}.
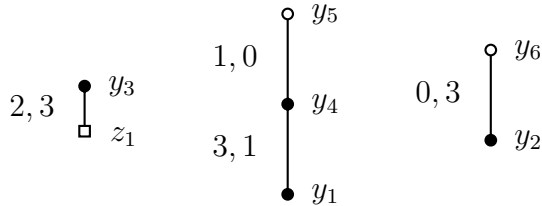
\begin{figure}[h]
	\centering
	\begin{tikzpicture}
	\def\l{1.2};
	\def\r{2};
	\def\s{2.7}
	\draw[thick, fill] (-\s,-0.8*\l)  circle (\r pt) node[xshift = 0.5cm] {$y_{3}$} -- node[xshift = -0.7cm] {$2, 3$} (-\s, -0.8*\l-0.5*\l);
	\draw[thick, fill = white] (-\s cm -\r pt, -0.8*\l cm -0.5*\l cm - \r pt ) rectangle (-\s cm + \r pt, -0.8*\l cm -0.5*\l cm + \r pt)  node[xshift = 0.45cm, yshift = - 2*\r pt]{$z_{1}$};
	\draw[thick, fill] (0,0) -- node[xshift = -0.7cm] {$1, 0$} (0,-\l) circle (\r pt) node[xshift = 0.5cm] {$y_{4}$} -- node[xshift = -0.7cm] {$3,1$} (0,-2*\l) circle (\r pt) node[xshift = 0.5cm] {$y_{1}$} ;
	\draw[thick, fill = white] (0,0) circle (\r pt) node[xshift = 0.5cm] {$y_{5}$};
	\draw[thick, fill] (\s,-0.4*\l) -- node[xshift = -0.7cm] {$0, 3$} (\s,-0.4*\l-\l) circle (\r pt) node[xshift = 0.5cm] {$y_{2}$} ;
	\draw[thick, fill = white] (\s,-0.4*\l) circle (\r pt) node[xshift = 0.5cm] {$y_{6}$};
	\end{tikzpicture}
	\caption{Direct product of three chain integrals}
	\label{fig:chains-product}
\end{figure}

The main technical result of this subsection is the following statement.
\begin{proposition}\label{propD1} For generic $\bo$, $g$ and $\bz_{2n}$ the function
	$G_m(y_{m+1},\ldots,y_n,\bz_{2n})$ defined by~\eqref{D4} is a sum of all possible direct products of chains $J_1(z_a| b,m^1,m^2)$ and $ I_k\left({y}_{i_0} | \bi_k,\bmm_k\right)$, such that parameters $z_a$ are distinct inside each summand and indices of integration variables do not exceed $m$.
\end{proposition}
\begin{figure}[h]
	\centering
	\begin{tikzpicture}
	\def\l{1.2};
	\def\r{2};
	\def\s{3};
	\draw (-1.8*\s, -0.5*\l) node {$G_2   = $};
	\draw (-2*\s, -2.3*\l) node {\begin{tabular}{c}
		number \\
		of terms
		\end{tabular}};
	\draw (-1.17*\s, -2.3*\l) node {$\binom{6}{2} \cdot 2$};
	\draw (-0.17*\s, -2.3*\l) node {$\binom{6}{1} \cdot 2$};
	\draw (0.9*\s, -2.3*\l) node {$2$};
	\draw[thick, fill] (-1.3*\s,-0.3*\l)  circle (\r pt) node[yshift = 0.35cm] {$y_{i}$} -- (-1.3*\s, -0.3*\l-0.5*\l);
	\draw[thick, fill = white] (-1.3*\s cm -\r pt, -0.3*\l cm -0.5*\l cm - \r pt ) rectangle (-1.3*\s cm + \r pt, -0.3*\l cm -0.5*\l cm + \r pt)  node[yshift = -0.45cm]{$z_{a}$};
	\draw[thick, fill] (-\s,-0.3*\l)  circle (\r pt) node[yshift = 0.35cm] {$y_{j}$} -- (-\s, -0.3*\l-0.5*\l);
	\draw[thick, fill = white] (-\s cm -\r pt, -0.3*\l cm -0.5*\l cm - \r pt ) rectangle (-\s cm + \r pt, -0.3*\l cm -0.5*\l cm + \r pt)  node[yshift = - 0.45cm]{$z_{b}$};
	\draw (-0.65*\s, -0.5*\l) node {$+$};
	\draw[thick, fill] (-0.3*\s,-0.3*\l)  circle (\r pt) node[yshift = 0.35cm] {$y_{i}$} -- (-0.3*\s, -0.3*\l-0.5*\l);
	\draw[thick, fill = white] (-0.3*\s cm -\r pt, -0.3*\l cm -0.5*\l cm - \r pt ) rectangle (-0.3*\s cm + \r pt, -0.3*\l cm -0.5*\l cm + \r pt)  node[yshift = -0.45cm]{$z_{a}$};
	\draw[thick, fill] (0,0) -- (0,-\l) circle (\r pt) node[xshift = 0.5cm] {$y_{j}$};
	\draw[thick, fill = white] (0,0) circle (\r pt) node[xshift = 0.5cm] {$y_{3}$};
	\draw (0.5*\s, -0.5*\l) node {$+$};
	\draw[thick, fill] (0.9*\s,0.5*\l) -- (0.9*\s,-0.5*\l) circle (\r pt) node[xshift = 0.5cm] {$y_{i}$} -- (0.9*\s,-1.5*\l) circle (\r pt) node[xshift = 0.5cm] {$y_{j}$};
	\draw[thick, fill = white] (0.9*\s,0.5*\l) circle (\r pt) node[xshift = 0.5cm] {$y_{3}$};
	\draw (2.3*\s, -0.6*\l) node {$\begin{aligned} \{i, j\} &= \{1, 2\}, \vspace{3pt} \\ \{a, b\} &\subset \{1, \dots, 6\} \end{aligned} $};
	\end{tikzpicture}
	\caption{\centering  All possible direct products for $n = 3, m = 2$. The sum over all possible edge parameters and indices $i, j, a, b$ is implied}
\end{figure}
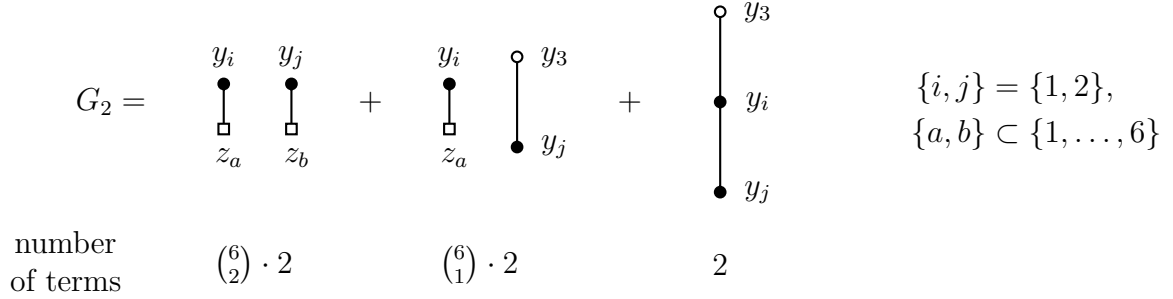
Note that the function  $G_n(\emptyset,\bz_{2n})$ coincides with the integral \eqref{J4}
\begin{equation}
G_n(\emptyset,\bz_{2n}) = Q_n(\bz_{2n};\l).
\end{equation}
An immediate corollary of Proposition \ref{propD1} is the following observation.
\begin{corollary}\label{corD1} The integral in the left hand side of \rf{J5} is the sum of all possible direct products of one-dimensional residues $J_1(z_a| b,m^1,m^2)$ with distinct $z_a$.
\end{corollary}
Indeed, in this case free parameters of possible chains could not be any integration variables $y_a$, so we have a sum of direct products of chain integrals $J_1(z_a| b,m^1,m^2)$.
The same result holds for the right hand side of \rf{J5}.
\medskip

For the proof of Proposition \ref{propD1} we need the following property of zeros location of the measure function $\mu(\by_n)$. Choose a pair of variables, say $y_1$ and $y_2$. Let
\beq \label{D5} \imath y_1=a+\ve+p^1\o_1+p^2\o_2,\qquad \imath y_2=a+q^1\o_1+q^2\o_2,\qquad p^i,q^i\in\Z.\eeq
Set \beqq  m^1=p^1-q^1,\qquad
m^2=p^2-q^2.\eeqq
Let the operator $\tau_{12}^{1,m^1}$ permute $p^1$ and $q^1$ and the operator $\tau^{2,m^2}_{12}$ permute $p^2$ and $q^2$, so that
\beqq\begin{split} \tau^{1,m^1}_{12}F(y_1,y_2,\ldots, y_n, \bz_{2n})= F(a+\ve+q^1\o_1+p^2\o_2,a+p^1\o_1+q^2\o_2,\ldots,y_n,\bz_{2n}),\\[4pt]
	\tau^{2,m^2}_{12}F(y_1,y_2,\ldots, y_n, \bz_{2n})= F(a+\ve+p^1\o_1+q^2\o_2,a+q^1\o_1+p^2\o_2,\ldots,y_n,\bz_{2n}).
\end{split}
\eeqq
In other words, the operators $\tau_{12}^{1,m^1}$ and $\tau_{12}^{2,m^2}$ are the following shift operators:
\beq\label{D5a}\tau_{12}^{1,m^1}\left(\begin{array}{cc} \imath y_1\\ \imath y_2\end{array}\right)=
\left(\begin{array}{cc} \imath y_1-m^1\o_1\\ \imath y_2+m^1\o_1\end{array}\right),
\qquad \tau_{12}^{2,m^2}\left(\begin{array}{cc} \imath y_1\\ \imath y_2\end{array}\right)=
\left(\begin{array}{cc} \imath y_1-m^2\o_2\\ \imath y_2+m^2\o_2\end{array}\right).\eeq

\begin{lemma}\label{lemmaD1} For generic $a$, variables $y_j$ and parameters $z_b$
	\beq\label{D6} (1+\tau^{1,m^1}_{12})(1+\tau^{2,m^2}_{12})F(\by_n,\bz_{2n})=O(\ve^2),\qquad \ve\to 0.\eeq
\end{lemma}
In other words, the sum of four  summands
\beq\label{D7} F(\by_n,\bz_{2n})+\tau^{1,m^1}_{12}F(\by_n,\bz_{2n})+\tau^{2,m^2}_{12}F(\by_n,\bz_{2n})+\tau^{1,m^1}_{12}\tau^{2,m^2}_{12}F(\by_n,\bz_{2n})\eeq
has zero of the second order at the hyperplane
\beq\label{D8} \imath (y_1-y_2)=m^1\o_1+m^2\o_2.\eeq

{\bf Proof} of Lemma \ref{lemmaD1}.
Note that shift operators in \eqref{D6} act only on $y_1, y_2$. Denote by~$A$ the part of function $F$ that doesn't depend on~$y_1, y_2$
\begin{equation}
F(\by_n,\bz_{2n}) = e^{\const \l (y_1 + y_2)} A(y_3, \dots, y_n, \bz_{2n})  B(\by_n,\bz_{2n}).
\end{equation}
The exponent is invariant under the shifts. Using reflection formula
\begin{equation}
S_2^{-1}(z) = S_2(\omega_1 + \omega_2 - z)
\end{equation}
we can write the remaining part $B$ in the form
\begin{equation}
\begin{aligned}
B &= S_2 ( \imath(y_1 - y_2)) \, S_2 ( \imath(y_2 - y_1)) \, S_2 ( \imath(y_1 - y_2) + g^*) \, S_2 ( \imath(y_2 - y_1) + g^*) \\[5pt]
&\times \prod_{(b, b')} S_2(\imath y_1 - b)\, S_2(\imath y_2 - b) \, S_2( b' - \imath y_1) \, S_2( b' - \imath y_2),
\end{aligned}
\end{equation}
where pairs $(b, b')$ contain all other variables $y_j, z_a$ ($j \not=1,2$) and constants. Moreover, the first two functions can be written as \eqref{trig4}
\begin{equation}\label{D11}
S_2 (\imath(y_1 - y_2))  S_2 (\imath(y_2 - y_1)) = -4 \sin \frac{\imath \pi}{\omega_1}(y_1 - y_2) \sin \frac{\imath \pi}{\omega_2}(y_1 - y_2).
\end{equation}
For the variables $y_1, y_2$ at the points \eqref{D5} we use the last formula with a factorization property \rf{trig4a}
\begin{equation}\label{D11a}
S_2(z + p^1 \omega_1 + p^2 \omega_2) ={(-1)^{p_1p_2}} \frac{S_2(z + p^1 \omega_1) \, S_2(z + p^2 \omega_2)}{S_2(z)}, \qquad p^j \in \mathbb{Z}
\end{equation}
to separate coordinates $p^j, q^j$ in the function $B$
\begin{equation}
B(p^1, q^1; p^2, q^2) = C \, B_1(p^1, q^1) \, B_2(p^2, q^2)
\end{equation}
where $C$ doesn't depend on any $p^j, q^j$. The signs coming from \rf{D11a} dissapear since each of them occurs an even number of times. Clearly, $B_j$ differ only by $\omega_j$. Therefore, it's sufficient to prove that
\begin{equation}\label{D13}
(1+\tau^{1,m^1}_{12}) B_1(p^1, q^1) = O(\ve).
\end{equation}
Evaluating $B_1$ at $\ve = 0$ we obtain a function antisymmetric with respect to $p^1, q^1$:
\begin{equation}
\begin{aligned}
B_1(p^1, q^1)\bigr\rvert_{\ve = 0} = (-1)^{p^1 + q^1} \sin \frac{\pi \omega_1}{\omega_2} (p^1 - q^1) \; S_2((p^1 - q^1) \omega_1 + g^*) \, S_2((q^1 - p^1) \omega_1 + g^*)\\[3pt]
\times \prod_{(b, b')} S_2(p^1 \omega_1 + a - b)\, S_2(q^1 \omega_1 + a - b) \, S_2( b' - a - p^1 \omega_1) \, S_2( b' - a - q^1 \omega_1).
\end{aligned}
\end{equation}
Since all functions in $B_1$ are analytic, the identity \eqref{D13} follows.
\hfill{$\Box$}

Now let $\bmm_k$ again denote two sequences of integers (without requiring them to be non-negative)
$$\bmm_k=(m^1_1,\ldots m^1_k; m^2_1,\ldots, m^2_k), \qquad m_j^i \in \mathbb{Z}.$$
We attach to this sequence $2k$ shift operators $\tau_{12}^{1,\bmm_k}$, $\tau_{12}^{2,\bmm_k}$, $\ldots $ $\tau_{2k-1,2k}^{1,\bmm_k}$, $\tau_{2k-1,2k}^{2,\bmm_k}$, so that
\beqq\tau_{2j-1,2j}^{1,\bmm_k}\left(\begin{array}{cc} \imath y_{2j-1}\\ \imath y_{2j}\end{array}\right)=
\left(\begin{array}{cc} \imath y_{2j-1}-m^1_j\o_1\\ \imath y_{2j}+m^1_j\o_1\end{array}\right),
\qquad \tau_{2j-1,2j}^{2,\bmm_k}\left(\begin{array}{cc} \imath y_{2j-1}\\ \imath y_{2j}\end{array}\right)=
\left(\begin{array}{cc} \imath y_{2j-1}-m^2_j\o_2\\ \imath y_{2j}+m^2_j\o_2\end{array}\right).\eeqq
Set
$$\ve_1=\imath (y_1-y_2)-(m^1_1\o_1+m^2_1\o_2), \qquad\ldots\qquad \ve_k=\imath (y_{2k-1}-y_{2k})-(m^1_k\o_1+m^2_k\o_2),$$
and denote by $F_k(\by_n,\bz_{2n})$ the sum of $4^k$ summands
\beq\label{D16a}F_k(\by_n,\bz_{2n})=\left(\prod_{j=1}^k\big(1+\tau_{2j-1,2j}^{1,\bmm_k}\big) \big(1+\tau_{2j-1,2j}^{2,\bmm_k}\big)
\right)F(\by_n,\bz_{2n})\eeq
The following statement is a direct consequence of Lemma \ref{lemmaD1}.
\begin{lemma}\label{lemmaD2} The function $F_k(\by_n,\bz_{2n})$ has zero of order $2k$ on the intersection of hyperplanes
	\beq\label{D17} \ve_1=\ldots=\ve_k=0.\eeq
	Moreover, its Taylor expansion in a generic point of the plane \rf{D17} starts from $\ve_1^2\cdots\ve_k^2$
	\beq\label{D17a}F_k(\by_n,\bz_{2n})=\ve_1^2\cdots\ve_k^2\cdot H_k(\by_n,\bz_{2n}),\eeq
	where $H_k(\by_n,\bz_{2n})$ is regular at generic point of \rf{D17}.
\end{lemma}
{\bf Proof}.
Using Lemma \ref{lemmaD1} for generic values of all the variables and parameters we have
\beqq\big(1+\tau_{2j-1,2j}^{1,\bmm_k}\big) \big(1+\tau_{2j-1,2j}^{2,\bmm_k}\big)
F(\by_n,\bz_{2n})=\ve_j^2 \cdot H_j(\by_n,\bz_{2n})\eeqq
for any $j=1,\ldots, k$, where $H_j(\by_n,\bz_{2n})$ is analytic function on the hyperplane $$\imath (y_{2j-1}-y_{2j})=m^1_j\o_1+m^2_j\o_2$$ and in particular on the plane \rf{D17}.
Since all operators $\tau_{2j-1,2j}^{1,\bmm_k}$ in the product \eqref{D16a} commute, the same is true for the total expression $F_k$. So, $F_k$ is analytic with respect to $\ve_j$ and has zero of the second order at $\ve_j = 0$ for all $j$. Then its Taylor expansion in $\ve_j$ starts with the $\ve_1^2\cdots\ve_k^2$ term.
\hfill{$\Box$}

\subsubsection{Induction step: fusion of chain integrals}
We are ready now to prove the induction step of Proposition \ref{propD1}. We regard the result $G_m(y_{m+1},\ldots,y_n,\bz_{2n})$ of the first $m$ integrations as an analytical function of parameters $y_{m+1},\ldots,y_n,\bz_{2n}$. Thus, during the integration over the variable $y_{m+1}$ we can assume that all the parameters $y_{m+2},\ldots,y_n,\bz_{2n}$ are generic so that there are no singularities between different factors in each summand of $G_m(y_{m+1},\ldots,y_n,\bz_{2n})$.
It means that the induction step reduces to the consideration of the fusions of chain integrals \rf{D3} and \rf{D2}. Namely we now consider one-dimensional integrals
	\begin{align}\label{D18}
	&\Res _{\imath y_{i_0}=\imath y_{j_0}+c} \, I_k \left({y}_{i_0}| \bi_k,\bmm_k\right) \times I_l \left({y}_{j_0}| \bj_l,\tilde{\bmm}_l\right), \\[4pt]
	&\Res _{\imath y_{i_0}=\imath z_j+c} \, I_k \left({y}_{i_0}| \bi_k,\bmm_k\right) \times J_1(z_a| b,\tilde{m}^1,\tilde{m}^2),\qquad \label{D19}\\[4pt]
	&\Res _{\imath y_{i_0}=\imath z_j+c} \, I_k \left({y}_{i_0}| \bi_k,{\bmm}_k\right).\label{D19a}\end{align}
Below we prove that all such residues cancel each other, except ones that form new chain integrals $I_{k + l + 1}$ and $J_1$. Having in mind that in the original integral \eqref{J4} all the parameters were real, we assume that in the first integral the variable $y_{j_0}$ is real and in the second and third integrals the parameter $z_a$ is also real.

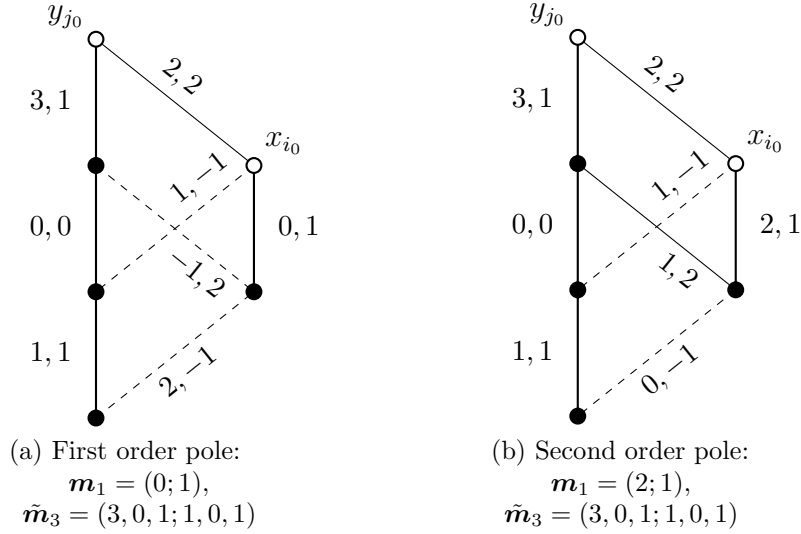
\begin{figure}[h]
	\centering
	\vspace{0.2cm}
	\subfloat[First order pole: \\
	$\begin{array}{c} \bmm_1 = (0;1),\\  \tilde{\bmm}_3 = (3, 0, 1; 1, 0, 1) \end{array} $]{
		\begin{tikzpicture}[scale =1.4]
		\def\l{1.2};
		\def\r{2};
		\def\s{3};
		\draw (0,0) -- node[above, pos=0.45, sloped] {\small $2, 2$} (0.5*\s,-\l);
		\draw[dashed] (0.5*\s,-\l) -- node[above, pos=0.25, sloped] {\small $1, -1$} (0,-2*\l);
		\draw[dashed] (0,-\l) -- node[below, pos=0.72, sloped] {\small $-1, 2$} (0.5*\s,-2*\l) -- node[below, pos=0.5, sloped] {\small $2, -1$} (0,-3*\l);
		
		\draw[thick, fill] (0.5*\s,-\l) -- node[xshift = 0.6cm] {\small$0, 1$} (0.5*\s,-2*\l) circle (\r pt) node[xshift = 0.5cm] {};
		\draw[thick, fill = white] (0.5*\s,-\l) circle (\r pt) node[xshift = 0.4cm, yshift = 0.3cm] {$x_{i_0}$};
		
		\draw[thick, fill] (0,0) -- node[xshift = -0.6cm] {\small $3, 1$} (0,-1*\l) circle (\r pt) -- node[xshift = -0.6cm] {\small $0, 0$} (0,-2*\l) circle (\r pt) -- node[xshift = -0.6cm] {\small $1, 1$} (0,-3*\l) circle (\r pt);
		\draw[thick, fill = white] (0,0) circle (\r pt) node[xshift = -0.4cm, yshift = 0.3cm] {$y_{j_0}$};
		\end{tikzpicture}} \hspace{2cm}
	\subfloat[Second order pole: \\
	$\begin{array}{c} \bmm_1 = (2;1),\\  \tilde{\bmm}_3 = (3, 0, 1; 1, 0, 1) \end{array} $]{
		\begin{tikzpicture}[scale =1.4]
		\def\l{1.2};
		\def\r{2};
		\def\s{3};
		\draw (0,0) -- node[above, pos=0.45, sloped] {\small $2, 2$} (0.5*\s,-\l);
		\draw[dashed] (0.5*\s,-\l) -- node[above, pos=0.25, sloped] {\small $1, -1$} (0,-2*\l);
		\draw (0,-\l) -- node[below, pos=0.72, sloped] {\small $1, 2$} (0.5*\s,-2*\l);
		\draw[dashed] (0.5*\s,-2*\l) -- node[below, pos=0.5, sloped] {\small $0, -1$} (0,-3*\l);
		
		\draw[thick, fill] (0.5*\s,-\l) -- node[xshift = 0.6cm] {\small $2, 1$} (0.5*\s,-2*\l) circle (\r pt) node[xshift = 0.5cm] {};
		\draw[thick, fill = white] (0.5*\s,-\l) circle (\r pt) node[xshift = 0.4cm, yshift = 0.3cm] {$x_{i_0}$};
		
		\draw[thick, fill] (0,0) -- node[xshift = -0.6cm] {\small $3, 1$} (0,-1*\l) circle (\r pt) node[xshift = -0.5cm] {} -- node[xshift = -0.6cm] {\small $0, 0$} (0,-2*\l) circle (\r pt) node[xshift = -0.5cm] {} -- node[xshift = -0.6cm] {\small $1, 1$} (0,-3*\l) circle (\r pt) node[xshift = -0.5cm] {};
		\draw[thick, fill = white] (0,0) circle (\r pt) node[xshift = -0.4cm, yshift = 0.3cm] {$y_{j_0}$};
		\end{tikzpicture}} \vspace{0.1cm}
	\caption{\centering Fusions of chains with different edge parameters. Solid lines correspond to singularities}
	\label{fig:chains-dashed-exmp}
\end{figure}

Consider the residue \rf{D18} first. For better convenience denote the variables
$y_{i_a}$ taking part in the integral $I_k \left({y}_{i_0}| \bi_k,\bmm_k\right)$ by letters
$x_{i_a}$. The residue \rf{D18} could be nonzero only if the integration variable $x_{i_0}$ meets the point corresponding to the pole of the measure function either
\beq\label{D20a} S(\imath(y_{j_b}-x_{i_a})+g^\ast) \qquad\text{or}\qquad S(\imath(x_{i_a}-y_{j_b})+g^\ast) .\eeq
This happens when the variables $\imath(y_{j_b}-x_{i_a})$ or $\imath(x_{i_a}-y_{j_b})$ in the arguments of the functions in \rf{D20a} equal to $g +m^1\o_1+m^2\o_2$ for some non-negative integers $m^1$ and $m^2$, see \rf{A1a}.  Moreover, a number of such singularities can appear together giving a multiple pole. A typical example is shown in Figure \ref{fig:chains-dashed-exmp}. The pairs of numbers on each edge indicate corresponding integers $m^1$ and $m^2$. When one of them is negative, the corresponding edge is dashed, which means the missing of the corresponding singularity. In the case of the multiple singularity  instead of single fusion integral we consider the fusion of several summands of the function $G_m(y_{m+1},\ldots,y_n,\bz_{2n})$ which fit using of Lemma \ref {lemmaD2}. We then justify that such sum either vanishes or it is given by a simple residue of the first order which we analyze further.

Consider each type of singularity \rf{D20a} separately.
In the first singularity imposed by the pole of $S(\imath(y_{j_b} - x_{i_a})+g^\ast)$ we have the relation
\beq\label{D20} \imath y_{j_b}=\imath x_{i_a}+g+p^1\o_1+p^2\o_2,\qquad p^1, p^2\geq 0. \eeq
In the second
\beq\label{D21} \imath x_{i_a}=\imath y_{j_b}+g+q^1\o_1+q^2\o_2,\qquad q^1, q^2\geq 0. \eeq
Consider the first case \rf{D20}, example is shown in Figure \ref{fig:chains-ex}.
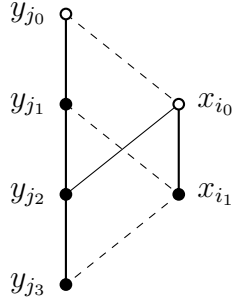
\begin{figure}[t]
	\centering
	\begin{tikzpicture}
	\def\l{1.2};
	\def\r{2};
	\def\s{3};
	\draw[dashed] (0,0) -- (0.5*\s,-\l);
	\draw (0.5*\s,-\l) -- (0,-2*\l);
	\draw[dashed] (0, -\l) -- (0.5*\s,-2*\l) -- (0,-3*\l);
	
	\draw[thick, fill] (0.5*\s,-\l) -- (0.5*\s,-2*\l) circle (\r pt) node[xshift = 0.5cm] {$x_{i_1}$};
	\draw[thick, fill = white] (0.5*\s,-\l) circle (\r pt) node[xshift = 0.5cm] {$x_{i_0}$};
	
	\draw[thick, fill] (0,0) -- (0,-1*\l) circle (\r pt) node[xshift = -0.5cm] {$y_{j_1}$} -- (0,-2*\l) circle (\r pt) node[xshift = -0.5cm] {$y_{j_2}$} -- (0,-3*\l) circle (\r pt) node[xshift = -0.5cm] {$y_{j_3}$};
	\draw[thick, fill = white] (0,0) circle (\r pt) node[xshift = -0.5cm] {$y_{j_0}$};
	\end{tikzpicture}
	\caption{\centering The pole \eqref{D20} with $a = 0$, $b = 2$}
	\label{fig:chains-ex}
\end{figure}
If $a<k$ then in the chain $I_k$ there is a variable $x_{i_{a+1}}$
such that
\beq\label{D21a} \imath x_{i_{a+1}}=\imath x_{i_a}+g+m^1_{a+1}\o_1+m^2_{a+1}\o_2,\qquad m^1_{a+1},m^2_{a+1}\geq 0,\eeq
and at the residue point we have the relation
\beq \label{D22}\imath  y_{j_b}=\imath x_{i_{a+1}}+n^1\o_1+n^2\o_2,\qquad n^i=p^i-m^i_{a+1}.\eeq
This relation coincides with the hyperplane from Lemma \ref{lemmaD1}, which we apply as follows. By definition \eqref{D5a} shift operators $\tau_{j_b, i_{a + 1}}^{i, n^i}$ act on the variables $y_{j_b}$, $y_{i_{a + 1}} \equiv x_{i_{a + 1}}$ as
\beq\tau_{j_b, i_{a + 1}}^{1, n^1}
\left(\!\! \begin{array}{cc} \imath y_{j_b}\\ \imath x_{i_{a + 1}}\end{array} \!\!\right)=
\left(\!\! \begin{array}{cc} \imath y_{j_b}-n^1\o_1\\ \imath x_{i_{a + 1}}+n^1\o_1\end{array} \!\!\right),
\qquad \tau_{j_b, i_{a + 1}}^{2, n^2}\left(\!\! \begin{array}{cc} \imath y_{j_b}\\ \imath x_{i_{a + 1}}\end{array} \!\!\right)=
\left(\!\! \begin{array}{cc} \imath y_{j_b}-n^2\o_2\\ \imath x_{i_{a + 1}}+n^2\o_2\end{array} \!\!\right).\eeq
In our case $y_{j_b}$, $x_{i_{a + 1}}$ are integration variables inside the residue integral \eqref{D18}. By action of shift operators on the residue integral we assume action on its residue points, for instance
\begin{equation}
\tau_{j_b, i_{a + 1}}^{1, n^1} \Res_{\imath x_{i_{a + 1}} = \imath x_{i_a} + g + m_{a + 1}^1 \omega_1 + m_{a + 1}^2 \omega_2} = \Res_{\imath x_{i_{a + 1}} = \imath x_{i_a} + g + (m_{a + 1}^1 - n^1) \omega_1 + m_{a + 1}^2 \omega_2}.
\end{equation}
Then instead of the single residue integral \eqref{D18} consider the sum of four integrals
\beq\label{D23}\Res _{\imath x_{i_0}=\imath y_{j_0}+c}\big(1+\tau^{1,n^1}_{j_b,i_{a+1}}\big)\big(1+\tau^{2,n^2}_{j_b,i_{a+1}}\big)I_k \left({x}_{i_0}| \bi_k,\bmm_k\right) \times I_l \left({y}_{j_0}| \bj_l,\tilde{\bmm}_l\right).\eeq
Here
\beq\label{D23a}
\begin{split}
	c=(b-a-1)g &+ (\tilde{m}_1^1 + \ldots + \tilde{m}_b^1-m_1^1 - \ldots - m_a^1-p^1)\o_1\\
	&+(\tilde{m}_1^2 + \ldots + \tilde{m}_b^2-m_1^2 - \ldots - m_a^2-p^2)\o_2,
\end{split}
\eeq
where $p^1$ and $p^2$ are given by \rf{D20}. The four residue integrals \eqref{D23} differ by parameters on the edges. At the same time by induction assumption the function $G_m$ \eqref{D4} equals to the sum of all possible chain integrals. In particular, it contains direct products $I_k \times I_l$ with all possible non-negative edge parameters
\begin{equation}\label{D23b}
\sum_{m_j^i, \tilde{m}_j^i \geq 0} I_k \left({x}_{i_0}| \bi_k,\bmm_k\right) \times I_l \left({y}_{j_0}| \bj_l,\tilde{\bmm}_l\right).
\end{equation}
Parameters on the edges in the shifted integrals from the sum \eqref{D23} could be negative, depending on the integers $\bmm_k$, $\tilde{\bmm}_l$, $p^i$. However, the chain integral with at least one negative edge parameter equals zero, since, as we noted earlier, the residues in it can be taken in any order and
\begin{equation}
\Res_{\imath y_i = \imath y_j + g + h^1 \omega_1 + h^2 \omega_2} F(\by_n, \bz_{2n}) = 0,
\end{equation}
unless both $h^i \geq 0$. Thus, all of the four residue integrals in \rf{D23} either are contained in the sum \rf{D23b} or equal to zero.

To apply Lemma 3 to the sum \eqref{D23} we shift the integration variables in each integral in order to remove the dependence on edge parameters from residue points. For the chain integral $I_k(x_{i_0}|\bi_k, \bmm_k)$ in \eqref{D23} we define new integration variables $\tilde{x}_{i_s}$ by the following shifts
\begin{equation}\label{D23c}
\begin{aligned}
\imath x_{i_0} &=\imath \tilde{x}_{i_0} + c, \\[4pt]
\imath x_{i_1} &= \imath \tilde{x}_{i_1} + g + m_1^1 \o_1 + m_1^2 \o_2 + c, \\[4pt]
\imath x_{i_2} &= \imath \tilde{x}_{i_2} + 2g + (m_1^1 + m_2^1) \o_1 + (m_1^2 + m_2^2) \o_2 + c, \\
& \;\;\vdots \\
\imath x_{i_k} &= \imath \tilde{x}_{i_k} + kg + (m_1^1 + \ldots + m_k^1) \o_1 + (m_1^2 + \ldots + m_k^2) \o_2 + c
\end{aligned}
\end{equation}
and similarly for all other chain integrals in the sum \eqref{D23}. Now only the integrands depend on edge parameters. Denote two tuples
\begin{equation}
\bx_{k+1} = (x_{i_0}, \ldots, x_{i_k}), \qquad \by_{l+1} = (y_{j_0}, \ldots, y_{j_l})
\end{equation}
with components given by the formulas \eqref{D23c} and analogous ones with $y_{j_d}$. Then we rewrite \eqref{D23} as one single residue integral
\begin{equation}
\begin{aligned}
\Res_{\imath \tilde{x}_{i_0} = \imath \tilde{y}_{j_0}} &\, \prod_{s = 1}^k \Res_{\imath \tilde{x}_{i_s} = \imath \tilde{x}_{i_{s - 1}}} \, \prod_{s = 1}^l \Res_{\imath \tilde{y}_{j_s} = \imath \tilde{y}_{j_{s - 1}}} \\[4pt]
&\quad \big(1+\tau^{1,n^1}_{j_b,i_{a+1}}\big)\big(1+\tau^{2,n^2}_{j_b,i_{a+1}}\big) F(\bx_{k+1}, \by_{l + 1}, \dots).
\end{aligned}
\end{equation}
Here by dots we mean all other variables of the integrand. Lemma \ref{lemmaD1} says that the integrand in the last formula has additional zero of the order two at the hyperplane~\rf{D22}. This double zero compensates two possible simple singularities along the hyperplanes
\beq\label{D23d} \imath y_{j_b}=\imath x_{i_a}+g+p^1\o_1+p^2\o_2,\qquad\text{and}\qquad
\imath y_{j_{b-1}}=\imath x_{i_{a+1}}+g+r^1\o_1+r^2\o_2.\eeq
The same statement holds for the singularity \rf{D21} once $b<k$.

Note also that due to inequalities $p^i\geq0$, $m_{i_{a + 1}}^i\geq 0$, $\Re g\geq0$ $\Re\o_i\geq 0$, all the new points the shifted variables $\imath y_{j_b}$ and $\imath x_{i_{a+1}}$ have positive real part once the variable $\imath x_{i_{a}}$ does have.

Suppose there are no singularities in the residue integral \eqref{D18} except \eqref{D23d}. Then we cancel this integral applying Lemma~\ref{lemmaD1} in the described way. Next assume there are other singularities besides \eqref{D23d}. Among all pairs with singularities let $y_{j_b}, x_{i_a}$ be the one with the smallest index $a$ (upper diagonal line). The corresponding singularity is of the type either \eqref{D20} or \eqref{D21}. Consider the first case~\eqref{D20}. Denote
\begin{equation}
r = \min(k - a - 1, l - b),
\end{equation}
see example in Figure \ref{fig:chains-T}.
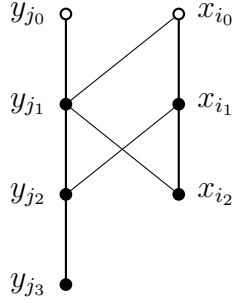
\begin{figure}[t]
	\centering
	\begin{tikzpicture}
	\def\l{1.2};
	\def\r{2};
	\def\s{3};
	\draw (0.5*\s,0*\l) -- (0,-1*\l) -- (0.5*\s,-2*\l);
	\draw (0.5*\s,-1*\l) -- (0,-2*\l);
	
	\draw[thick, fill] (0.5*\s,0*\l) -- (0.5*\s,-1*\l) circle (\r pt) node[xshift = 0.5cm] {$x_{i_1}$} -- (0.5*\s,-2*\l) circle (\r pt) node[xshift = 0.5cm] {$x_{i_2}$};
	\draw[thick, fill = white] (0.5*\s,0*\l) circle (\r pt) node[xshift = 0.5cm] {$x_{i_0}$};
	
	\draw[thick, fill] (0,0) -- (0,-1*\l) circle (\r pt) node[xshift = -0.5cm] {$y_{j_1}$} -- (0,-2*\l) circle (\r pt) node[xshift = -0.5cm] {$y_{j_2}$} -- (0,-3*\l) circle (\r pt) node[xshift = -0.5cm] {$y_{j_3}$};
	\draw[thick, fill = white] (0,0) circle (\r pt) node[xshift = -0.5cm] {$y_{j_0}$};
	\end{tikzpicture}
	\caption{\centering Third order pole with $a = 0$, $b = 1$ and $r = 1$}
	\label{fig:chains-T}
\end{figure}
Then there are $r + 1$ pairs of variables $y_{j_{b + \alpha}}, x_{i_{a + 1 + \alpha}}$ with \mbox{$\alpha \in \{0, \ldots, r \}$}, for which we have relations
\begin{equation}\label{D23e}
\imath y_{j_b} = \imath x_{i_{a + 1}} + n^1_{a,b}\omega_1 + n^2_{a,b}\omega_2, \quad \ldots \quad \imath y_{j_{b + r}} = \imath x_{i_{a + 1 + r}} + n^1_{a+r,b+r}\omega_1 + n^2_{a+r,b+r}\omega_2.
\end{equation}

Now we apply Lemma \ref{lemmaD2} for the intersection of hyperplanes \eqref{D23e} repeating the procedure described above for Lemma \ref{lemmaD1}. Let us consider the sum of $4^{r+1}$ integrals
\beq\label{D24}
\Res _{\imath x_{i_0}=\imath y_{j_0}+c}T_{kl} \, I_k \left({x}_{i_0}| \bi_k,\bmm_k\right)I_l \left({y}_{j_0}| \bj_l,\tilde{\bmm}_l\right)
\eeq
where $T_{kl}$ denotes the sum of shift operators
\beq \label{D24a} T_{kl}=
\prod_{\a = 0}^r \bigl(1+\tau^{1,n^1_{a+\a,b+\a}}_{j_{b+\a},i_{a+1+\a}}\bigr)\bigl(1+\tau^{2,n^2_{a+\a,b+\a}}_{j_{b+\a},i_{a+1+\a}}\bigr).
\eeq
This operator shifts the parameters on the edges of chain integrals $I_k, I_l$. The arguments above show that all such integrals are either contained in the sum \eqref{D23b} or equal to zero. Therefore, we can consider the sum \eqref{D24} instead of the single residue~\eqref{D18}.

Making linear changes of variables analogous to \rf{D23c} we may regard this sum of integrals as a residue integral
\begin{equation}
\Res_{\imath \tilde{x}_{i_0} = \imath \tilde{y}_{j_0}} \, \prod_{s = 1}^k \Res_{\imath \tilde{x}_{i_s} = \imath \tilde{x}_{i_{s - 1}}} \, \prod_{s = 1}^l \Res_{\imath \tilde{y}_{j_s} = \imath \tilde{y}_{j_{s - 1}}} T_{kl}\,  F(\bx_{k+1}, \by_{l + 1}, \dots),
\end{equation}
where $\bx_{k+1}$ components are given by \eqref{D23c} and similarly for $\by_{l+1}$. By Lemma \ref{lemmaD2} the number of poles in this one-dimensional integral could exceed the number of zeros of integrand by one only in two cases.
\begin{itemize}
	\item[\bfseries{I.}] We have the singularity
	\beq\label{D25} \imath y_{j_d}=\imath x_{i_k}+g+p^1\o_1+p^2\o_2,\qquad p^1, p^2\geq 0. \eeq
	\item[\bfseries{II.}] We have the singularity
	\beq\label{D26} \imath x_{i_d}=\imath y_{j_l}+g+q^1\o_1+q^2\o_2,\qquad q^1, q^2 \geq 0. \eeq
\end{itemize}
Otherwise the sum \eqref{D24} vanishes. Similarly, for the residue between pair of variables $x_{i_a}, y_{j_b}$ of the second type \eqref{D21} we have vanishing sum, unless the same two cases.

Consider the case {\bf I}. Note that $d>0$ since otherwise $\Im x_{i_0} > 0$. Define two new chain integrals $I_{k+l-d+1}\left({x}_{i_0}| \bi'_{k+l-d+1},\bmm'_{k+l-d+1}\right)$ and $I_{d-1} \left({y}_{j_0}| \bj'_{d-1},\tilde{\bmm}'_{d-1}\right)$ as follows
	\beq\label{D27}\begin{array}{llll}
	s\leq k\colon \hspace{1.5cm}  & i'_s=i_s, &{m'}^i_s=m^i_s, \\[4pt]
	s = k + 1 \colon & i'_{k + 1}=j_{d},  & {m'}^i_{k + 1}=p^i, \\[4pt]
	s > k + 1 \colon & i'_s=j_{d-1+s-k}, \hspace{0.8cm}  & {m'}^i_s=\tilde{m}^i_{d-1+s-k},\\[4pt]
   	s<d\colon & j'_s=j_s, & {\tilde{m}'}\-^i_s={\tilde{m}}^i_s.\end{array}
   	\eeq
   	\color{black}
   	This definition is rather simple in terms of pictures, see example in Figure \ref{fig:chains-simple-pole}. The following lemma describes cancellation mechanism for such integrals.
	\begin{figure}[t]
	\centering
	\begin{tikzpicture}
		\def\l{1.2};
		\def\r{2};
		\def\s{3};
		\draw (0.5*\s,0*\l) -- (0,-1*\l);
		\draw (0, 0) -- (0.5*\s,-1*\l) -- (0,-2*\l);
		
		\draw[thick, fill] (0.5*\s,0*\l) -- (0.5*\s,-1*\l) circle (\r pt) node[xshift = 0.5cm] {$x_{i_1}$};
		\draw[thick, fill = white] (0.5*\s,0*\l) circle (\r pt) node[xshift = 0.5cm] {$x_{i_0}$};
		
		\draw[thick, fill] (0,0) -- (0,-1*\l) circle (\r pt) node[xshift = -0.5cm] {$y_{j_1}$} -- (0,-2*\l) circle (\r pt) node[xshift = -0.5cm] {$y_{j_2}$} -- (0,-3*\l) circle (\r pt) node[xshift = -0.5cm] {$y_{j_3}$};
		\draw[thick, fill = white] (0,0) circle (\r pt) node[xshift = -0.5cm] {$y_{j_0}$};
	\end{tikzpicture}\hspace{2cm}
	\begin{tikzpicture}
		\def\l{1.2};
		\def\r{2};
		\def\s{3};
		\draw (0.5*\s,0*\l) -- (0,-1*\l) -- (0.5*\s,-2*\l);
		\draw (0,0) -- (0.5*\s,-1*\l);
		
		\draw[thick, fill] (0.5*\s,0*\l) -- (0.5*\s,-1*\l) circle (\r pt) node[xshift = 0.5cm] {$x_{i_1}$} -- (0.5*\s,-2*\l) circle (\r pt) node[xshift = 0.5cm] {$y_{j_2}$} -- (0.5*\s,-3*\l) circle (\r pt) node[xshift = 0.5cm] {$y_{j_3}$};
		\draw[thick, fill = white] (0.5*\s,0*\l) circle (\r pt) node[xshift = 0.5cm] {$x_{i_0}$};
		
		\draw[thick, fill] (0,0) -- (0,-1*\l) circle (\r pt) node[xshift = -0.5cm] {$y_{j_1}$};
		\draw[thick, fill = white] (0,0) circle (\r pt) node[xshift = -0.5cm] {$y_{j_0}$};
\end{tikzpicture}
	\caption{\centering Singularity of the type \eqref{D25} with $d = 2$ and the corresponding pair of new chains from the right}
	\label{fig:chains-simple-pole}
	\end{figure}
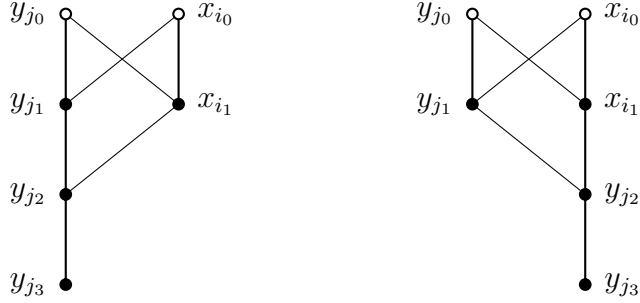
   	
\begin{lemma}\label{lemmaD3}
   	\beq \label{D28}\begin{split}&\Res _{\imath x_{i_0}=\imath y_{j_0}+c}T_{kl} \, I_k \left({x}_{i_0}| \bi_k,\bmm_k\right)I_l \left({y}_{j_0}| \bj_l,\tilde{\bmm}_l\right)+\\[4pt]
   	&\Res _{\imath x_{i_0}=\imath y_{j_0}+c}T_{kl} \, I_{k+l-d+1}\left({x}_{i_0}| \bi'_{k+l-d+1},\bmm'_{k+l-d+1}\right)I_{d-1} \left({y}_{j_0}| \bj'_{d-1},\tilde{\bmm}'_{d-1}\right)=0,
   		\end{split}\eeq
   	 where the operator $T_{kl}$ is defined in \rf{D24a}.
\end{lemma}

   	{\bf Proof}. Indeed, by the arguments above both integrals are simple residue integrals with the same integrand and singularities. In the shifted variables $\tx_{i_s}$ and $\ty_{j_s}$, all the singularities are at the diagonals
   	\begin{equation}
   	\begin{aligned}\label{D28a}
	&\tx_{i_s}=\tx_{i_{s+1}},  &s=0,\ldots, k-1, & \hspace{1.3cm} \ty_{j_s}=\ty_{j_{s+1}}, & s=0,\ldots, l-1,\\[3pt]
	&\tx_{i_s}=\ty_{j_{s+d-k}}, & s=a,\ldots,k, & \hspace{1.3cm} \tx_{i_{s+1}}=\ty_{j_{s+d-k-1}}, \;\; & s=a,\ldots,k-1,
	\end{aligned}
   	\end{equation}
   	where $\ty_{j_0} = y_{j_0}$. Lemma \ref{lemmaD2} says that we may present the integrand in the form
   	\beqq\frac{\big(\tx_{i_{a+1}}-\ty_{j_{a+d-k}}\big)^2 \cdots\big(\tx_{i_{k}}-\ty_{j_{d-1}}\big)^2}
   		{\prod\limits_{s=0}^{k-1}	\big(\tx_{i_s}-\tx_{i_{s+1}}\big)\prod\limits_{s=0}^{l-1}\big(\ty_{j_s}-\ty_{j_{s+1}}\big)
   		\prod\limits_{s=a}^k\big(\tx_{i_s}-\ty_{j_{s+d-k}}\big)	\prod\limits_{s=a}^{k-1}\big(\tx_{i_{s+1}}-\ty_{j_{s+d-k-1}}\big)} H\big(\tilde{\bx}_{k+1},\tilde{\by}_{l+1}\big)\eeqq
   	where $H\big(\tilde{\bx}_{k + 1},\tilde{\by}_{l + 1}\big)$ does not have singularities on integration contour.
   	Replacing each factor $\big(\tx_{i_{s+1}}-\ty_{j_{s+d-k}}\big)^2$ of the nominator by
   	\beqq \big((\tx_{i_{s}}-\tx_{i_{s-1}})+(\tx_{i_{s-1}}-\ty_{j_{s+d-k}})\big)
   	\big((\tx_{i_{s}}-\ty_{j_{s+d-k-1}})+(\ty_{j_{s+d-k-1}} - \ty_{j_{s+d-k}})\big)\eeqq
   	we obtain the sum of simple fractions such that the number of factors in denominator of each of them  equals the number of integration. One can then also note that only one  fraction
   	\beq \label{D29}\frac{1}{\big(\tx_{i_k}-\ty_{j_{d}}\big)\prod\limits_{s=0}^{k-1}	\big(\tx_{i_s}-\tx_{i_{s+1}}\big)\prod\limits_{s=0}^{l-1}\big(\ty_{j_s}-\ty_{j_{s+1}}\big)}
   		H\big(\tilde{\bx}_{k+1},\tilde{\by}_{l+1}\big)\eeq
   	gives nontrivial contribution to  the both integrals \eqref{D28}. The corresponding integrals can be computed. The first one equals to $$H\big(\tilde{\bx}_{k+1},\tilde{\by}_{l+1}\big) \bigr|_{\tx_{i_s}=\ty_{j_0},\,s=0,\ldots, k; \;\, \ty_{j_s}=\ty_{j_0}, \, s=1,\ldots, l} \;,$$ the second to $$ -H\big(\tilde{\bx}_{k+1},\tilde{\by}_{l+1}\big) \bigr|_{\tx_{i_s}=\ty_{j_0},\,s=0,\ldots, k; \;\, \ty_{j_s}=\ty_{j_0},\, s=1,\ldots, l} \;,$$ so that their sum equals zero. \hfill{$\Box$}	
   	\medskip
   	
   	The same arguments hold for fusions of type {\bf II} unless $d\not=0$. For $d = 0$ the corresponding fusion of chain integrals is by definition the chain integral
   	\beq\label{D30}\frac{1}{-2\pi \imath}I_{k+l+1}\big(y_{i_0}|\bi'_{k+l+1},\bmm'_{k+l+1}\big)\eeq
   	where
	\beqq
	\begin{array}{lll}
   	s\leq l \colon  & i'_s=j_s,   &{m'}^i_s={\tilde{m}}^i_s,\\[4pt]
   	s = l+1 \colon  & i'_{l + 1}=i_{0},   &{m'}^i_{l + 1}=q^i,\\[4pt]
   	s>l + 1 \colon \qquad & i'_s=i_{s-l-1}, \quad &{m'}^i_s = {m}^i_{s-l-1}.
   \end{array}
	\eeqq
	\color{black} 	
   	Here the integers $q^i$ are given by the relation \rf{D26}. See example in Figure \ref{fig:chains-new}.
   	\begin{figure}[h]
   		\centering
   		\begin{tikzpicture}
   		\def\l{1.2};
   		\def\r{2};
   		\def\s{3};
   		\draw (0,-\l) -- (0.5*\s,-2*\l);
   		
   		\draw[thick, fill] (0.5*\s,-2*\l) -- (0.5*\s,-3*\l) circle (\r pt) node[xshift = 0.5cm] {$x_{i_1}$};
   		\draw[thick, fill = white] (0.5*\s,-2*\l) circle (\r pt) node[xshift = 0.5cm] {$x_{i_0}$};
   		
   		\draw[thick, fill] (0,0) -- (0,-1*\l) circle (\r pt) node[xshift = -0.5cm] {$y_{j_1}$};
   		\draw[thick, fill = white] (0,0) circle (\r pt) node[xshift = -0.5cm] {$y_{j_0}$};
   		\end{tikzpicture}
   		\caption{\centering Two chain integrals give a new one}
   		\label{fig:chains-new}
   	\end{figure}
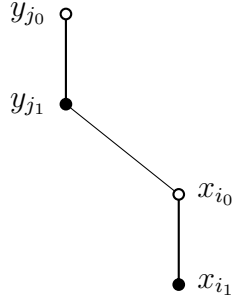

   	This ends the consideration of fusions of chains integrals \rf{D18}. All of them cancel, except new chains \rf{D30}.
   	\medskip
   	
   	Consider now the fusion \rf{D19}
   	\beq\label{D31}\Res _{\imath x_{i_0}=\imath z_j+c}I_k \left({y}_{i_0}| \bi_k,\bmm_k\right)J_1(z_j| b,\tilde{m}^1,\tilde{m}^2).\eeq
   	Here we again change symbols $y$ to $x$ in the chain $I_k$.
   	There could appear such singularities, labeled by a fixed number $a$:
   	\begin{align}
   	&\label{D32} \imath x_{i_a}= \imath z_j+\frac{g^\ast}{2}+p^1\o_1+p^2\o_2, & p^1,p^2\geq 0,\\[4pt]
   	&\label{D32a} \imath x_{i_{a+1}}=\imath y_b+g+q^1\o_1+q^2\o_2, &q^1,q^2\geq 0,\\[5pt]
   	&\label{D32b} \imath y_b=\imath x_{i_{a-1}}+g+r^1\o_1+r^2\o_2,& r^1,r^2\geq 0. \end{align}
   	See example in Figure \ref{fig:chains-J}. Once there is a pole \rf{D32a}, there is a zero  at the point
   	\beq\label{D33} \imath x_{i_{a+1}}= \imath z_j-\frac{g^\ast}{2}+s^1\o_1+s^2\o_2,\qquad s^i = \tilde{m}^i +q^i + 1> 0\eeq
   	given by the function $S^{-1}(\imath x_{i_{a + 1}} - \imath z_j + g^*/2)$.
   	Once we have a singulariry \rf{D32} or \rf{D32b}, we are in the position of Lemma \ref{lemmaD1} with respect to the variables $y_b$ and $x_{i_a}$ and consider instead of the residue  \rf{D31} the corresponding sum of four integrals with shift operators, for which we have additional zero of the second order.
   		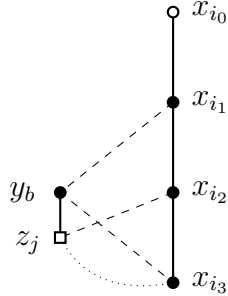
\begin{figure}[h]
   		\centering
   		\begin{tikzpicture}
   		\def\l{1.2};
   		\def\r{2};
   		\def\s{3};
   		\draw[dashed] (0,-\l) -- (-0.5*\s,-2*\l) -- (0,-3*\l);
   		\draw[dashed] (0, -2*\l) -- (-0.5*\s,-2*\l-0.5*\l);
   		\draw[dotted] (-0.5*\s,-2*\l-0.5*\l) .. controls (-0.4*\s,-3*\l) and (-0.2*\s,-3.1*\l) .. (0,-3*\l);
   		
   		\draw[thick, fill] (-0.5*\s,-2*\l)  circle (\r pt) node[xshift = -0.5cm] {$y_b$} -- (-0.5*\s, -2*\l-0.5*\l);
   		\draw[thick, fill = white] (-0.5*\s cm -\r pt, -2*\l cm -0.5*\l cm - \r pt ) rectangle (-0.5*\s cm + \r pt, -2*\l cm -0.5*\l cm + \r pt)  node[xshift = -0.5cm, yshift = - 2*\r pt]{$z_j$};
   		
   		\draw[thick, fill] (0,0) -- (0,-1*\l) circle (\r pt) node[xshift = 0.5cm] {$x_{i_1}$} -- (0,-2*\l) circle (\r pt) node[xshift = 0.5cm] {$x_{i_2}$} -- (0,-3*\l) circle (\r pt) node[xshift = 0.5cm] {$x_{i_3}$};
   		\draw[thick, fill = white] (0,0) circle (\r pt) node[xshift = 0.5cm] {$x_{i_0}$};
   		\end{tikzpicture}
   		\caption{\centering Singularities \eqref{D32}, \eqref{D32a}, \eqref{D32b} are shown with dashed lines ($a = 2$). Dotted line shows a possible zero \eqref{D33}}
   		\label{fig:chains-J}
   	\end{figure}
   	
   	Analyzing the balance of poles and zeros we conclude that the residue \rf{D31} could be nontrivial only when
   	$a=k+1$, so that the variables $x_{i_a}$ and $x_{i_{a+1}}$ are missing and the residue \rf{D31} is taken along
   		the only singularity
   		\beq\label{D34} \imath y_b=\imath x_{i_{k}}+g+r^1\o_1+r^2\o_2,\qquad r^1,r^2\geq 0.\eeq
   		
   	Finally, consider the residue \eqref{D19a}. In the calculation
   	\beqq \Res _{\imath x_{i_0}=\imath z_j+c}I_k \left({y}_{i_0}| \bi_k,{\bmm}_k\right)\eeqq
   	we could meet only one pole of the form \rf{D32} together with zero \rf{D33} if the variable
   	$x_{i_{a+1}}$ exists, see Figure \ref{fig:chains-z}. Thus, the only nontrivial result could be only when $k=a$ and the residue is taken along
   	the singularity
   	\beq\label{D35} \imath x_{i_k}=\imath z_j+\frac{g^\ast}{2}+p^1\o_1+p^2\o_2,\qquad p^1,p^2\geq 0.\eeq
   	One can note that the resulting integrals obtained in the calculations of the fusion integrals
   	\rf{D19} and \rf{D19a} coincide up to a sign and cancel each other except for the case $k=0$ in \rf{D19a}, see Figure \ref{fig:chains-z}.
   		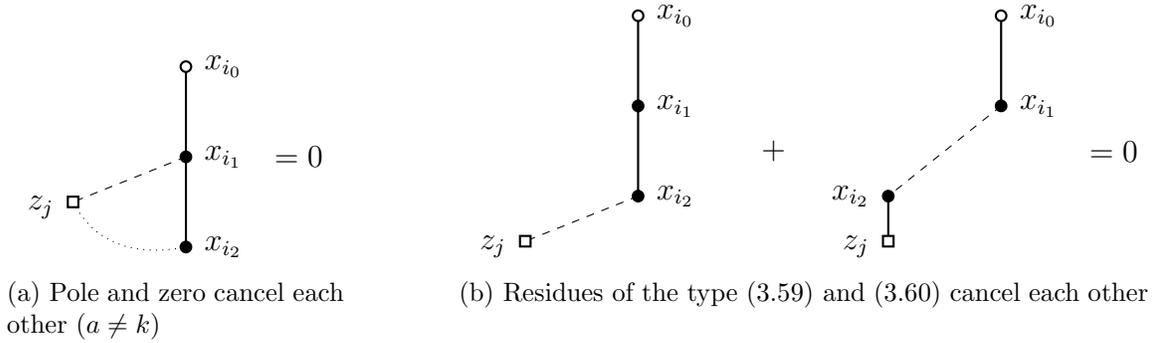
\begin{figure}[h]
   		\centering
   		\subfloat[Pole and zero cancel each other ($a\not=k$)]{
   			\begin{tikzpicture}
   			\def\l{1.2};
   			\def\r{2};
   			\def\s{3};
   			\draw (0.5*\s, -2*\l) node {$ = 0$};
   			\draw[dashed] (0, -2*\l) -- (-0.5*\s,-2*\l-0.5*\l);
   			\draw[dotted] (-0.5*\s,-2*\l-0.5*\l) .. controls (-0.4*\s,-3*\l) and (-0.2*\s,-3.1*\l) .. (0,-3*\l);
   			
   			\draw[thick, fill = white] (-0.5*\s cm -\r pt, -2*\l cm -0.5*\l cm - \r pt ) rectangle (-0.5*\s cm + \r pt, -2*\l cm -0.5*\l cm + \r pt)  node[xshift = -0.5cm, yshift = - 2*\r pt]{$z_j$};
   			
   			\draw[thick, fill] (0,-1*\l) -- (0,-2*\l) circle (\r pt) node[xshift = 0.5cm] {$x_{i_1}$} -- (0,-3*\l) circle (\r pt) node[xshift = 0.5cm] {$x_{i_2}$};
   			\draw[thick, fill = white] (0,-1*\l) circle (\r pt) node[xshift = 0.5cm] {$x_{i_0}$};
   			\end{tikzpicture}
   		}\hspace{1.4cm}
   		\subfloat[Residues of the type \eqref{D19} and \eqref{D19a} cancel each other]{
   			\begin{tikzpicture}
   			\def\l{1.2};
   			\def\r{2};
   			\def\s{3};
   			\draw[dashed] (0, -3*\l) -- (-0.5*\s,-3*\l-0.5*\l);
   			
   			\draw[thick, fill = white] (-0.5*\s cm -\r pt, -3*\l cm -0.5*\l cm - \r pt ) rectangle (-0.5*\s cm + \r pt, -3*\l cm -0.5*\l cm + \r pt)  node[xshift = -0.5cm, yshift = - 2*\r pt]{$z_j$};
   			
   			\draw[thick, fill] (0,-1*\l) -- (0,-2*\l) circle (\r pt) node[xshift = 0.5cm] {$x_{i_1}$} -- (0,-3*\l) circle (\r pt) node[xshift = 0.5cm] {$x_{i_2}$};
   			\draw[thick, fill = white] (0,-1*\l) circle (\r pt) node[xshift = 0.5cm] {$x_{i_0}$};
   			
   			\draw (0.6*\s, -2.5*\l) node {$+$};
   			\end{tikzpicture}\hspace{0.2cm}
   			\begin{tikzpicture}
   			\def\l{1.2};
   			\def\r{2};
   			\def\s{3};
   			\draw[dashed] (0, -2*\l) -- (-0.5*\s,-3*\l);
   			
   			\draw[thick, fill] (-0.5*\s,-3*\l) circle (\r pt) node[xshift = -0.5cm] {$x_{i_2}$} -- (-0.5*\s, -3.5*\l);
   			\draw[thick, fill = white] (-0.5*\s cm -\r pt, -3*\l cm -0.5*\l cm - \r pt ) rectangle (-0.5*\s cm + \r pt, -3*\l cm -0.5*\l cm + \r pt)  node[xshift = -0.5cm, yshift = - 2*\r pt]{$z_j$};
   			
   			\draw[thick, fill] (0,-1*\l) -- (0,-2*\l) circle (\r pt) node[xshift = 0.5cm] {$x_{i_1}$};
   			\draw[thick, fill = white] (0,-1*\l) circle (\r pt) node[xshift = 0.5cm] {$x_{i_0}$};
   			\draw (0.5*\s, -2.5*\l) node {$= 0$};
   			\end{tikzpicture}}
   		\caption{\centering Cancellation of residues}
   		\label{fig:chains-z}
   	\end{figure}
   But for $k = 0$ this gives precisely the new chain $J_1\big(z_j|i_0,p^1,p^2\big)$. For their direct products $J_1(z_a| b,m^1,m^2) \times J_1(z_c| d,n^1,n^2)$ the indices $a$ and $c$ could not coincide since otherwise we can apply Lemma \ref{lemmaD1} to the variables $y_b$ and $y_d$ and cancel the sum of four corresponding terms.

   	This ends the proof of Proposition \ref{propD1}. \hfill{$\Box$}
   	
   	\subsection{Simple poles calculations}
   	In the previous subsections we proved that
   	\begin{enumerate}
   		\item Both integrals in \rf{J5} can be calculated by residues, moving the contours of integration either to the lower or to the upper half planes depending on the sign of~$\Re\lambda$;
   		
   		\item In residue calculations only direct products of simple poles $J_1\big(z_a|b,m^1,m^2\big)$ \eqref{D3} with distinct $z_a$ contribute to the integrals,
   		see Corollary \ref{corD1} to Proposition \ref{propD1}.
   	\end{enumerate}   	
   	
   	Now consider the integral $Q_{n}(\bz_{2n}; \l)$ in the left hand side of \eqref{J5}. By Corollary~\ref{corD1} it equals to the sum of all possible direct products   	
   	\beq\label{H0} J_1\big(z_{a_1}|b_1,m^1_1,m^2_1\big)\times\ldots\times J_1\big(z_{a_n}|b_n,m^1_n,m^2_n\big) \eeq
   	with distinct indices $a_1,\ldots, a_n$. These indices form a subset $I\subset [2n]=\{1,\ldots, 2n\}$ of cardinality $n$. Collecting the direct products with a given choice of the set $I$ we arrive at the sum of $\binom{2n}{n}$ series
   	\beq\label{H1} Q_{n}(\bz_{2n}; \l) = \sum_{\substack{I\subset [2n]\\|I|=n}}e^{{2\pi\l}\bigl(\frac{ng^\ast}{2}+\imath\sum\limits_{i\in I}z_i\bigr)}L^I(u,v)\eeq
   	over
   	\beq\label{H1a} u=e^{{2\pi  \l}{\o_1}} \qquad \text{and}\qquad v= e^{{2\pi  \l\o_2}}.\eeq
   	The series converges for sufficiently big negative $\Re\l$. More precisely, denote two sequences of non-negative integers
   	\begin{equation}\label{H1b}
   	\bmm^1 = (m_1^1, \dots, m_n^1), \qquad \bmm^2 = (m_1^2, \dots, m_n^2)
   	\end{equation}
   	and the sum of their components in a standard way
   	\begin{equation}
   	|\bmm^i | = m_1^i + \ldots + m_n^i.
   	\end{equation}
   	Then the function $L^I(u, v)$ equals
   	\beq\label{H3}L^I(u,v)=n!\, {(-2\pi\imath)}^n\sum_{M,K\geq 0}L^I_{M,K}u^Mv^K
   	\eeq
   	where $L^I_{M,K}$ is the following sum of multiple residues
   	\beq\label{H4}\begin{split}L^I_{M,K}=
   		\sum_{\substack{m_i^1\geq 0\colon |\bmm^1|=M\\[3pt] m_i^2\geq 0\colon | \bmm^2|=K}} L^I_{\bmm^1,\bmm^2}\end{split}\eeq
   	with \beq\label{H5}\begin{split}
   		L^I_{\bmm^1,\bmm^2}&=\Res_{\imath y_1=\imath z_{i_1}+\frac{g^\ast}{2}+m^1_1\o_1+m_1^2\o_2} \cdots\\[4pt]
   		&\cdots\, \Res_{\imath y_n=\imath z_{i_n}+\frac{g^\ast}{2}+m^1_n\o_1+m_n^2\o_2}\, \mu(\by_n) \prod_{a=1}^{2n} \prod_{j=1}^n K(y_j - z_a).
   	\end{split}
   	\eeq
   	
   	In the same way we compute the integral in the right hand side of \rf{J5}, but moving the integration contours to the upper half plane. Again it is expressed via the sum of chain integrals with different sign due to the opposite orientation of the contours. Collecting the terms where the indices $a_j$ of the parameters $z_{a_j}$ belong to a given subset $J\subset [2n]$ of cardinality $n$ we get a sum of $\binom{2n}{n}$ series
   	\beq\label{H6} Q_n(\bz_{2n};-\l)= \sum_{\substack{J\subset [2n]\\|J|=n}}e^{{-2\pi\l}\bigl(-\frac{ng^\ast}{2}+\imath\sum\limits_{i\in J}z_i\bigr)}R^J(u,v)\eeq
   over the same variables $u$ and $v$ defined in \rf{H1a}. Here
   	\beq\label{H7}R^J(u,v)=n!\,(2\pi\imath)^n\sum_{M,K\geq 0}R^J_{M,K}u^Mv^K
   	\eeq
   	where $R^J_{M,K}$ is the following sum of multiple residues
   	\beq\label{H8}\begin{split}R^J_{M,K}=
   		\sum_{\substack{m_i^1\geq 0\colon |\bmm^1|=M\\[3pt] m_i^2\geq 0\colon | \bmm^2|=K}} R^J_{\bmm^1,\bmm^2}\end{split}\eeq
   	with \beq\label{H9}\begin{split}
   		R^J_{\bmm^1,\bmm^2} &=\Res_{\imath y_1=\imath z_{j_1}-\frac{g^\ast}{2}-m^1_1\o_1-m_1^2\o_2} \cdots\\[4pt]
   		&\cdots\, \Res_{\imath y_n=\imath z_{j_n}-\frac{g^\ast}{2}-m^1_n\o_1-m_n^2\o_2} \, \mu(\by_n) \prod_{a=1}^{2n} \prod_{j=1}^n K(y_j - z_a).
   	\end{split}
   	\eeq
   	For any subset $I\subset[2n]$ of cardinality $n$ denote by $\bar{I}$  the complement
   	\beqq \bar{I} =[2n]\setminus I\eeqq
   	of $I$ in the set $[2n]$.
   	\begin{proposition}\label{propH1} For any $I\subset[2n]$, $|I|=n$ we have the equality of series
   		\beq \label{H10} L^I(u,v)=(-1)^nR^{\bar{I}}(u,v)\eeq
   		Equivalently,
   		\beq\label{H11} L^I_{M,K}=(-1)^n R^{\bar{I}}_{M,K}\qquad\text{for any}\qquad M,K\geq0.\eeq \end{proposition}
   	
   	Proposition \ref{propH1} immediately implies the equality \rf{J5} due to \rf{H1} and \rf{H6}. Thus it also implies the commutativity of $Q$-operators \rf{J1}.
   	\medskip
   	
   	{\bf Proof} of Proposition \ref{propH1}.  It is clear by symmetry arguments, that it is sufficient to prove the equalities \rf{H11} for the set $I_0=\{1,\ldots,n\}$. For the sake of convenience below we change the notation for variables $z_a$ in the following way
   	\beq \label {H12} \imath z_a \rightarrow z_a,\qquad a=1,\ldots,n,\qquad \imath z_{n + i} \rightarrow x_i,\qquad i=1,\ldots,n.\eeq
   	Using \rf{trig5} we get the precise value of the multiple residue \rf{H5},
   	\beq\label{H13}\begin{split}
   		L^{I_0}_{\bmm^1,\bmm^2}=&\frac{(\o_1\o_2)^{n/2}}{(-2\pi \imath)^n}\prod_{a=1}^n\frac{(-1)^{m^1_a m^2_a+m_a^1+m_a^2}
   			\,\S^{-1}(g^\ast+m^1_a\o_1+m^2_a\o_2)}{\prod_{j=1}^{m^1_a}2\sin\frac{\pi j\o_1}{\o_2}\prod_{l=1}^{m_a^2}2\sin\frac{\pi l\o_2}{\o_1}}\times\\
   		&\prod_{\substack{a,b=1 \\ a\not=b}}^n\Bigl[\S(z_a-z_b+(m^1_a-m^1_b)\o_1+(m_a^2-m^2_b)\o_2)\\
   		&\S(z_a-z_b+g^\ast+(m^1_a-m^1_b)\o_1+(m^2_a-m^2_b)\o_2)
   		\\
   		&\S^{-1}(z_a-z_b+g^\ast+m_a^1\o_1+m^2_a\o_2)\,\S^{-1}(z_a-z_b-m_b^1\o_1-m^2_b\o_2)\Bigl]\times\\
   		&\prod_{i,a=1}^n \S^{-1}(z_a-x_i+g^\ast+m_a^1\o_1+m^2_a\o_2)\,\S^{-1}(x_i-z_a-m_a^1\o_1-m_a^2\o_2).	
   	\end{split}\eeq
   	For a variable $x$ and integer $m,k$ denote by $\Pochab{x}{m}{k}$ the hyperbolic analog of the Pochhammer symbol:
   	\beq\label{trig6}\Pochab{x}{m}{k}=(-1)^{mk}\frac{S_2(x)}{S_2(x+m\o_1+k\o_2)}.
   	\eeq
   	For non-negative $m$ and $k$
   	\beq\Pochab{x}{m}{k}=\prod_{s=0}^{m-1}2\sin\pi\frac{x+ s\o_1}{\o_2}\prod_{l=0}^{k-1}2\sin\pi\frac{x+ l\o_2}{\o_1}.\eeq
   	In these notations the expression \rf{H13} can be rewritten as follows
   	\begin{align}\nonumber
   		&L^{I_0}_{\bmm^1,\bmm^2}=\frac{(\o_1\o_2)^{n/2}}{(-2\pi \imath \S(g^\ast))^n}
   		\prod_{i,a=1}^n \S^{-1}(z_a-x_i+g^\ast)\, \S^{-1}(x_i-z_a)	\times\\  \nonumber
   		&\prod_{a=1}^n \frac{\Pochab{g^\ast}{m_a^1}{m_a^2}}{\Pochab{\o_1+\o_2}{m_a^1}{m_a^2}}
   		\prod_{a\not=b}\frac{\Pochab{z_a-z_b+g^\ast+(m^1_a-m^1_b)\o_1+(m^2_a-m^2_b)\o_2}{m^1_b}{m^2_b}}
   		{\Pochab{z_a-z_b-m^1_b\o_1-m^2_b\o_2}{m^1_a}{m^2_a}}\times\\  \label{H14a}
   		&\prod_{i,a=1}^n\frac{\Pochab{z_a-x_i+g^\ast}{m_a^1}{m_a^2}}{\Pochab{x_i-z_a-m_a^1\o_1-m_a^2\o_2}{m_a^1}{m_a^2}}.\qquad\qquad\qquad\qquad\qquad\qquad
   	\end{align}
 The inversion formula \rf{trig4b} implies the following symmetry of the hyperbolic Pochhammer symbol:
   	\beq \Pochab{x-m\o_1-k\o_2}{m}{k}=\Pochab{\o_1+\o_2-x}{m}{k}.\eeq
   	With its use we simplify the relation \rf{H14a} as follows
   	\begin{align}\nonumber
   		L^{I_0}_{\bmm^1,\bmm^2}=&\frac{(\o_1\o_2)^{n/2}}{(-2\pi \imath \S(g^\ast))^n}
   		\prod_{i,a=1}^n \S^{-1}(z_a-x_i+g^\ast)\, \S^{-1}(x_i-z_a)	\times\\  \nonumber
   		&	\prod_{a=1}^n \frac{\Pochab{g^\ast}{m_a^1}{m_a^2}}{\Pochab{\o_1+\o_2}{m_a^1}{m_a^2}}
   		\prod_{a\not=b}\frac{\Pochab{z_a-z_b+g-m^1_b\o_1-m^2_b\o_2}{m^1_a}{m^2_a}}
   		{\Pochab{z_a-z_b-m^1_b\o_1-m^2_b\o_2}{m^1_a}{m^2_a}}\times\\  \label{H16}
   		&\prod_{i,a=1}^n\frac{\Pochab{z_a-x_i+g^\ast}{m_a^1}{m_a^2}}{\Pochab{z_a-x_i+\o_1+\o_2}{m_a^1}{m_a^2}}.
   	\end{align}
   	Analogous calculations for the multiple residue \rf{H9} for $J=\bar{I}_0$ give
   	\beq\label{H17}\begin{split}
   		R^{\bar{I}_0}_{\bmm^1,\bmm^2}=&\frac{(\o_1\o_2)^{n/2}}{(2\pi \imath)^n}\prod_{i=1}^n\frac{(-1)^{m^1_i m^2_i+m_i^1+m_i^2}
   			\, \S^{-1}(g^\ast+m^1_i\o_1+m^2_i\o_2)}{\prod_{j=1}^{m^1_i}2\sin\frac{\pi j\o_1}{\o_2}\prod_{l=1}^{m_i^2}2\sin\frac{\pi l\o_2}{\o_1}}	\times\\
   		&\prod_{\substack{i,j=1 \\ i\not=j}}^n\Bigl[\S(x_i-x_j+(m^1_j-m^1_i)\o_1+(m_j^2-m^2_i)\o_2)\\ &\S(x_i-x_j+g^\ast+(m^1_j-m^1_i)\o_1+(m^2_j-m^2_i)\o_2)
   		\\ &
   		\S^{-1}(x_i-x_j+g^\ast+m_j^1\o_1+m^2_j\o_2)\, \S^{-1}(x_i-x_j-m_i^1\o_1-m^2_i\o_2)\Bigr]\times\\
   		&\prod_{i,a=1}^n \S^{-1}(z_a-x_i+g^\ast+m_i^1\o_1+m^2_i\o_2)\,\S^{-1}(x_i-z_a-m_i^1\o_1-m_i^2\o_2),
   	\end{split}\eeq  	
   so that
   	\beq\label{H18}\begin{split}
   		R^{\bar{I}_0}_{\bmm^1,\bmm^2}=&\frac{(\o_1\o_2)^{n/2}}{(2\pi \imath S(g^\ast))^n}
   		\prod_{i,a=1}^n S^{-1}(z_a-x_i+g^\ast)S^{-1}(x_i-z_a)	\times\\
   		&	\prod_{i=1}^n \frac{\Pochab{g^\ast}{m_i^1}{m_i^2}}{\Pochab{\o_1+\o_2}{m_i^1}{m_i^2}}
   		\prod_{i\not=j}\frac{\Pochab{x_i-x_j+g-m^1_i\o_1-m^2_i\o_2}{m^1_j}{m^2_j}}
   		{\Pochab{x_i-x_j-m^1_i\o_1-m^2_i\o_2}{m^1_j}{m^2_j}}\times\\&
   		\prod_{i,a=1}^n\frac{\Pochab{z_a-x_i+g^\ast}{m_i^1}{m_i^2}}{\Pochab{z_a-x_i+\o_1+\o_2}{m_i^1}{m_i^2}}.
   	\end{split}\eeq
   	Comparing \rf{H16} and \rf{H18} we see that the equality \rf{H11} is equivalent to the relation
   	\beq\label{H19}\begin{split}\sum_{\substack{ |\bmm^1|=M\\  |\bmm^2|=K}}	&	\prod_{a=1}^n \frac{\Pochab{g^\ast}{m_a^1}{m_a^2}}{\Pochab{\o_1+\o_2}{m_a^1}{m_a^2}}
   		\prod_{a\not=b}\frac{\Pochab{z_a-z_b+g-m^1_b\o_1-m^2_b\o_2}{m^1_a}{m^2_a}}
   		{\Pochab{z_a-z_b-m^1_b\o_1-m^2_b\o_2}{m^1_a}{m^2_a}}\times\\
   		&\prod_{i,a=1}^n\frac{\Pochab{z_a-x_i+g^\ast}{m_a^1}{m_a^2}}{\Pochab{z_a-x_i+\o_1+\o_2}{m_a^1}{m_a^2}}=
   		\\[8pt] \sum_{\substack{ |\bmm^1|=M\\  |\bmm^2|=K}}
   		&	\prod_{i=1}^n \frac{\Pochab{g^\ast}{m_i^1}{m_i^2}}{\Pochab{\o_1+\o_2}{m_i^1}{m_i^2}}
   		\prod_{i\not=j}\frac{\Pochab{x_i-x_j+g-m^1_i\o_1-m^2_i\o_2}{m^1_j}{m^2_j}}
   		{\Pochab{x_i-x_j-m^1_i\o_1-m^2_i\o_2}{m^1_j}{m^2_j}}\times\\&
   		\prod_{i,a=1}^n\frac{\Pochab{z_a-x_i+g^\ast}{m_i^1}{m_i^2}}{\Pochab{z_a-x_i+\o_1+\o_2}{m_i^1}{m_i^2}}.
   	\end{split}\eeq
   	Here the sums in both sides of the relation are taken over two sequences $\bmm^1$ and $\bmm^2$~\eqref{H1b} of non-negative integers with their fixed sums equal to $M$ and $K$,
   	\beq\label{H20}m_i^j \geq 0, \qquad |\bmm^1| = \sum_{i=1}^n m^1_i=M,\qquad |\bmm^2|=\sum_{i = 1}^n m^2_i=K.\eeq
   	Make the change of variables
   	\beq\label{H21} x_i\mapsto x_i+\o_1+\o_2.\eeq
   	Then the relation \rf{H19} looks as
   	\beq\label{H22}\begin{split}\sum_{\substack{ |\bmm^1|=M\\  |\bmm^2|=K}}	&	\prod_{a=1}^n \frac{\Pochab{\o_1+\o_2-g}{m_a^1}{m_a^2}}{\Pochab{\o_1+\o_2}{m_a^1}{m_a^2}}
   		\prod_{\substack{a,b=1 \\ a\not=b}}^n\frac{\Pochab{z_a-z_b+g-m^1_b\o_1-m^2_b\o_2}{m^1_a}{m^2_a}}
   		{\Pochab{z_a-z_b-m^1_b\o_1-m^2_b\o_2}{m^1_a}{m^2_a}}\times\\
   		&\prod_{i,a=1}^n\frac{\Pochab{z_a-x_i-g}{m_a^1}{m_a^2}}{\Pochab{z_a-x_i}{m_a^1}{m_a^2}}=
   		\\[8pt] \sum_{\substack{ |\bmm^1|=M\\  |\bmm^2|=K}}
   		&	\prod_{i=1}^n \frac{\Pochab{\o_1+\o_2-g}{m_i^1}{m_i^2}}{\Pochab{\o_1+\o_2}{m_i^1}{m_i^2}}
   		\prod_{\substack{i,j=1 \\ i\not=j}}^n\frac{\Pochab{x_i-x_j+g-m^1_i\o_1-m^2_i\o_2}{m^1_j}{m^2_j}}
   		{\Pochab{x_i-x_j-m^1_i\o_1-m^2_i\o_2}{m^1_j}{m^2_j}}\times\\&
   		\prod_{i,a=1}^n\frac{\Pochab{z_a-x_i-g}{m_i^1}{m_i^2}}{\Pochab{z_a-x_i}{m_i^1}{m_i^2}}.
   	\end{split}\eeq
   	The factorization formula \rf{trig4a} is equivalent to the factorization of the hyperbolic Pochhammer symbol:
   	\beq \label{H14}\Pochab{x}{m}{k}=\Pocha{x}{m}\cdot\Pochb{x}{k}.\eeq
   	Here \beq\label{H15}\Pocha{x}{m}=\frac{S_2(x)}{S_2(x+m\o_1)},\qquad \Pochb{x}{k}=\frac{S_2(x)}{S_2(x+k\o_2)}.\eeq
   	For non-negative $m$ and $k$
   	\beq\label{H23}\Pocha{x}{m}=\prod_{i=0}^{m-1}2\sin\pi\frac{x+i\o_1}{\o_2},\qquad
   	\Pochb{x}{k}=\prod_{j=0}^{k-1}2\sin\pi\frac{x+j\o_2}{\o_1}.
   	\eeq
   	By using \rf{H14} and canceling the appearing sings we can factorize each ratio in  \rf{H22} into the product over periods:
   	\beqq
   	\begin{split}& \frac{\Pochab{\o_1+\o_2-g}{m_a^1}{m_a^2}}{\Pochab{\o_1+\o_2}{m_a^1}{m_a^2}}=
   		\frac{\Pocha{\o_1-g}{m_a^1}}{\Pocha{\o_1}{m_a^1}}\times
   		\frac{\Pochb{\o_2-g}{m_a^1}}{\Pocha{\o_2}{m_a^2}},\\[5pt]
   		&	\frac{\Pochab{z_a-z_b+g-m^1_b\o_1-m^2_b\o_2}{m^1_a}{m^2_a}}
   		{\Pochab{z_a-z_b-m^1_b\o_1-m^2_b\o_2}{m^1_a}{m^2_a}}=
   		\frac{\Pocha{z_a-z_b+g-m^1_b\o_1}{m^1_a}}
   		{\Pocha{z_a-z_b-m^1_b\o_1}{m^1_a}}\times\\[5pt]	&\frac{\Pochb{z_a-z_b+g-m^2_b\o_2}{m^2_a}}
   		{\Pochb{z_a-z_b-m^2_b\o_2}{m^2_a}},
   		\\[5pt]
   		&\frac{\Pochab{z_a-x_i-g}{m_a^1}{m_a^2}}{\Pochab{z_a-x_i}{m_a^1}{m_a^2}}=
   		\frac{\Pocha{z_a-x_i-g}{m_a^1}}{\Pocha{z_a-x_i}{m_a^1}}\times
   		\frac{\Pochb{z_a-x_i-g}{m_a^2}}{\Pochb{z_a-x_i}{m_a^2}}.
   	\end{split}\eeqq
   	Thus, the relation \rf{H22} decouples into two independent identities
   	\beq\label{H25}
   	\begin{split}\sum_{ |\bmm^1|=M}	&	\prod_{a=1}^n \frac{\Pocha{\o_1-g}{m_a^1}}{\Pocha{\o_1}{m_a^1}}
   		\prod_{\substack{a,b=1 \\ a\not=b}}^n\frac{\Pocha{z_a-z_b+g-m^1_b\o_1}{m^1_a}}
   		{\Pocha{z_a-z_b-m^1_b\o_1}{m^1_a}}
   		\prod_{i,a=1}^n\frac{\Pocha{z_a-x_i-g}{m_a^1}}{\Pocha{z_a-x_i}{m_a^1}}=
   		\\[4pt] \sum_{ |\bmm^1|=M}
   		&	\prod_{i=1}^n \frac{\Pocha{\o_1-g}{m_i^1}}{\Pocha{\o_1}{m_i^1}}
   		\prod_{\substack{i,j=1 \\ i\not=j}}^n\frac{\Pocha{x_i-x_j+g-m^1_i\o_1}{m^1_j}}
   		{\Pocha{x_i-x_j-m^1_i\o_1}{m^1_j}}
   		\prod_{i,a=1}^n\frac{\Pocha{z_a-x_i-g}{m_i^1}}{\Pocha{z_a-x_i}{m_i^1}}
   	\end{split}\eeq
   and
   	\beq\label{H26}\begin{split}\sum_{{|\bmm^2|=K}}&\prod_{a=1}^n \frac{\Pochb{\o_2-g}{m_a^2}}{\Pochb{\o_2}{m_a^2}}
   		\prod_{\substack{a,b=1 \\ a\not=b}}^n\frac{\Pochb{z_a-z_b+g-m^2_b\o_2}{m^2_a}}
   		{\Pochb{z_a-z_b-m^2_b\o_2}{m^2_a}}
   		\prod_{i,a=1}^n\frac{\Pochb{z_a-x_i-g}{m_a^2}}{\Pochb{z_a-x_i}{m_a^2}}=
   		\\ \sum_{\substack{|\bmm^2|=K}}
   		&	\prod_{i=1}^n \frac{\Pochb{\o_2-g}{m_i^2}}{\Pochb{\o_2}{m_i^2}}
   		\prod_{\substack{i,j=1 \\ i\not=j}}^n\frac{\Pochb{x_i-x_j+g-m^2_i\o_2}{m^2_j}}
   		{\Pochb{x_i-x_j-m^2_i\o_2}{m^2_j}}
   		\prod_{i,a=1}^n\frac{\Pochb{z_a-x_i-g}{m_i^2}}{\Pochb{z_a-x_i}{m_i^2}}.
   	\end{split}\eeq
   	These are precisely hypergeometric identities \rf{I20} written in additive form. Their proof is given in the next section. Using it we complete the proof of Proposition \ref{propH1} and of the main statement of commutativity of Baxter $Q$-operators. \hfill{$\Box$}
   	\setcounter{equation}{0}
   	  \section{Proof of hypergeometric identities}\label{AppX}
   	The relations  \rf{H25} and \rf{H26} are equivalent modulo the interchange of the periods. We choose \rf{H26}.
   	 Rewrite it in the common multiplicative notations  of basic hypergeometry. Set
   	\beq\label{p1} q=e^{\frac{2\pi\imath\o_2}{\o_1}},\qquad t=e^{\frac{-2\pi\imath g}{\o_1}},\qquad u_i=e^{\frac{2\pi\imath z_i}{\o_1}}, \qquad v_a=e^{\frac{2\pi\imath x_a}{\o_1}}.\eeq	
   	 Denote by $(z;q)_k$ and $[z;q]_k$ nonsymmetric and symmetric $q$-analogs of Pochhammer symbols,
   	\beq\label{p3}\begin{split}(z;q)_k&=(1-z)(1-qz)\cdots(1-q^{k-1}z),\\[4pt] [z;q]_k&=(z^{1/2}-z^{-1/2})(q^{1/2}z^{1/2}-q^{-1/2}z^{-1/2})\cdots (q^{(k-1)/2}z^{1/2}-q^{(-k+1)/2}z^{-1/2}).
   	\end{split}\eeq
   	Then \rf{H26} becomes
   	 \beq\label{p4}\begin{split}  &\sum_{|\bk|=K}\prod_{i=1}^n\frac{[qt;q]_{k_i}}{[q;q]_{k_i}}	
   		\times \prod_{\substack{i,j=1 \\ i\not=j}}^n
   		\frac{[t^{-1}q^{-k_j}u_i/u_j;q]_{k_i}}{[q^{-k_j}u_i/u_j;q]_{k_i}}
   		\times
   		\prod_{a,j=1}^n\frac{[tu_j/v_a;q]_{k_j}}{[u_j/v_a;q]_{k_j}}
   		=\\
   		&\sum_{|\bk|=K}\prod_{a=1}^n\frac{[qt;q]_{k_a}}{[q;q]_{k_a}}	
   		\times \prod_{\substack{a,b=1 \\ a\not=b}}^n
   		\frac{[t^{-1}q^{-k_a}v_a/v_b;q]_{k_b}}{[q^{-k_a}v_a/v_b;q]_{k_b}}
   		\times
   		\prod_{a,j=1}^n\frac{[tu_j/v_a;q]_{k_a}}{[u_j/v_a;q]_{k_a}}
   	\end{split}\eeq
   	in terms of symmetric $q$-Pochhammers. Here the sum in both sides of the equality is taken over $n$-tuples of non-negative integers with total sum equal to $K$
   	\beqq \bk=(k_1,\ldots, k_n), \qquad  k_i\geq0,\qquad k_1+\ldots+k_n=K.\eeqq
   	It has the same form in terms of traditional nonsymmetric
   	$q$-Pochhammer symbols:
   	\beq\label{p4a}\begin{split}  &\sum_{{|\bk_n|=K}}\prod_{i=1}^n\frac{(qt;q)_{k_i}}{(q;q)_{k_i}}	
   		\times \prod_{\substack{i,j=1 \\ i\not=j}}^n
   		\frac{(t^{-1}q^{-k_j}u_i/u_j;q)_{k_i}}{(q^{-k_j}u_i/u_j;q)_{k_i}}
   		\times
   		\prod_{a,j=1}^n\frac{(tu_j/v_a;q)_{k_j}}{(u_j/v_a;q)_{k_j}}
   		=\\
   		&\sum_{{|\bk_n|=K}}\prod_{a=1}^n\frac{(qt;q)_{k_a}}{(q;q)_{k_a}}	
   		\times \prod_{\substack{a,b=1 \\ a\not=b}}^n
   		\frac{(t^{-1}q^{-k_a}v_a/v_b;q)_{k_b}}{(q^{-k_a}v_a/v_b;q)_{k_b}}
   		\times
   		\prod_{a,j=1}^n\frac{(tu_j/v_a;q)_{k_a}}{(u_j/v_a;q)_{k_a}}.
   	\end{split}\eeq
   	However, it is more convenient for us to prove the symmetric version of identity \rf{p4}.
   	
   	The proof follows the standard line of complex analysis: in a rather tricky way we check
   	that the difference of the left and right hand sides has zero residues at all possible simple poles. Thus, both sides are the  Laurent polynomials symmetric over the variables $u_i$ and over the variables  $v_j$. Then the  asymptotic analysis of these polynomials shows that their difference is actually equal to zero.
   	
   	  The crucial step --- calculation of the residues of both sides of the equality --- divides into two parts. First we show that each side is regular at the diagonals $u_i=q^pu_j$ and $v_a=q^sv_b$ between the variables of the same group, see Lemma \ref{lemmap1}. In this calculation we actually observe the canceling of  terms grouped in corresponding pairs. Then we show that residues at mixed diagonals $u_i=q^pv_a$ vanish. This is done by induction, using the nontrivial relation between such residues stated in Lemma \ref{lemma2p}.  Below we give  a brief proof of both lemmas, all technical details are presented in our paper \cite{BDKK2}.
   	
   	  It is not difficult to verify that all the poles in \rf{p4} are simple. Consider the left hand side of \rf{p4} as the function of $u_1$ and calculate the residue of this function at the point
   	\beq \label{p21} u_1=u_2q^p,\qquad p\in\Z.\eeq
   	For each $\bk$, $\sum_{j=1}^nk_j=K$ denote by $\XL_{\bk}=\XL_{\bk}(\bu;\bv)$ the corresponding summand of the left hand side of \rf{p4}, and by $\YR_{\bk}=\YR_{\bk}(\bu;\bv)$ the corresponding summand of the right hand side of \rf{p4},
   	\beqq\begin{split} \XL_{\bk}&=\prod_{i=1}^n\frac{[qt;q]_{k_i}}{[q;q]_{k_i}}	
   	\times \prod_{\substack{i,j=1 \\ i\not=j}}^n
   	\frac{[t^{-1}q^{-k_j}u_i/u_j;q]_{k_i}}{[q^{-k_j}u_i/u_j;q]_{k_i}}
   	\times
   	\prod_{a,j=1}^n\frac{[tu_j/v_a;q]_{k_j}}{[u_j/v_a;q]_{k_j}},\\
   	\YR_{\bk}&=\prod_{a=1}^n\frac{[qt;q]_{k_a}}{[q;q]_{k_a}}	
   	\times \prod_{\substack{a,b=1 \\ a\not=b}}^n
   	\frac{[t^{-1}q^{-k_a}v_a/v_b;q]_{k_b}}{[q^{-k_a}v_a/v_b;q]_{k_b}}
   	\times
   	\prod_{a,j=1}^n\frac{[tu_j/v_a;q]_{k_a}}{[u_j/v_a;q]_{k_a}}.
   	\end{split}\eeqq
   	The summands $\XL_{\bk}$, which contribute to the residue at the point \rf{p21}, are divided into two groups. The denominators of the terms $\XL_{\bk}$ from the first group $\bk\in I_p$ contain Pochhammer symbol
   	\beqq [q^{-k_2}u_1/u_2;q]_{k_1} \eeqq
   	which vanishes at the point \rf{p21}. It happens when
   	\beqq
   	k_2-k_1+1\leq p\leq k_2,
   	\eeqq
   	so that
   	\beqq
   	I_p=\{\bk,\, |\bk|=K\colon k_1\geq k_2+1-p,\ k_2\geq p\}.\eeqq
   	The denominators of the terms $\XL_{\bll}$ in the second group $\bll\in II_p$ contain Pochhammer
   	\beqq [q^{-l_1}u_2/u_1;q]_{l_2} \eeqq
   	which vanishes at the point \rf{p21}. It happens when
   	\beqq
   	-l_1\leq p\leq l_2-l_1-1,
   	\eeqq
   	so that
   	\beqq
   	II_p=\{\bll,\, |\bll|=K\colon l_1\geq-p, l_2\geq l_1+1+p\}.\eeqq
   	Define the maps of sets $\vf_p:I_p\to II_p$ and $\psi_p:II_p\to I_p$ by the same formulas
   	\beqq
   	\begin{split}
   		&\phi_p\colon I_p\to II_p\qquad  \phi_p(k_1,k_2,\bk')=(k_2-p,k_1+p,\bk'),\\
   		&\psi_p\colon II_p\to I_p\qquad  \psi_p(k_1,k_2,\bk')=(k_2-p,k_1+p,\bk')\end{split}\eeqq
   	where $\bk' = (k_3, \dots, k_n)$.
   	\begin{lemma}\label{lemmap1}${}$
   		\begin{enumerate}
   			\item Maps $\phi_p$ and $\psi_p$ establish bijections between the sets $I_p$ and $II_p$;
   			
   			\item For any $\bk\in I_p$
   			\begin{align}\label{p25b} \Res_{u_1=u_2q^p}\XL_{\bk}(\bu;\bv)&+\Res_{u_1=u_2q^p}\XL_{\phi_p(\bk)}(\bu;\bv)=0,\\
   			\Res_{v_2=v_1q^p}\YR_{\bk}(\bu;\bv)&+\Res_{v_2=v_1q^p}\YR_{\phi_p(\bk)}(\bu;\bv)=0.	
 \label{p25a}\end{align}
   		\end{enumerate}
   	\end{lemma}
   	{\bf Proof} of Lemma \ref{lemmap1}. The first part is purely combinatorial and can be checked directly. Let us prove the second part.
   	
   	Note first that each summand $\XL_{\bk}(\bu;\bv)$ of the left hand side of \rf{p4} has the following structure
   	\beq\label{p10} \XL_{\bk}(\bu;\bv)=\frac{\X_{\bk}(\bu;\bv;t)}{\X_{\bk}(\bu;\bv;1)}\eeq
   	where
   	\beq
   	\X_{\bk}(\bu;\bv;t)=\prod_{i=1}^n{[qt;q]_{k_i}}	
   	\times \prod_{\substack{i,j=1 \\ i\not=j}}^n
   	{[t^{-1}q^{-k_j}u_i/u_j;q]_{k_i}}
   	\times
   	\prod_{a,j=1}^n{[tu_j/v_a;q]_{k_j}}.\eeq
   	The following  identity
   	\beq\label{p13} \X_{k_1,k_2,\bk'}(\bu;\bv;t)|_{u_1=q^pu_2}=\X_{k_2-p,k_1+p,\bk'}(\bu;\bv;t)|_{u_1=q^pu_2},
   	\qquad \bk=(k_1,k_2,\bk')\in I_p
   	\eeq
   	valid for any  $\bk=(k_1,k_2,\bk')\in I_p$ is established with a help of an explicit bijection between linear factors of the products in both sides of equality \rf{p13}. Then this equality implies the statement \rf{p25b} about zero sum of  the residues. Indeed, the relation \rf{p13} establishes a bijection between all nonzero factors of the denominators $\X_{k_1,k_2,\bk'}(\bu;\bv;1)|_{u_1=q^pu_2}$ and $\X_{k_2-p,k_1+p,\bk'}(\bu;\bv;1)$ and the equality of their products. Factors in denominators of $\XL_{k_1,k_2,\bk'}(\bu;\bv)$ and
   	$\XL_{k_2-p,k_1+p,\bk'}(\bu;\bv)$ which tend to zero when $u_1$ tends to $q^pu_2$ are
   	\beq\label{p13a} q^{-p/2}u_1/u_2-q^{p/2}u_2/u_1 \qquad\text{and}\qquad q^{p/2}u_2/u_1-q^{-p/2}u_1/u_2.\eeq
   	They give inputs into residues, which just differ by sign. Thus we arrive at \rf{p25b}. For the proof of \rf{p25a} we note that the involution
   	\beq\label{p14} \tau\colon u_i\mapsto v_i^{-1}, \qquad v_i\mapsto u_i^{-1}\eeq
   	exchanges each $\XL_{\bk}$ with $\YR_{\bk}$, as well as the left and right hand sides of \rf{p4}.
   	\hfill{$\Box$}
   	\begin{corollary}\label{corollaryp1}
   		Both sides of \rf{p4} have no poles of the form $u_i=q^pu_j$ and $v_a=q^pv_b$.
   		\end{corollary}
   	
   	For any non-negative integer $p$ denote  by $\vf_p(\bu;\bv)$ the following rational function
   	of $\bu=(u_1,\ldots,u_n)$ and $\bv=(v_1,\ldots,v_n)$:
   		\beq\label{p15b}	
   	\vf_p(\bu;\bv)=
   	(-1)^p \frac{[tq;q]_{2p}}{[q;q]_p[q;q]_{p-1}}\prod_{j=2}^n\frac{[tu_j/v_1;q]_p}{[u_1/u_j;q]_p}\prod_{b=2}^n\frac{[tu_1/v_b;q]_p}{[v_b/v_1;q]_p}
   	\eeq
   	\begin{lemma}\label{lemma2p} For any $1\leq p\leq k_1$ and $\bk'\in\Z_{\geq 0}^{n-1}$
   		\begin{align}\label{p15}
   			\Res_{v_1=q^{p-1}u_1}\frac{1}{v_1}\YR_{k_1,\bk'}(\bu;\bv)=&\vf_p(\bu;\bv)\times \YR_{k_1-p,\bk'}(qv_1,\bu';q^{-1}u_1,\bv'),\\
   			\label{p15a}
   			\Res_{v_1=q^{p-1}u_1}\frac{1}{v_1}\XL_{k_1,\bk'}(\bu;\bv)=&\vf_p(\bu;\bv)\times
   			\XL_{k_1-p,\bk'}(qv_1,\bu';q^{-1}u_1,\bv').\end{align}
   	\end{lemma}
   	{\bf Proof} of Lemma \ref{lemma2p} is a direct computation which uses the following properties of $q$-Pochhammer symbols:
   	\begin{align}\label{p8a}[q^pu;q]_m\times[u]_n&=[q^pu;q]_{n-p}\times[u;q]_{m+p},\\
   		\label{p8b}[qu;q]_m\times [q^{-(m+p)}u^{-1};q]_n&=(-1)^p[qu;q]_{m+p}\times[q^{-m} u^{-1};q]_{n-p}
   	\end{align}
   	which are valid for any $u$ and integer $m,n,p$.
   	Here we assume that
   	\beq
   	[z;q]_{-n}=(q^{1/2}z^{1/2}-q^{-1/2}z^{-1/2})^{-1}\cdots (q^{n/2}z^{1/2}-q^{-n/2}z^{-1/2})^{-1}, \qquad  n>0.
   	\eeq For more technical details see \cite{BDKK2}.
   	\hfill{$\Box$}
   	
   	   	{\bf Proof} of Theorem \ref{theorem2}. We are ready now to prove the equality \rf{p4} and thus Theorem \ref{theorem2} by induction over $K$. Denote the difference of the left and right hand sides of \rf{p4} by $W_K(\bu;\bv)$. Assume that $W_K(\bu,\bv)=0$ for all $K<N$ and any $m$-tuples of
   	   	variables $\bu= (u_1,\ldots,u_m)$, $\bv=(v_1,\ldots,v_m)$ for arbitrary $m$.
   	   	Summing up the difference of \rf{p15} and \rf{p15a} over all $\bk$ with $|\bk|=K$ we get the
   	   	 relation
   	   	\beq\label{p25}\Res_{v_1=q^{p-1}u_1}\frac{1}{v_1}W_K(\bu;\bv)=\vf_p(\bu;\bv)\times W_{K-p}(\bu^*,\bv^*),
   	   	\eeq
   	   	where
   	   	\beq\label{p26}\bu^*=(qv_1,\bu'),\qquad \bv^*=(q^{-1}u_1,\bv').
   	   	\eeq
   	   	 By the induction assumption the right hand side of \rf{p25} equals zero. Taking in mind the symmetricity of $W_K(\bu;\bv)$  with respect to permutation of $u_i$ and of $v_j$ we conclude that it has no poles at all.  Since $W_K(\bu;\bv)$ is a  homogeneous rational function of the variables $u_i$ and $v_j$ of total degree zero, it is equal to a constant, which could depend on $q$ and $t$. To compute this constant we consider the behavior of this function in asymptotic zone
   	   	 \beq\label{p27} u_1\ll u_2\ll\ldots\ll u_n\ll v_n\ll v_{n-1}\ll \ldots \ll v_1.\eeq
   	   	  Here both sides of \rf{p4} tend to
   	   	  \beq\sum_{{|\bk_n|=K}}\prod_{i=1}^n\frac{[qt;q]_{k_i}}{[q;q]_{k_i}}\times t^{\frac{1}{2}\big( (n-1)k_1+(n-3)k_2+\ldots+(3-n)k_{n-1}+(1-n)k_n\big)}\times t^{-\frac{nK}{2}}.\eeq
 	   	Therefore, $W_K(\bu;\bv)$ tends to zero in this asymptotic zone and so equals zero identically. This completes the induction step, the proof of the identity \rf{p4} and of Theorem \ref{theorem2}. \hfill{$\Box$}

	\section*{Acknowledgments}
	We are grateful to Ole Warnaar and Hjalmar Rosengren for communicating to us their
	results. We also thank referees for their helpful comments and remarks.
	
	The work of N. Belousov (Section 3) was supported by the Euler International Mathematical Institute, grant No. 075-15-2022-289. The work of S. Derkachov was supported by the Theoretical Physics and Mathematics Advancement Foundation «BASIS». The  work of S. Kharchev (Section 2) was supported by the Russian Science Foundation (Grant No. 23-41-00049). The work of S. Khoroshkin (Section 4) was supported by the International Laboratory of Cluster Geometry of	National Research University Higher School of Economics, Russian Federation Government grant, ag. No. 075-15-2021-608 dated 08.06.2021. He also thanks the Weizmann Institute of Science for the kind hospitality during the summer of 2022. A big part of this work was done during his stay there. 
   	
   	\setcounter{equation}{0}
   		\section*{Appendix}
   	\appendix
   {	\section{Double Gamma and sine functions}\label{AppendixA}
   	The Barnes double Gamma function $\Gamma_2(z|\bo)$ \cite{B} is defined by the relation
   	\beqq  \Gamma_2(z|\bo)=\exp{\left(\frac{\d}{\d s} \zeta_2(s,z|\bo)\right)}\Big|_{s=0},\eeqq
   	where $\zeta_2(s,z|\bo)$ is the analytical continuation of the series
   	\beqq  \zeta_2(s,z|\bo)=\sum_{n_1,n_2\geq 0} (z+n_1\o_1+n_2\o_2)^{-s}, \qquad \Re s>2\eeqq
   	which under assumptions \rf{I0a} and $\Re z>0$ can be presented by the integral
   	\beqq \zeta_2(s,z|\bo)=\Gamma(1-s)\int_C \frac{e^{-zt}(-t)^{s}}{\left(1-e^{-\o_1 t}\right)\left(1-e^{-\o_2 t}\right)}\frac{dt}{2\pi\imath t}\eeqq
   	over the Hankel contour $C$ enclosing the ray $\{t\geq 0\}$ counterclockwise.
   	Under the same assumptions analogous integral presentation of  $\ln \Gamma_2(z|\bo)$ looks as follows
   	\beq \label{A1} \ln  \Gamma_2(z|\bo)= \frac{\g}{2}B_{2,2}(z|\bo)+ \int_C \frac{e^{-zt}\ln(-t)}{\left(1-e^{-\o_1 t}\right)\left(1-e^{-\o_2 t}\right)}\frac{dt}{2\pi\imath t}.\eeq
   	Here
   	\beq\label{A2} B_{2,2}(z|\bo)=\frac{z^2}{\omega_1\omega_2}-
   	\frac{\omega_1+\omega_2}{\omega_1\omega_2}\,z\,+\,
   	\frac{\omega_1^2+3\omega_1\omega_2+\omega_2^2}{6\omega_1\omega_2}\eeq
   	is a particular multiple Bernoulli polynomial, $\gamma$ is the Euler constant.
   	
   The double sine function $S_2(z):=S_2(z|\bo)$, see \cite{Ku} and references therein, is then defined as
   \beq\label{A3} S_2(z|\bo)=\Gamma_2(\o_1+\o_2-z|\bo)\Gamma_2^{-1}(z|\bo).
   \eeq	 }  	
   	   It	 satisfies functional relations
   	\beq\label{trig3}  \frac{S_2(z)}{S_2(z+\o_1)}=2\sin \frac{\pi z}{\o_2},\qquad \frac{S_2(z)}{S_2(z+\o_2)}=2\sin \frac{\pi z}{\o_1}
   	\eeq
   	and inversion relation
   	\beq \label{trig4} S_2(z)S_2(-z)=-4\sin\frac{\pi z}{\o_1}\sin\frac{\pi z}{\o_2}, \eeq
   	or equivalently
   	\beq\label{trig4b} S_2(z)S_2(\o_1+\o_2-z)=1.\eeq
   	 The double sine function is a homogeneous function of all its argumets
   	\beq\label{A0} S_2(\g z|\g\o_1,\g\o_2)= S_2(z| \o_1, \o_2),\qquad \g\in(0, \infty)\eeq
   	and is invariant under permutation of periods
   	\beq S_2(z|\o_1, \o_2) = S_2(z|\o_2, \o_1). \eeq
   	The relation \rf{trig3} has a useful corollary
   	\beq\label{trig3a}
   	\begin{aligned}
		\frac{S_2(z)}{S_2(z+m\o_1+k\o_2)}&=(-1)^{mk}\prod_{j=0}^{m - 1}2\sin\frac{\pi}{\o_2}(z + j \o_1)\prod_{j=0}^{k - 1} 2\sin\frac{\pi}{\o_1} (z + j \o_2), \\[4pt]
		\frac{S_2(z-m\o_1-k\o_2)}{S_2(z)}&=(-1)^{mk}\prod_{j=1}^{m}2\sin\frac{\pi}{\o_2}(z - j \o_1)\prod_{j=1}^{k} 2\sin\frac{\pi}{\o_1} (z - j \o_2)
   	\end{aligned}
   	\eeq
   	that holds for $m, k \geq 0$. The latter relations also imply the following factorization formula
   	\beq\label{trig4a} {S_2(z)}{S_2(z+m\o_1+k\o_2)}=(-1)^{mk}{S_2(z+m\o_1)}{S_2(z+k\o_2)}\eeq
   	for $m,k\in \mathbb{Z}$.
   	The function $S_2(z)$ is a meromorphic function of $z$ with poles at
   	\beq\label{A1a} z_{m,k} = m\o_1 + k\o_2, \qquad m,k \geq 1\eeq
  	and zeros at
   	\beq\label{A1b} z_{-m,-k}=-m\o_1-k\o_2,\qquad m,k\geq 0. \eeq
    For $\o_1/\o_2 \not\in \mathbb{Q}$ all poles and zeros are simple. The residues of $S_2(z)$ and $S^{-1}_2(z)$ at these points are
   	\begin{align}
   	\underset{z = z_{m,k}}{\Res} \, S_2(z) = \frac{\sqrt{\o_1\o_2}}{2\pi}\frac{(-1)^{mk}}{\prod\limits_{s=1}^{m - 1}2\sin\dfrac{\pi s\o_1}{\o_2}\prod\limits_{l=1}^{k - 1}2\sin\dfrac{\pi l\o_2}{\o_1}},
   	\\[10pt]
   	\label{trig5} \underset{z = z_{-m,-k}}{\Res} \, S^{-1}_2(z) = \frac{\sqrt{\o_1\o_2}}{2\pi}\frac{(-1)^{mk+m+k}}{\prod\limits_{s=1}^m2\sin\dfrac{\pi s\o_1}{\o_2}\prod\limits_{l=1}^k2\sin\dfrac{\pi l\o_2}{\o_1}}.
   	\end{align}
   	The integral representation for the logarithm of double sine function
   	\begin{equation}\label{S2-int}
   		\ln S_2 (z) = \int_0^\infty \frac{dt}{2t} \left( \frac{\sh \left[ (2z - \omega_1 - \omega_2)t \right]}{ \sh (\omega_1 t) \sh (\omega_2 t) } - \frac{ 2z - \omega_1 - \omega_2 }{ \omega_1 \omega_ 2 t } \right)
   	\end{equation}
   	holds true for $ \Re z \in ( 0, \Re (\omega_1 + \omega_2 ))$. 
   	
   	The double dine function also can be written in terms of Ruijsenaars hyperbolic Gamma function $G(z|\bo)$ \cite{R2}
   	\beq G(z|\bo) = S_2\Bigl(\imath z + \frac{\o_1+\o_2}{2} \,\Big|\, \bo\Bigr)\eeq 
   	or Faddeev quantum dilogarithm $\gamma(z|\bo)$ \cite{F}
   	\beq \gamma(z|\bo) = S_2\Bigl(-\imath z + \frac{\o_1+\o_2}{2} \,\Big|\, \bo\Bigr) \exp \Bigl( \frac{\imath \pi}{2\o_1 \o_2} \Bigl[z^2 + \frac{\o_1^2+\o_2^2}{12} \Bigr]\Bigr). \eeq
	Both functions $G(z|\bo)$ and $\gamma(z|\bo)$ were investigated independently.
   	
   	In what follows we use the same notations as in Section \ref{sectioneEstimates}. Denote by $\sigma_i$ the arguments of the periods~$\o_i$, $|\sigma_i|<\pi/2$. Since the double sine function is invariant under permutation of $\o_1,\o_2$, suppose for definiteness that $\sigma_1 \geq \sigma_2$. Let $D_+$ and $D_-$ be the cones of poles and zeros of the double
   	sine function $S_2(z|\bo)$:
   	\beqq D_+=\{ z\colon \sigma_2< \arg z<\sigma_1\},\qquad D_-=\{ z\colon \pi+\sigma_2< \arg z<\pi+\sigma_1\},\qquad D=D_+\cup D_-.\eeqq
   	Denote by $d(z,D_+)$ and $d(z,D_-)$ the distances between a point $z$ and the cones $D_{\pm}$. Then the Barnes' Stirling formula for the logarithm of the double Gamma function, see \cite[§§85--86]{B}, with error term suggested by E. Rains \cite[Theorem 2.6]{R} looks as
   	\beq\begin{split}\label{J6}
   		\ln\Gamma_2(z|\bo)
   		= -\frac{1}{2} B_{2,2}(z|\bo) \ln z+ \,\frac{3}{4\omega_1\omega_2}\,z^2-
   		\frac{\omega_1+\omega_2}{2\omega_1\omega_2}\,z+\,O\Big(d^{-1}(z,D_-)\Big)\,.
   	\end{split}\eeq
   	Here $z\in\C\setminus D_-$. Moreover, the estimates for the error term given in \cite[§57]{B2} are uniform on compact subsets of parameters $\bo$ separated from zero.
   	Then for $z\in \C\setminus (D_+\cup D_-)$
   	\beq \begin{split}\label{J7} \ln S_2(z|\bo)=  \ln\Gamma_2(\o_1+\o_2-z|\omega_1,\omega_2)- \ln\Gamma_2(z|\omega_1,\omega_2)=\\[4pt]
   		\pm\frac{\pi\imath}{2} B_{2,2}(z|\bo)
   		+\,O\Big(d^{-1}(z,D)\Big).
   	\end{split}\eeq
   	where the sign $+$ is taken for $z$ in the upper half plane, and the sign $-$ for $z$ in the  lower half plane (and not in $D$). Finally, in the same notations,
   	\beq \begin{split}\label{J8} \ln \frac{S_2(z|\bo)}{S_2(z+g|\bo)}=\mp\pi\imath \frac{g}{\o_1\o_2}\left(z-\frac{g^\ast}{2}\right)\,+O\Big(d^{-1}(z,D)\Big).
   	\end{split}\eeq	
   	Equivalently, for $z\in \C\setminus D$
   	\beq\label{J9}  \frac{S_2(z|\bo)}{S_2(z+g|\bo)}=  e^{\mp\pi\imath \frac{g}{\o_1\o_2}\left(z-\frac{g^\ast}{2}\right)}\Big(1+\,O\Big(d^{-1}(z,D)\Big)\Big). \eeq
   	Using the asymptotics \rf{J9}, we can derive the following bounds which we use for the study of integrals convergence throughout the paper.
   	
   	Let $K\subset \C$ be a closed subset of a complex plane satisfying the following conditions:
   	\begin{enumerate}
   		\item $K$ is  inside the domain of analyticity of ${S_2(z|\bo)}{S^{-1}_2(z+g|\bo)}$;
   		\item There exists $R>0$ and $\rho>|g|$ such that  $K\cap \{|z|>R\}$ does not intersect with  $D$  and
   		 \beq\label{A5} d(K\cap \{|z|>R\}, D)\geq\rho.
   		 \eeq
   		\end{enumerate}
   	\begin{proposition} \label{propA1} Under the conditions $1$ and $2$ above we have a bound
   				\beq\label{A6}\left| {S_2(z|\bo)}{S^{-1}_2(z+g|\bo)}   \right|<Ce^{\mp \Re \frac{\pi \imath gz}{\o_1\o_2}},\qquad z\in K.\eeq   		
   		\end{proposition}
   	The constant $C$ can be stated uniform as the parameters $g,\o_1,\o_2$ range in a compact domain separated from zero values of periods.
   	
   	{\bf Proof}. Due to \rf{J9} and condition 2 there exists $R_1>R$ and  $C_1$ such that
   		\beq\label{A7}\left| {S_2(z|\bo)}{S^{-1}_2(z+g|\bo)}   \right|<C_1e^{\mp \Re \frac{\pi \imath gz}{\o_1\o_2}},  \qquad z\in K,\ |z|> R_1.\eeq
   	 On the other hand, the set
   	\beq\label {A8}  K\cap\{|z|\leq R_1\}\eeq
   	 is compact and belongs to the region of analyticity of the function   ${S_2(z|\bo)}{S^{-1}_2(z+g|\bo)}$. Thus this function is bounded on the set \rf{A8},   	
   		\beq\label{A9}\left| {S_2(z|\bo)}{S^{-1}_2(z+g|\bo)}   \right|<C_2,\qquad z\in K,\qquad |z|<R_1.\eeq
   	At the same time both real functions $e^{\mp \Re \frac{\pi \imath gz}{\o_1\o_2}}$ are analytic and positive on the compact set \rf{A8}. Thus, they are bounded from below on this set
   	\beq\label{A10} e^{\mp \Re \frac{\pi \imath gz}{\o_1\o_2}}>C_3>0.\eeq
   	Combining \rf{A9} and \rf{A10} we conclude that there exists a positive constant $C_4$ such that
   		\beq\label{A11}\left| {S_2(z|\bo)}{S^{-1}_2(z+g|\bo)}   \right|<C_4e^{\mp \Re \frac{\pi \imath gz}{\o_1\o_2}}, \qquad z\in K,\ |z|\leq R_1.\eeq
   	Combining  \rf{A7} and \rf{A11} we arrive at the proof of Proposition \ref{propA1}. \hfill{$\Box$}

   	There are two straightforward corollaries of Proposition \ref{propA1}. First, since $|\Re z|<|z|$, \rf{A6} implies that the function ${S_2(z|\bo)}{S^{-1}_2(z+g|\bo)}$ grows at most exponentially
   	\beq\label{A12}\left| {S_2(z|\bo)}{S^{-1}_2(z+g|\bo)}   \right|<Ce^{a|z|},  \qquad z\in K .\eeq
   	Second, assume that $K$ is contained in a strip $|\Re z|<b$ for some $b>0$. Then \rf{A6} implies the bound
   	\beq\label{A13} \left| {S_2(z|\bo)}{S^{-1}_2(z+g|\bo)}\right|<\tilde{C}e^{\Re \frac{\pi  g}{\o_1\o_2}|y|}, \qquad z=x+iy\in K.
   	\eeq
   	
   	The same  statement holds for the inverse ratio. Namely, 	Let $K'\subset \C$ be a closed subset of a complex plane satisfying the following conditions:
	\begin{enumerate}
		\item[1$^\prime$.] $K'$ is  inside the domain of analyticity of ${S_2^{-1}(z|\bo)}{S_2(z+g|\bo)}$;
		
		\item[2$^\prime$.] There exists $R'>0$ and $\rho'>|g|$ such that  $K'\cap \{|z|>R'\}$ does not intersect with  $D$  and
		\beq\label{A14} d(K'\cap \{|z|>R'\}, D)\geq\rho'.
		\eeq
	\end{enumerate}

   	\begin{proposition} \label{propA2} Under the conditions $1'$ and  $2'$ above we have a bound
   		\beq\label{A15}\left| {S_2^{-1}(z|\bo)}{S_2(z+g|\bo)}   \right|<C'e^{\pm \Re \frac{\pi \imath gz}{\o_1\o_2}}.\eeq   		
   	\end{proposition}
   	 In particular, the function  ${S_2^{-1}(z|\bo)}{S_2(z+g|\bo)}$ grows at most exponentially
   	 \beq\label{A16}\left| {S_2^{-1}(z|\bo)}{S_2(z+g|\bo)}   \right|<C'e^{a|z|},  \qquad z\in K .\eeq
   	 If $K'$ is contained in a strip $|\Re z|<b'$ for some $b'>0$ then
   	 \beq\label{A17} \left| {S_2^{-1}(z|\bo)}{S_2(z+g|\bo)}\right|<\tilde{C}'e^{-\Re \frac{\pi  g}{\o_1\o_2}|y|}, \qquad z=x+iy\in K'.
   	 \eeq}
   	\setcounter{equation}{0}

   	\section{Bounds for integrals} \label{AppendixB}
   	Both functions $\mu(z)$ and $K(z)$ can be presented as the ratios of double sine functions
   	that appear in Propositions \ref{propA1}  and \ref{propA2}
   	\beq\label{B1}\begin{split}\mu(z)& =S_2(\imath z)S_2^{-1} (\imath z+g),\\[6pt]
   	 K(z)& =  S_2\left(\imath z+\frac{\o_1+\o_2}{2}+\frac{g}{2}\right)S_2^{-1}\left(\imath z+\frac{\o_1+\o_2}{2}-\frac{g}{2}\right).\end{split} \eeq
   	The conditions \rf{I0a} and \rf{I0b} imply that both functions have a strip of analiticity which include
   	the real line of the parameter $z$.	For brevity, we also denote by $\nu_g$ the constant in the assumption \eqref{I17a}
   	\beq\label{B2} \nu_g= \Re \frac{g}{\o_1\o_2} >0.\eeq
   	Then by \rf{A13} and  \rf{A17} we have
   	\beq\label{B3}\begin{split} |K(y)|&<Ce^{-\pi\nu_g|y|},\\[4pt]
   		|\mu(y)|&<Ce^{\pi\nu_g|y|},\end{split} \qquad y\in\R,\eeq
   	where $C$ is a positive constant uniform for a compact subset of parameters $\o_1,\o_2$ and $g$ preserving the conditions above.
   	Assume also the condition
   	\beq\label{B5} |\Im \l|\leq\delta<\nu_g \eeq
   	with some positive $\delta$.
\begin{proposition}\label{propB1}
	The integral \eqref{J4} corresponding to the kernel of $Q$-operators product
	\begin{equation}
	Q_n(\bm{z}_{2n}; \lambda ) = \int_{\mathbb{R}^n} d\bm{y}_n \, \prod_{\substack{i, j =1 \\ i\not=j}}^n \mu(y_i - y_j) \prod_{a = 1}^{2n} \prod_{i = 1}^n K(y_i - z_a) \, e^{2 \pi \imath  \lambda \bby_n}
	\end{equation}
	converges uniformly with respect to the  parameters  $\lambda$, $z_a$, $\bo$, $g$, while the parameters  $z_a$, $\bo$, $g$ range over compact sets preserving the conditions \rf{I0a}, \rf{I0b} and \rf{B2} and the parameter $\l$ varies satisfying the condition \rf{B5}
	\end{proposition}

	{\bf Proof}. Denote integrand by $F$. Using \rf{B3} we arrive at the following bound
	\begin{equation}\label{B6}
	| F | \leq C \, \exp \Biggl(\pi \nu_g \sum_{\substack{i, j =1 \\ i \not= j}}^n |y_i - y_j|
	-\pi \nu_g\sum_{a=1}^{2n} \sum_{i=1}^n |y_i - z_a | + \Re \Bigl(2\pi \imath \lambda \sum_{i=1}^n y_i\Bigr) \Biggr),
		\end{equation}
	where constant $C$ depends on $g, \bo$. Using for the first two sums inequalities
	\beq \label{B6a}
	|y_i - y_j| \leq |y_i| + |y_j|, \qquad |y_i - z_a| \geq |y_i| - |z_a|
	\eeq
	together with $|z_a| \leq M$ (since all $z_a$ vary over compact set) and for the last sum inequality
	\beqq \biggl|\sum_{i=1}^n y_i \biggr|\leq \sum_{i = 1}^n|y_i|\eeqq
	we arrive at
	\beq \label{B7}
	|F| \leq C \, \exp \left(2\pi\nu_g n^2 M +2\pi(| \Im \l|-\nu_g) \sum_{i=1}^n |y_i| \right).
	\eeq
	Since $\l$ satisfies \eqref{B5} the bound \rf{B7} implies the statement of the proposition.	
	\hfill{$\Box$}
	
	{\bf Remark}. As it can be seen from the bound \eqref{B6}, the second inequality from the line \eqref{B6a} and the bound on the measure function
	\begin{equation}
	| \mu(\bz_n) | \leq C \exp\Biggl(\pi \nu_g \sum_{\substack{i, j =1 \\ i \not= j}}^n |z_i - z_j|\Biggr) \leq C \exp\Biggl( 2\pi \nu_g (n - 1) \sum_{i = 1}^n |z_i| \Biggr)
	\end{equation} 
	the product of two $Q$-operators $Q_n(\l)Q_n(\rho)$ is well-defined on fast decreasing functions $f(\bz_n)$ bounded as
	\begin{equation}
	|f(\bz_n)| \leq C \exp \left( - \pi \nu_g \biggl[ \, 3n - 2 + \frac{2 |\Im \rho|}{\nu_g} + \ve \, \biggr] \sum_{i = 1}^n |z_i| \right)
	\end{equation}
	with any $\ve > 0$. In the case $\Im \rho = 0$ the bound doesn't depend on the $Q$-operators parameters
	\begin{equation}
	|f(\bz_n)| \leq C \exp \left( - \pi \nu_g ( 3n - 2 + \ve ) \sum_{i = 1}^n |z_i| \right).
	\end{equation}
   	
   	Besides the integral \rf{J4} over the real plane in Section \ref{sectioneEstimates} we consider  the iterated integral with the same kernel over big semicircles. The study of its convergence and vanishing in the limit splits into three parts: the behavior near real plane where the integrand should rapidly vanish; the total exponential bound of the integrand
   	 \beq  \label{B8} \tilde{F}=\mu(\bm{y}_n) \prod_{a = 1}^{2n} \prod_{i = 1}^n K(y_i - z_a) \eeq
   	 with the exponent  $e^{{2 \pi \imath} \lambda \bby_n}$
   	  in the regular domain $\C\setminus D$; and the exponential bound of the integrand  \rf{B8} in the
   	  irregular domain $D$. The second part is performed by using inequalities \rf{B6a} and exponential bounds
   	  \rf{A12} and \rf{A16}. The third part follows from the same inequalities \rf{B6a} and the results of Section  \ref{sectioneEstimates}.
   	
   	  Finally, for the first part we need a bound similar to \rf{B7} but for the arguments in a cone around a real line.  This can be  done for the parameter $\lambda$ with negative real part, so that
   	  \beq\label{B9} 2\pi\l=-R+\imath\theta,\qquad R>0, \qquad |\theta|\leq 2\pi \delta < 2\pi\nu_g,\eeq
   	  and the integration variables on the cone around a real line with negative imaginary parts
   	  \beq\label{B10} y_i=\bar{y}_i(1\pm\imath\tg \vf_i),\qquad \bar{y}_i\in\R,\qquad \pm\bar{y}_i\tg \vf_i<0, \qquad 0<\vf_i<\sigma.\eeq
   	  Here $\sigma$ is the angle of the cone. The sign $+$ (or $-$) corresponds to $\bar{y}_i < 0$ (or $\bar{y}_i > 0$).
   	  Denote also
   	  \beq \label{B10a} \frac{g}{\o_1 \o_2} = \nu_g(1 + \imath \tg \vf_g), \qquad \a=|\tg\vf_g\tg\sigma|.
   	  \eeq
   	  Suppose the following inequality is satisfied
   	  \beq\label{B11}
   	  	2\pi\nu_g (1-(2n-1)\a) - |\theta|>\ve
   	   \eeq
   	   for some $\ve>0$. For the fixed parameters $g, \theta$ this inequality tells us, how small should be $\sigma$, that is how narrow should be the cone around a real line, in order to have the following bound.
   \begin{proposition}\label{propB2} Under the conditions \rf{B9}, \rf{B10}, \rf{B11} we have the bound
   	\beq\label{B13} | F | \leq C \, \exp {\left( -\ve \sum_{i=1}^n |\bar{y}_i| \right)}.
   	\eeq
   	\end{proposition}
   {\bf Proof.} For the variables on the cone \eqref{B10} use the bound \eqref{A6} for measure function together with \eqref{B10a}
   \beq
	|\mu(y_i - y_j)| \leq C \exp\left( \Bigl| \Re \frac{\pi g}{\o_1 \o_2} (y_i - y_j) \Bigr| \right) \leq C \exp \biggl( \pi \nu_g (1 + \alpha) (|\bar{y}_i| + |\bar{y}_j|) \biggr).
	\eeq
	In the same spirit we use the bound \eqref{A15} for kernel function assuming big enough values of $|y_i|$ (compared to $z_a$) and, as before, $|z_a| \leq M$
	\beq
	|K(y_i - z_a) | \leq C' \exp \Biggl( \mp \Re \frac{\pi \imath g}{\o_1 \o_2} \left[ \imath(y_i - z_a) + \frac{g^*}{2} \right] \Biggr) \leq \tilde{C}' \exp \Bigl( -\pi\nu_g (1 - \alpha) |\bar{y}_i| \Bigr).
	\eeq
   	Therefore, for the whole integrand we have the bound
   		\begin{equation}
   		| F | \leq C \, \exp \Biggl(\pi\nu_g (1 + \a) \sum_{\substack{i,j =1 \\ i \not= j}}^n  (|\bar{y}_i|+ |\bar{y}_j|)
   		-\pi\nu_g (1 - \alpha)\sum_{a=1}^{2n} \sum_{i=1}^n |\bar{y}_i| + |\theta| \, \biggl| \sum_{i = 1}^n \bar{y}_i \biggr| \Biggr),
   	\end{equation}
   	which implies
   		\begin{equation}
   		| F | \leq C \, \exp \left( \bigl[ 2\pi\nu_g ((2n - 1)\a - 1) + |\theta| \bigr] \sum_{i = 1}^n |\bar{y}_i| \right).
   	\end{equation}
   	Then the proposition follows from the condition \rf{B11}. \hfill{$\Box$}

\end{document}